\documentclass[%
 aip,
 amsmath,amssymb,
 reprint,%
]{revtex4-1}

\usepackage{graphicx}
\usepackage{dcolumn}
\usepackage{bm}

\usepackage[utf8]{inputenc}
\usepackage[T1]{fontenc}
\usepackage{mathptmx}
\usepackage{xcolor}
\graphicspath{{./figures/}}
\usepackage{tabularx} 
\usepackage{booktabs} 
\usepackage{subcaption} 

\usepackage{listings}

\definecolor{codegreen}{rgb}{0,0.6,0}
\definecolor{codegray}{rgb}{0.5,0.5,0.5}
\definecolor{codepurple}{rgb}{0.58,0,0.82}
\definecolor{backcolour}{rgb}{0.95,0.95,0.98}

\lstdefinestyle{mystyle}{
    backgroundcolor=\color{backcolour},   
    commentstyle=\color{codegreen},
    keywordstyle=\color{magenta},
    numberstyle=\tiny\color{codegray},
    stringstyle=\color{codepurple},
    basicstyle=\ttfamily\footnotesize,
    breakatwhitespace=false,         
    breaklines=true,                 
    captionpos=b,                    
    keepspaces=true,                 
    showspaces=false,                
    showstringspaces=false,
    showtabs=false,                  
    tabsize=2
}

\begin{document}

\preprint{AIP/123-QED}

\title[Estimating fractal dimensions]{Estimating fractal dimensions:\\a comparative review and open source implementations}

\author{George Datseris}
\email{g.datseris@exeter.ac.uk}
\affiliation{Department of Mathematics and Statistics, University of Exeter, EX4 4QF, United Kingdom}

\author{Inga Kottlarz}

\affiliation{Max Planck Institute for Dynamics and Self-Organization, Am Fassberg 17, 37077  G\"ottingen, Germany}
\affiliation{University Medical Center Göttingen (UMG),
Department of Pharmacology and Toxicology,
Robert-Koch-Str. 40, 37075 Göttingen, Germany}
\affiliation{Institute for the Dynamics of Complex Systems, University of G\"ottingen, Friedrich-Hund-Platz 1, 37077 G\"ottingen, Germany}

\author{Anton P. Braun}
\affiliation{Max Planck Institute for Dynamics and Self-Organization, Am Fassberg 17, 37077 G\"ottingen, Germany}
\affiliation{Institute for the Dynamics of Complex Systems, University of G\"ottingen, Friedrich-Hund-Platz 1, 37077 G\"ottingen, Germany}

\author{Ulrich Parlitz}
\affiliation{Max Planck Institute for Dynamics and Self-Organization, Am Fassberg 17, 37077  G\"ottingen, Germany}
\affiliation{Institute for the Dynamics of Complex Systems, University of G\"ottingen, Friedrich-Hund-Platz 1, 37077 G\"ottingen, Germany}

\date{\today}


\begin{abstract}
The fractal dimension is a central quantity in nonlinear dynamics and can be estimated via several different numerical techniques. In this review paper we present a self-contained and comprehensive introduction to the fractal dimension. 
We collect and present various numerical estimators and focus on the three most promising ones: generalized entropy, correlation sum, and extreme value theory. 
We then perform an extensive quantitative evaluation of these estimators, comparing their performance and precision using different datasets and comparing the impact of features like length, noise, embedding dimension, falsify-ability, among many others. 
Our analysis shows that for synthetic noiseless data the correlation sum is the best estimator with extreme value theory following closely. 
For real experimental data we found the correlation sum to be more strongly affected by noise versus the entropy and extreme value theory. 
The recent extreme value theory estimator seems powerful as it has some of the advantages of both alternative methods. 
However, using four different ways for checking for significance, we found that the method yielded ``significant' low-dimensional results for inappropriate data like stock market timeseries. 
This fact, combined with some ambiguities we found in the literature of the method applications, have implications for both previous and future real world applications using the extreme value theory approach, as, for example, the argument for small effective dimensionality in the data cannot come from the method itself. 
All algorithms discussed are implemented as performant and easy to use open source code via the DynamicalSystems.jl library.

\end{abstract}

\maketitle

\begin{quotation}
When chaotic dynamical systems evolve in time, they typically create sets in the state space that have fractal properties.
One of the major ways to characterize these chaotic sets is through a computationally feasible version of a \emph{fractal dimension}.
In the field of nonlinear dynamics, the correlation sum and the generalized (R\'enyi) entropy are the two most commonly used approaches.
One attempts to find a scaling exponent of these quantities versus a size parameter, and this exponent approximates the fractal dimension. A third method based on extreme value theory is a promising alternative, but it has been developed only recently and hence has not undergone the same amount of scrutiny as the previous two methods.
Here we provide a comprehensive, up to date, and self-contained analysis of available methods, comparing across every conceivable scenario. We also provide open source implementations to compute each method.
\end{quotation}


\section{Introduction}

Fractal geometry deals with geometric objects (or sets) called \emph{fractals}~\cite{mandelbrot1982, Falconer2003}, that are ``irregular'' in terms of traditional Euclidean geometry.
Their most striking property arguably is that they possess structure at all scales, which typically remains invariant in one form or another, no matter how much one zooms into the set.
Because of this, the traditional topological dimension is not fit to describe these sets adequately~\cite[Chap. 5]{DatserisParlitz}.
The concept of a \emph{fractal dimension}, a generally non-integer number, is therefore employed to characterize such objects~\cite{Falconer2003, Ott2008}.
The fractal dimension, which will be shortened to FD in the rest of the article, can be used to quantify the complexity of the geometry, its scaling properties and self-similarity, and the effective ratio of surface areas to volumes~\cite{Falconer2003}. It has been applied in a vast array of different scenarios, from the archetypal measurement of coastlines~\cite{Richardson1961, Mandelbrot1967} to being suggested as a tool for validating abstract art, e.g., that of Pollock~\cite{Taylor1999, Taylor2007}.

The evolution of chaotic dynamical systems results in sets in the state space that are typically fractal~\cite{Kress1986, Ott2008} and thus can be characterized by a FD~\cite{Grassberger1983, Grassberger1983b, Frederickson1983, Takens1985, Paladin1987, Ruelle1989, Theiler1990}, done to our knowledge for the first time by Russel, Hanson, and Ott~\cite{Russell1980}.
For dissipative systems, these sets are called \emph{strange} or \emph{chaotic attractors}~\cite{Russell1980, Ott2012} (boundaries of basins of attraction and chaotic saddles can also be fractal~\cite{Grebogi1983, Banerjee1997}).
For conservative systems they often have special properties and are (typically) called \emph{fat fractals}~\cite{Tel2007}.
For fractal sets resulting from evolving dynamical systems an estimate of FD provides the additional crucial information of the effective degrees of freedom of the time evolution.
This means that calculating the FD for an experimentally obtained dataset can be used to guide the modelling process~\cite{DatserisParlitz, KantzSchreiber2003, Abarbanel1996}, as the ceiling of the FD is the minimum number of independent variables that can model the system. Alternatively, if one has a model that confidently approximates the real system, due to physical arguments, then the FD of the model output can be used to tune model parameters: since FD is a dynamic invariant~\cite{DatserisParlitz, KantzSchreiber2003}, one can compare FDs of the model simulations to the FD obtained by observed data, and change parameters until those two values match.

These reasons have motivated many researchers to compute the FD of many real world systems by delay embedding measured timeseries\cite{KantzSchreiber2003}. Examples include global climate\cite{Nicolis1984, Grassberger1986, Tsonis1993}, physiology~\cite{Babloyantz1989}, lasers~\cite{Kantz1993}, but many more exist. Unfortunately, some of these studies have been challenged, because calculating the fractal dimension in practice is a difficult, error prone process, with limited means of providing confidence to the obtained result. This is, for example, highlighted well in the controversy of estimating FDs of ``climate attractors'' (see e.g., Ref.\cite{Tsonis1993} and references therein), where, FD estimates were incorrectly used to claim very low FD of $\approx 3$ for the whole climate system. Researchers thus often need to interpret results partly subjectively, due to the lack of objective measures.

In this paper we compare as many computationally feasible estimators of FDs as possible, in as objective manner as possible, and across as many scenarios as possible. With this we provide an objective baseline with which researchers can check against their results, reducing the amount of interpretation and subjectivity. It is important to stress the separation between computing the FD of ``the deterministic dynamics'', i.e., the dynamic invariant characterizing the flow in the state space, and the FD of ``the graph of a timeseries $f(t)$ vs $t$''. The latter is relevant for stochastic perspectives, is often related with the Hurst exponent~\cite{Hurst1951, kantelhardt_fractal_2009}, and is typically used as a quantifier of timeseries more suitable for classification tasks versus typical statistics-based quantifiers (see e.g., Refs. \cite{CEPS2021, CEPS2023}). These two FD versions are very different, and unrelated in the general sense. Our paper focuses exclusively on the first version. 

Hence, our main operating assumption is that we have a \emph{multivariate} (sometimes also called \emph{multidimensional}) dataset, obtained from a dynamical system from which we do not know the dynamic rule (equations of motion). If one has a timeseries, it first must be reconstructed into a state space set via delay embedding or other means~\cite{DatserisParlitz, KantzSchreiber2003, Abarbanel1996}. Our operating assumption is on the one hand motivated by the increase of interest in observed or measured multivariate data, and higher accessibility of sensors and experimental data in preparation for a nonlinear-dynamics-based analysis~\cite{Bradley2015}.
On the other hand, there is also noticeable recent progress regarding better attractor reconstruction techniques based on delay embeddings, which can also utilize multivariate measurements, yielding a higher quality reconstruction overall, see e.g., Refs.~\cite{Kraemer2021, Kraemer2022} by Kraemer et al., and references therein. We point out that our goal here is estimating FDs of chaotic sets \emph{knowing that these exist}, in order to separate the problem of the FD estimation from the scientific questions surrounding the interpretation of the data. The later requires following best practices, e.g., various data pre- and post- processing, de-noising, or surrogate tests comparing the FD value of the real data with those of surrogates. These steps are entirely skipped here, and the reader can find more information in standard timeseries analysis textbooks or review articles such as Refs.~\cite{Kantz1993, Abarbanel1996}.

Since the first review on the topic of FD by Theiler in 1990~\cite{Theiler1990}, computers and software have improved and several new algorithms to estimate fractal dimensions have been proposed.
It is thus timely to revisit the subject, and here we will provide a comparison and evaluation of FD estimators across a range of topics much larger than what has been done so far.
Beyond comparing and evaluating various fractal dimension estimators, we provide optimized, easy-to-use, extensively tested, open source implementations for all algorithms discussed in this review in the  DynamicalSystems.jl software library~\cite{Datseris2018}.

\begin{figure*}[!t]
\centering 
\includegraphics[width=\textwidth]{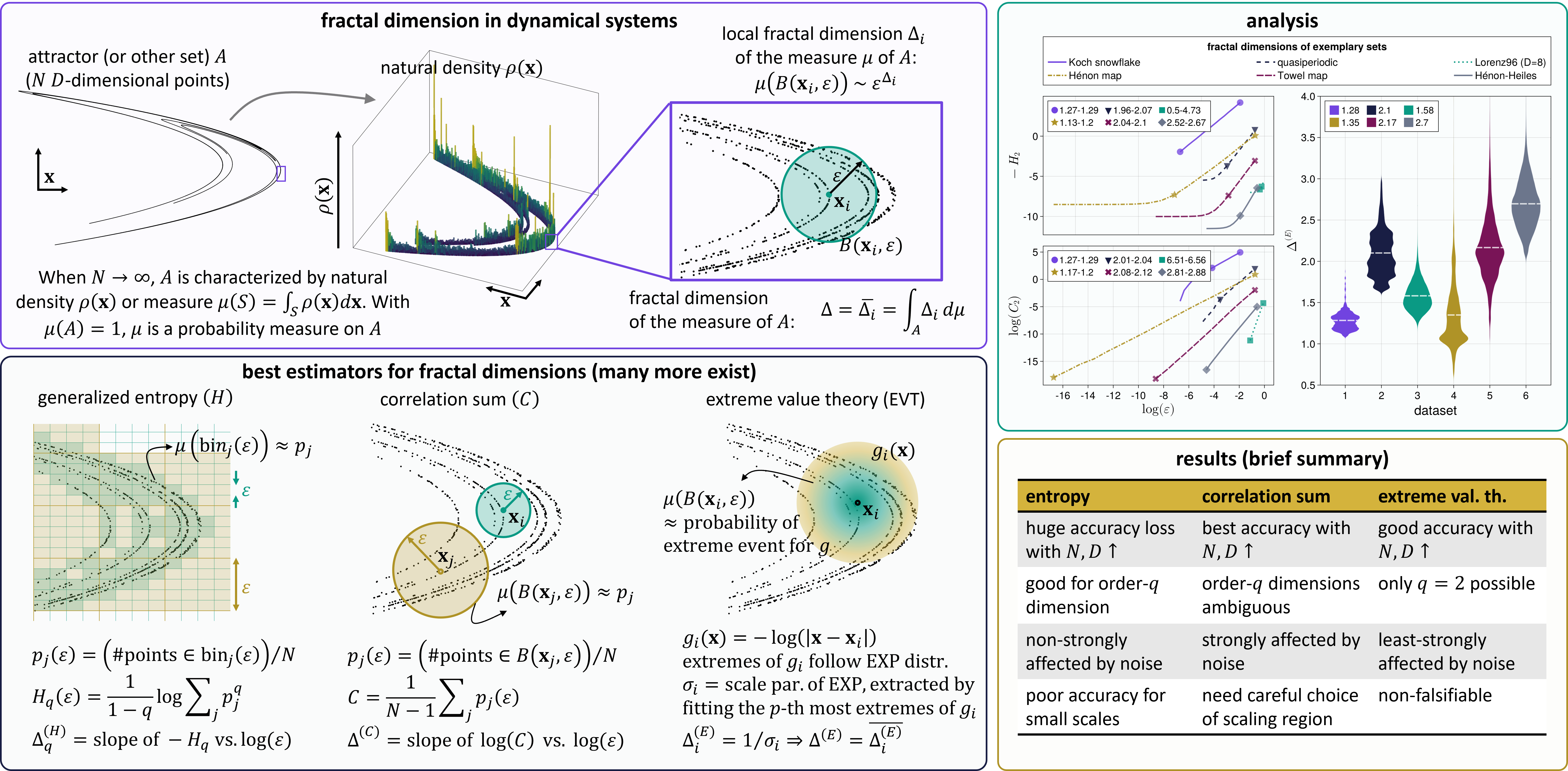}
\caption{Summary and overview of this paper. Sec.~\ref{sec:fd_def} discusses top-left panel. Secs.~\ref{sec:entropy} to \ref{sec:evt_def} discuss bottom left panel. Secs.~\ref{sec:main} and \ref{sec:mainevt} discuss top right panel and lead to the results summarized in the table of the bottom right panel (which are also stated in the conclusions, Sec.~\ref{sec:conclusions}).}
\label{fig:summary}
\end{figure*}

This review and comparison paper is structured as follows (and see also Fig.~\ref{fig:summary} for a summary of the paper). 
In Sec. \ref{sec:status_quo} we provide a self-contained concise definition of the major methods used to compute FDs and how they relate with the natural density and with each other.
Connected with this section is Appendix~\ref{app:all}, which presents all (computationally feasible) algorithms we have found for estimating a fractal dimension. The core of the paper are Sec. ~\ref{sec:main} and Sec. \ref{sec:mainevt}, which compare in detail the three best and most popular estimators of FDs: scaling of entropy, scaling of correlation sum, and an extreme value theory approach.
We compare across data dimensionality, data length, different kinds of dynamical systems, noise levels, real world data, various embeddings of timeseries, the order $q$ of the fractal dimension, among others.
We close the paper with a summary of our findings in Sec. \ref{sec:conclusions}.
Every result, plot, and method that we present in this paper is fully reproducible via open source code, and adjustable to other input datasets, see Appendix~\ref{app:software} for details.

\section{State of the art}
\label{sec:status_quo}
An easy to understand introduction and review regarding the fractal dimension was given by Theiler in 1990~\cite{Theiler1990}.
The theoretical background of the fractal dimension and the methods (known until 1990) to compute it are summarized, and a plethora of historic references is given there.
A more recent publication on fractal dimensions in a style of a review is given by by Lopes and Betrouni in 2009~\cite{Lopes2009}.
However, it is focused on image analysis and pattern recognition, and does not do any quantitative comparison.
The most recent detailed source that provides quantitative information on the limitations and pitfalls of fractal dimension estimates is the well known textbook by Kantz and Schreiber~\cite{KantzSchreiber2003}.

A number of ways to estimate a FD, that are applicable to dynamical systems, have been devised in the literature.
For this comparison we found, implemented, and compared the methods that we briefly summarize in Table~\ref{tab:all_estimators}.
From these, our main analysis focuses on the three most prominent ones. In the rest of this section we will provide a concise, yet self-contained, summary of the concept of a fractal dimension in dynamical systems. Then, we introduce the three main estimators that we use in the extensive comparisons done in Secs. \ref{sec:main} and \ref{sec:mainevt} and illustrate how the different estimators connect to the natural density and with each other. 
In the following, we assume that we have a set $X$ that contains $N$ $D$-dimensional points representing a (possibly observed) multivariate timeseries of a dynamical system.
We will use the letter $\Delta$ to denote various versions of a fractal dimension, and we will use a superscript in parenthesis, such as $\Delta^{(S)}$ to denote the particular \emph{estimator} used to estimate $\Delta$.

\begin{table*}[t]
    \centering
    \begin{tabularx}{\textwidth}{lll}\toprule
        Estimator &   Brief description & Main refs. \\  \midrule
        Natural measure entropy &  Scaling of generalized entropy $H_q$ of amplitude binning, versus box size & \cite{Russell1980} \\
        Molteno's histogram optimization & Optimized algorithm for amplitude binning with restricted size $\varepsilon$ & \cite{Molteno1993} \\
        Correlation sum & Scaling of correlation sum $C_q$ versus radius $\epsilon$  & \cite{Grassberger1983,Grassberger2007} \\
        Box/prism-assisted correlation sum & Optimized algorithm to calculate the correlation sum & \cite{Theiler1987} \\
        Performance-optimized box size & Optimized for performance box-assisted algorithm & \cite{BuenoOrovio2007} \\
        Logarithmic correction & Better converging fit instead of the standard least square fit of $\log(C_2)$ vs. $\log(\varepsilon)$ & \cite{Sprott2001}\\
        Takens' estimator & Maximum likelihood estimation of scaling exponent, $C_2(\varepsilon)\propto \varepsilon^{\Delta}$ for $\varepsilon\in(0,\varepsilon_\text{max}]$ & \cite{Takens1985,borovkova1999} \\
        Judd's estimator & Binned MLE with additional degrees of freedom from polynomial & \cite{Judd1992,Judd1994} \\
        Mean return times &  Logarithmic scaling of mean return time to an $\varepsilon$-sphere, versus $\varepsilon$ & \cite{Jensen1985, Theiler1990} \\
        Lyapunov dimension & Kaplan and Yorke's linear interpolation for sum of Lyapunov exponents = 0 & \cite{Kaplan1979} \\
        Lyapunov dimension via fits & Higher-order interpolations and other fits for sum of Lyapunov exponents = 0 & \cite{Chlouverakis2005} \\
        Extreme value theory based & Rare events follow a Pareto distribution whose parameter is the fractal dim. & \cite{Lucarini2016} \\
        Persistent homology & Quantify how the topology of shape changes as it is thickened & \cite{Jaquette2020} \\
        \bottomrule
    \end{tabularx}
    \caption{Description of various estimators for fractal dimension considered in this paper, see also Appendix~\ref{app:all}.}
    \label{tab:all_estimators}
\end{table*}

\subsection{What does a fractal dimension really quantify?}
\label{sec:fd_def}

In the introduction we discussed how the FD is useful for practical matters, and hence worthy of being estimated. However, before discussing any estimators for a FD, it is useful to conceptualize what the FD truly characterizes, as this is the origin of all practical estimations. 

For this discussion we ignore the presence of noise existing in real data, and the fact that observed timeseries need to be delay-embedded to yield higher dimensional data. The starting assumption therefore is that the set $X = \{\mathbf{x_1}, \ldots, \mathbf{x}_N\}$ we have at hand is a faithful sampling of some sort of $D$-dimensional invariant set (typically a chaotic attractor) of a dynamical system, i.e., $X \subset A$. $A$ is itself characterized by its \emph{natural, or invariant, measure} $\mu(\mathbf{x})$ which defines a probability space on top of $A$ by requiring $\mu(A)=1$. Equivalently we may use the natural density $\rho(\mathbf{x}) d\mathbf{x} = d\mu$, which in practical terms is the $D$-dimensional histogram of $X$. To learn more on the natural measure one should consult various textbooks on nonlinear dynamics~\cite{DatserisParlitz, Tel2007, Ott2012} or the review by Theiler~\cite{Theiler1990}. In essence, while the samples $\mathbf{x}_i \in X$ form a time sequence of points with deterministic origin, they can be also be thought of as points sampled ``at random'' from $A$ according to the measure $\mu$. Note that throughout this FD description we make the fundamental assumption that $A$ (with measure $\mu$) is ergodic.

A FD is a number characterizing the scaling of $\mu$ with the ``scale'' one looks at $\mu$ at. Let $\mathcal{B}(\mathbf{x}, \varepsilon)$ be a $D$-dimensional sphere\footnote{Sphere is used due to the usage of the Euclidean norm to calculate distances. If using the maximum norm, we would use a hypercube instead. The fractal dimension is invariant to the choice of the distance norm.} centered at $\mathbf{x}$ with radius $\varepsilon$. For sufficiently small $\varepsilon$, and for almost all $\mathbf{x}_i \in A$, it is assumed that the scaling of $\mu$ with $\varepsilon$ is exponential, so that
\begin{align}
    &\mu(\mathcal{B}(\mathbf{x}_i, \varepsilon)) \sim \varepsilon^{\Delta_i} \Rightarrow  \nonumber  \\  &\Delta_i = \lim_{\varepsilon \to 0} \frac{\log \mu(\mathcal{B}(\mathbf{x}_i, \varepsilon)) }{\log \varepsilon}.
\label{eq:origins1}
\end{align}
Here $\Delta_i$ is labelled the \emph{local dimension} at $\mathbf{x}_i$, and by construction $\Delta_i \le D$. The assumption in Eq.~\eqref{eq:origins1} is a reflection of our intuitive notion of dimension, which describes that the ``bulk'' or the ``amount'' of a set scales with a linear size (scale) exponentiated to the dimension. E.g., the amount of a cube scales with its side length to the 3rd power. here we measure ``amount'' of a subset of $A$ by its natural measure, i.e., its relative probability mass. From $\Delta_i$ a fractal dimension characterizing $A$ as a whole can be obtained as the mean, 
\begin{equation}
    \Delta = \int_A \Delta_i \, d\mu.
    \label{eq:origins2}
\end{equation}

This intuition-based definition of a local and attractor dimension is further motivated by Theiler in his review article on the basis of the Hausdorff dimension~\cite{Theiler1990}. 
Eq.~\eqref{eq:origins1} however is non computable because $\mu$ is unknown, since in practice we have only partial knowledge of $\mu$ due to the finite observations of $X\subset A$. Besides, for the overwhelming majority of dynamical systems, $\mu$ does not have an analytic expression anyways, irrespectively of finite data. 
An estimator of a FD therefore attempts to measure either the local scaling of Eq.~\eqref{eq:origins1} or the global scaling by averaging in Eq.~\eqref{eq:origins2}.

Before going into specific estimators we need to stress that none of the estimators we considered in this review yield the Hausdorff dimension~\cite{Hausdorff1918, Falconer2003}. Often in the literature researchers use the term ``Hausdorff dimension'' to refer to the output of the estimators, but no-one has provided any formal proof of the equivalence between the estimates and the Hausdorff dimension (which itself has a very precise and rigorous mathematical definition).
This is especially so for the box counting dimension, where it is easy to prove that it cannot be the same as the Hausdorff dimension\footnote{Proof: Consider the rationals in the interval $[0, 1]$. The rationals are dense in this interval, and as such the box counting dimension of the rationals is exactly 1, as, no matter the box size $\varepsilon$, the whole number line will always be covered by boxes (i.e., there will never be a box without a rational in it). The rationals however are a countable set, and the Hausdorff dimension of any countable set is 0.}.

\subsection{Entropy of natural density}
\label{sec:entropy}
The first way to define and compute $\Delta$ is based on an approximation of the natural density $\rho$ of the set.
By discretizing the state space in boxes of size $\varepsilon$ one can assign a probability $p_i, \, i = 1,\ldots,M$ to each box, which is simply the count of points in the box divided by the total amount of points $N$.
These probabilities approximate $\rho$, and from there the R\'enyi (also called generalized) entropy can be obtained as~\cite{Renyi1959}
\begin{equation}
    H_q = \frac{1}{1-q} \log \left(\sum_{i=1}^{M} p_i^q\right).
    \label{eq:renyi_entropy}
\end{equation}
Here $q$ is the order of the entropy, and it allows putting more ($q<1$) or less ($q>1$) weight to boxes with relatively smaller visitation frequency by the trajectory (and hence smaller $p_i$).
For $q = 1$ $H_q$ reduces to the known Shannon entropy while for $q = 0$ it becomes $\log(M)$ with $M$ the minimum number of non-overlapping boxes needed to cover the set $X$.
More than one way exists for estimating the probabilities $p_i$, see Appendix~\ref{app:all}.
In the main comparison of Sec. \ref{sec:main} we will use the algorithm~\ref{app:histogram} which works for any values of $D, N, \varepsilon$ while having performance scaling of $D\cdot N \cdot \log(N)$.

To connect $H_q$ with $\mu$ we re-write $\sum_i p_i^q = \sum_i p_i p_i^{(q-1)}$. This is a weighted average, equaling to $\langle p^{(q-1)} \rangle$, since $\sum p_i = 1$.
We note that $\langle p \rangle$ is our notion of ``amount'', as we measure amount by the probability mass. Regularizing the expression by its exponent, the quantity $\langle p^{(q-1)} \rangle^{1/(q-1)} \equiv \exp (-H_q)$ is the ``average bulk'' or ``average amount'' in a hypercube of linear size (i.e., scale) $\varepsilon$. The number $q$ settles the way we average: for $q=2$ we have the arithmetic (typical) average, for $q=3$ a root mean square and for $q\to 1$ a geometric average. 

By following the same intuition that led to Eq.~\eqref{eq:origins1} (that ``amount'' $\sim$ scale $\string^$ exponent), the so-called \emph{generalized dimension of order $q$}~\cite{Grassberger1983q, Hentschel1983}, is defined as
\begin{equation}
    \Delta_q^{(H)} = \lim_{N\to \infty} \lim_{\epsilon \to 0}\left( \frac{- H_q(\epsilon)}{\log \varepsilon }\right).
    \label{eq:generalized_dim}
\end{equation}
This definition was used, to our knowledge, for the first time for $q=1$ by Russel, Hanson and Ott~\cite{Russell1980}.
$\Delta_0^{(H)}$ is called the \emph{box counting or capacity dimension}, while $\Delta_1^{(H)}$ is called the \emph{information dimension}. The box counting dimension can also be thought as the fractal dimension of the \emph{support} of the attractor, as it disregards the values of $\mu$.

In Eq.~\eqref{eq:generalized_dim} both limits are theoretical and cannot be realized in practice. As a result, $\Delta_q^{(H)}$ is estimated by plotting $-H_q$ vs. $\log(\varepsilon)$ and estimating the slope of a linear scaling region (for sufficiently large $N$, more on this in Sec. \ref{sec:estimating_slope}).

If the fractal dimension depends on $q$, the set is called multi-fractal.
This is the case when the natural measure underlying the attractor is strongly non-uniform.
Large positive values of $q$ then put more emphasis on regions of the attractor with higher natural measure.
In dynamical systems theory, almost all chaotic sets are multi-fractal~\cite{Paladin1987}.
The dependence of $\Delta_q^{(H)}$ on $q$ is connected with another concept, the so-called singularity spectrum, or multifractality spectrum $f(\alpha)$, however we will not be calculating $f(\alpha)$ here and refer to other sources for more (see e.g. chapter 9 of Ref.~\cite{Ott2012} by Ott).

\subsection{Correlation sum}
\label{sec:correlationsum}
The second way of defining and estimating a fractal dimension uses the correlation sum~\cite{Grassberger1983}. The correlation sum is an alternative way of estimating the average amount at a given scale, based on the points nearby a given point. Specifically, we associate an amount $S$ at point $\mathbf{x}_i \in X$ and at scale $\varepsilon$ as 
\begin{equation}
    S(\mathbf{x}_i, \varepsilon) = \frac{1}{N-1} \sum_{j\ne i} B(||\mathbf{x}_i - \mathbf{x}_j|| < \varepsilon)
    \label{eq:pointwise}
\end{equation}
with $||\cdot||$ some distance norm (here always the Euclidean, but different norms have little practical impact and theoretically exactly no impact) and $B(\cdot) = 1$ if its argument is true and 0 otherwise. I.e., we count the $\varepsilon$-close neighbors of $\mathbf{x}_i$ in $X$. The exponential scaling of $S(\mathbf{x}_i, \varepsilon)$ versus $\varepsilon$ is called the \emph{pointwise  (local) dimension}~\cite{Hidaka2014}. The average ``amount'' $S$ over $X$ is called the \emph{correlation sum}, given by $C(\varepsilon) = \sum_i S(\mathbf{x}_i, \varepsilon)/N$.

Using the same reasoning as in the preceding subsection, this average amount is expected to scale with the linear size $\varepsilon$ exponentiated to the FD. And, again similarly with the preceding section, we do not have to limit ourselves to the typical arithmetic average when defining the correlation sum, but can introduce an order $q$ that adjusts how we are averaging the ``amount''. This leads to the definition of the $q$-order correlation sum ~\cite{Grassberger1983b,KantzSchreiber2003,Grassberger2007}
\begin{align}
\label{eq:correlationsum}
        C_q(\varepsilon) & = \left[ \sum_{i=1}^{N} \mathcal{N}_i \left[\sum_{\substack{j=1\\|i - j| > w}}^{N} B(||\mathbf{x}_i - \mathbf{x}_j|| < \varepsilon)\right]^{q-1} \right] ^{1/(q-1)} \\
        \mathcal{N}_i & =  \frac{1}{N\left(\max(N-w, i) - \min(w+1,i)\right)^{q-1}} \nonumber
\end{align}
Here we also added $w\ge 0$, the correction by Theiler~\cite{Theiler1986} (known as Theiler window) which excludes as neighbors points that are temporally close. This removes spurious correlations due to dense sampling of continuous dynamical systems. 
We choose $w$ as follows: for each timeseries present in the multi-dimensional input dataset $X$ we calculate the first minimum of its self-mutual information~\cite{DatserisParlitz}. The maximum of the time shifts corresponding to these minima is chosen as $w$.

Originally the version explicitly having $q=2$ was used to define the correlation dimension~\cite{Grassberger1983}, and the process of defining a FD was in fact the same as in $H_q$ for any $q$.
A linear scaling region is estimated from the curve of $\log(C_q)$ versus $\log(\varepsilon)$.
Then the correlation-sum-based FD $\Delta^{(C)}_q$ is the slope of that linear region. 

We may leverage two potential improvements here. 
First, to calculate $C_q(\varepsilon)$ we used a box-assisted method\cite{Theiler1987, BuenoOrovio2007}. We modify this method as discussed in App.~\ref{app:boxassisted}, because otherwise it fails for data with even a small amount of noise (see discussions in Sec. \ref{sec:noise} and App.~\ref{app:boxassisted}). Second, besides the standard least squares fit $\log(C_q) \sim a + \Delta_q^{(C)}\log(\varepsilon)$, we also used the correction by Sprott and Rowlands~\cite{Sprott2001} when possible (i.e. when at least half the range has $\log(\varepsilon)<0$). Ref.\cite{Sprott2001} optimizes the fit $\log(C_2) \sim a + \Delta_2^{(C)}\log(\varepsilon) + b \log(-\log(\varepsilon))$ ($a,b$ are parameters to be optimized in parallel with $\Delta_2^{(C)}$). It is intended to give better fits for sets that have a slowly-converging fractal dimension estimate, but as we will show later, it is best to not use it in practice.

In the rest of the manuscript we will use $H_q$ or $C_q$ to refer to the methods of estimating FD via the (generalized) entropy or (generalized) correlation sum. We will explicitly use a subscript $H_2, C_2$ when we make statements that apply only to this particular order $q=2$. 

\subsection{Extreme value theory}
\label{sec:evt_def}
The third major way of defining and estimating a FD from a set $X$ is based on extreme value theory (EVT) applied to dynamical systems~\cite{Lucarini2016}. The method estimates a local dimension $\Delta^{(E)}_i$, as in Eq.~\eqref{eq:origins1}, and then provides the FD as the average. Interestingly, the method utilizes exactly the same information as the correlation sum: all inter-point distances. However, $\Delta^{(E)}_i$ is not estimated directly via an exponential scaling relationship in contrast to the generalized entropy and correlation sum methods. 

To the best of our knowledge, the method has been developed using progress across several papers~\cite{Felici2007, Freitas2010, Faranda2011, Lucarini2012, Caby2019, Pons2020}, see also Ref~\cite[Chap. 4 and 9]{Lucarini2016}. 
Even though relatively recent, this method has been applied already to a plethora of real-world cases, see e.g., Refs. \cite{Faranda2017, Buschow2018, Faranda2019, Brunetti2019, Rodrigues2018, Gualandi2020, Messori2021, Giamalaki2021, Hochman2021, Falasca2022}, and many more, see Ref.~\cite{Flavio2023} for a summary of recent applications.
Despite this plethora of applications, we have noticed that some applications have ambiguities with the basic theory that connects EVT with FD, and we discuss these issues in App.~\ref{app:evt}.

First, let us summarize the computational algorithm to estimate a FD via EVT, and then discuss how it connects to the natural measure $\mu$.
Let $g(\varepsilon) = -\log(\varepsilon)$ be a function of a distance (or radius) in state space. For the $i$-th point in $X$, we estimate
\begin{equation}
    g_i = -\log(||\mathbf{x}_i - \mathbf{x}_j|| ) \; \forall j \ne i
    \label{eq:evt_distances}
\end{equation}
with $||\cdot||$ the Euclidean distance. Note that $g_i$ is a real-valued vector of length $N-1$.
Next, we choose an ``extreme'' probability $p$ for a quantile of the distribution of $g_i$ (e.g., $p=0.99$). We found no reference that clarified what ``extreme'' means in a mathematically precise way, but we discuss in high detail how the choice of $p$ impacts the results in Sec.~\ref{sec:evt_quantile}.

In any case, we then compute $g_p$ as the $p$ quantile of $g_i$. Then, we collect the \emph{exceedances} of $g_i$, defined as 
\begin{equation}
    E_i = \{ g_i - g_p: g_i \ge g_p \},
    \label{eq:exceedances}
\end{equation}
i.e., all values of $g_i$ larger or equal to $g_p$, also shifted by $g_p$. $E_i$ is also a real-valued vector with $n = \lfloor N(1-p) \rfloor$ values in total. 
Now, according to extreme value theory~\cite{Lucarini2012, Leadbetter2012}, in the limit $N \to \infty, p \to 1$, and for the particularly chosen form of the function $g$, the values $E_i$ follow a generalized Pareto Distribution (GPD) with parameters $\sigma, \xi$ and shift parameter 0 (see App.~\eqref{app:evt} for more on GPD).
However, if the measure $\mu$ and attractor $A$ satisfy the criterion of Eq.~\eqref{eq:origins1}, then the GPD is reduced to an exponential distribution (EXPD) with parameter $\sigma$, $E \sim \exp(E/\sigma)$. 
Within this extreme value theory approach, the local dimension $\Delta^{(E)}_i$ assigned to state space point $\textbf{x}_i$ is given by the inverse of the $\sigma_i$ parameter of the EXPD fit to the exceedances, i.e.
\begin{align}
    \Delta^{(E)}_i &=  1/\sigma_i \\
    \sigma_i &= \bar{E_i} \nonumber = (N-1) / \sum E_i .
\end{align}
In the above expression we explicitly used the maximum likelihood estimator for fitting the $\sigma$ parameter, which is simply the sample mean for an exponential distribution.
Additionally, the above expression places one additional assumption on the local scaling of the natural measure $\mu$, see discussion below around  Eq.~\eqref{eq:evt_additional_assumption}.
The FD for $A$ follows as the arithmetic mean,
\begin{equation}
    \Delta^{(E)} = \frac{1}{N}\sum_{i = 1}^{N} \Delta^{(E)}_i,\label{eq:evt}
\end{equation}
which means that the EVT FD corresponds to a FD of order $q=2$.

Let us now discuss how this theory connects to Eq.~\eqref{eq:origins1} and hence relates $\Delta_i$ to the $\sigma_i$ parameter.
Due to the invertibility of the $g$ function, the exceedances $E_i$ correspond to when the orbit of the dynamical system comes closer than $\varepsilon^*$ to the reference point $\mathbf{x}_0$, with $\varepsilon^* = \exp(-g_p)$.
This clarifies why we focus on extremes of $g_i$: we are interested in what happens around a small radius $\varepsilon$ around a reference point $\mathbf{x}_0$, in order to connect with the fundamental definition of the local dimension of Eq.~\eqref{eq:origins1}.

Now, let us discuss the probability $\pi_i(E)$ that \emph{given} that an exceedance has occurred (i.e., a state $\mathbf{x}_j$ is at least $\varepsilon^*$-close to $\mathbf{x}$), there is an exceedance of $E$. This probability is by construction given by 1 minus the cumulative distribution function of the fitted EXPD, i.e., $\pi_i(E) = \exp(-E/\sigma_i)$. However, the same probability can be constructed in terms of the natural measure $\mu$ to be 
\begin{equation}
    \pi_i(E) = \frac{\mu\left(\mathcal{B}(\mathbf{x_i}, \varepsilon)\right)}{\mu\left(\mathcal{B}(\mathbf{x_i}, \varepsilon^*)\right)}
    \label{eq:prob_evt_exceedance}
\end{equation}
with 
\begin{equation}
    \varepsilon = \exp(-g_p - E) \le \varepsilon^* = \exp(-g_p).
    \label{eq:exceedances_radii_relation}
\end{equation}
We can relate $\Delta_i$ with $\sigma_i$ if we place one more assumption on $\mu$. We assume that locally around $\mathbf{x}_i$ 
\begin{equation}
    \mu\left(\mathcal{B}(\mathbf{x_i}, \varepsilon)\right) = f_i(\varepsilon) \varepsilon^{\Delta_i} 
    \label{eq:evt_additional_assumption}
\end{equation}
where $f_i(\varepsilon)$ is a \emph{slowly varying function of $\varepsilon$} as $\varepsilon \to 0$, which may or may not depend on reference point $\mathbf{x}_i$.
With this assumption, both numerator and denominator of Eq.~\eqref{eq:prob_evt_exceedance} scale exponentially with $\Delta_i$ and the remaining factors cancel-out even though $\varepsilon \ne \varepsilon^*$. By taking into account also the expressions of Eq.~\eqref{eq:exceedances_radii_relation}, and that $\pi_i(E) = \exp(-E/\sigma_i)$, we put everything together and have
\begin{align}
      \exp(-E/\sigma_i) &=  (\varepsilon / \varepsilon^*)^{\Delta_i} \nonumber \\
      &= \frac{\exp(-(g_p + E)\Delta_i)}{\exp(-g_p\Delta_i)} \nonumber \\
      &= \exp(-E\Delta_i) \label{eq:evt_equivalence}
\end{align}
from which $\Delta_i = 1/\sigma_i$.

\subsection{Lyapunov (Kaplan-Yorke) dimension}
\label{sec:lyapunov}
Lastly, a FD estimate proposed by Kaplan and Yorke~\cite{Kaplan1979} is based on the Lyapunov exponents characterizing an attractor.
Let $ \{ \lambda_i \}$ denote the Lyapunov spectrum, with $\lambda_1 \ge \lambda_2 \ge \ldots \ge \lambda_D$.
Then the Lyapunov dimension is defined as
\begin{equation}
    \Delta^{(L)} = \ell + \frac{\sum_{i=1}^\ell \lambda_i}{|\lambda_{\ell+1}|},\quad \ell = \max_j \left[ \sum_{i=1}^j \lambda_i > 0 \right].
    \label{eq:kaplan}
\end{equation}
In simple terms, it is the (linearly interpolated) index value where the sum of the Lyapunov exponents of the set first crosses zero. Table~\ref{tab:kaplan} provides estimates for the systems we use in this paper.
It is conjectured that $\Delta^{(L)} \approx \Delta^{(H)}_1$, see also~\cite{Ledrappier1981}, and generally one does find similar values in practice, however there is no formal proof yet.
An extension to Eq.~\eqref{eq:kaplan} has been proposed in~\cite{Chlouverakis2005} that is not included in the main comparison but discussed in Appendix~\ref{app:all}.
Keep in mind that Eq.~\eqref{eq:kaplan} is defined only for dissipative systems, where $\sum \lambda_i < 0$. Applying it to conservative systems does not make much sense. E.g., Hamiltonian systems satisfy $\lambda_i = - \lambda_{D-i+1}$ and obtain $\Delta^{(L)} = D$ always, even though the motion might happen in a lower-dimensional manifold. For the 4D H\'enon-Heiles system (Fig.~\ref{fig:different_ds}) energy conservation limits the dynamics in a 3D manifold, and thus its FD cannot be greater than 3.

$\Delta^{(L)}$ has a huge advantage when compared to the previous definitions of fractal dimensions: it can be computed with very  high precision, even for high dimensional systems (where the other methods typically suffer from accuracy, as we will show below).
But it also has a huge disadvantage: practically it can be computed only if the dynamic rule (equations of motion) is known. Only using the dynamical rule and its linearization one can estimate the entire Lyapunov spectrum with satisfactory precision, for example by means of the known algorithm due Shimada and Nagashima~\cite{Shimada1979} and Benettin et al.~\cite{Benettin1980}.
From a finite, and often noisy real world dataset, calculating the entire spectrum of exponents is a very challenging task that requires for higher dimensional attractors very large data sets~\cite{Parlitz_2016}. Therefore one is in most cases better off calculating a fractal dimension directly from data, or instead fit an explicit model to the data, e.g.~Ref.\cite{Pathak2017}.

Unfortunately, it is not straightforward to connect $\Delta^{(L)}$ to Eq.~\eqref{eq:origins1}. One may find some intuition in Sec. 11.5.2. of Ref.~\cite{KantzSchreiber2003}, but this discussion relies on expansion and contraction rates in state space and hence falls outside the scope of this review.

\begin{table}[t]
    \begin{tabular*}{\columnwidth}{p{120pt}p{30pt}l}\toprule
        System & $D$ & $\Delta^{(L)}$ \\  \midrule
        Lorenz96 & 4 & 2.99 \\
        Lorenz96 & 6 & 4.93 \\
        Lorenz96 & 8 & 6.91 \\
        Lorenz96 & 10 & 8.59 \\
        Lorenz96 & 12 & 10.35 \\
        Lorenz96 & 14 & 12.10 \\
        Lorenz96 & 32 & 27.68 \\
        R\"ossler (chaotic) & 3 & 1.9 \\
        H\'enon map & 2 & 1.26 \\
        Kaplan-Yorke map & 2 & 1.43 \\
        Towel map & 3 & 2.24\\
        Coupled logistic maps & 8 & 8 \\
        Kuramoto-Sivashinsky & 101 & 31.76 \\
        \bottomrule
    \end{tabular*}
    \caption{Lyapunov (Kaplan-Yorke) dimensions for systems with known dynamic rule (listed in Table~\ref{tab:all_systems_rules}). }
    \label{tab:kaplan}
\end{table}

\section{Correlation sum vs. Entropy}
\label{sec:main}
In this section we perform a quantitatively rigorous and exhaustive comparison of the methods based on entropy $H_q$ and correlation sum $C_q$. Even though fundamentally different, both rely on estimating the scaling of some quantity versus some size $\varepsilon$. For a given $q$ they both (in theory) approximate the same quantity, the exponential scaling of the $q$-average of ``amount'' of measure versus the scale, as we illustrated in Sec~\ref{sec:status_quo}. To compute $H_q$ we use the method \ref{app:histogram}, and for $C_q$ the method \ref{app:boxassisted} for most cases, and the straightforward implementation \ref{app:vanilla} for very high dimensional data. For $C_2$ we also tested the logarithmic correction of Ref.~\cite{Sprott2001}.
The motivation of choosing these methods, and why an exhaustive comparison of other methods is not presented, is explained in Appendix~\ref{app:all}. 
Before any numerical analysis we normalize input data $X$ so that each of its columns is transformed to have 0 mean and standard deviation of 1.
This linear transformation leaves dynamic invariants (like the FD) unaffected, however makes all numerical methods more accurate and faster to converge.

In the following subsections all results will be presented with the same plot type, as e.g. shown in Fig.~\ref{fig:analytic}.
The legend shows the different datasets used in the plot and a description of the plot's purpose.
Top panel is entropy estimate while bottom is correlation sum.
To estimate $\Delta_q^{(H)}, \Delta_q^{(C)}$, for each curve we identify automatically, and objectively, a linear scaling region as discussed in Sec. \ref{sec:estimating_slope}.
This region is denoted by markers of the same color on each curve.
The secondary legends inside the panels provide the 5-95\% confidence intervals for the estimated slopes of these segments.
Unless otherwise stated, all datasets used have length $N=10^5$, which is for many experiments a typical upper bound for the amount of data one has access to. Continuous time systems are sampled with approximately 10 points per characteristic oscillation period.

\subsection{Benchmark sets with known dimensions}
\begin{figure}
    \centering
    \includegraphics[width=\columnwidth]{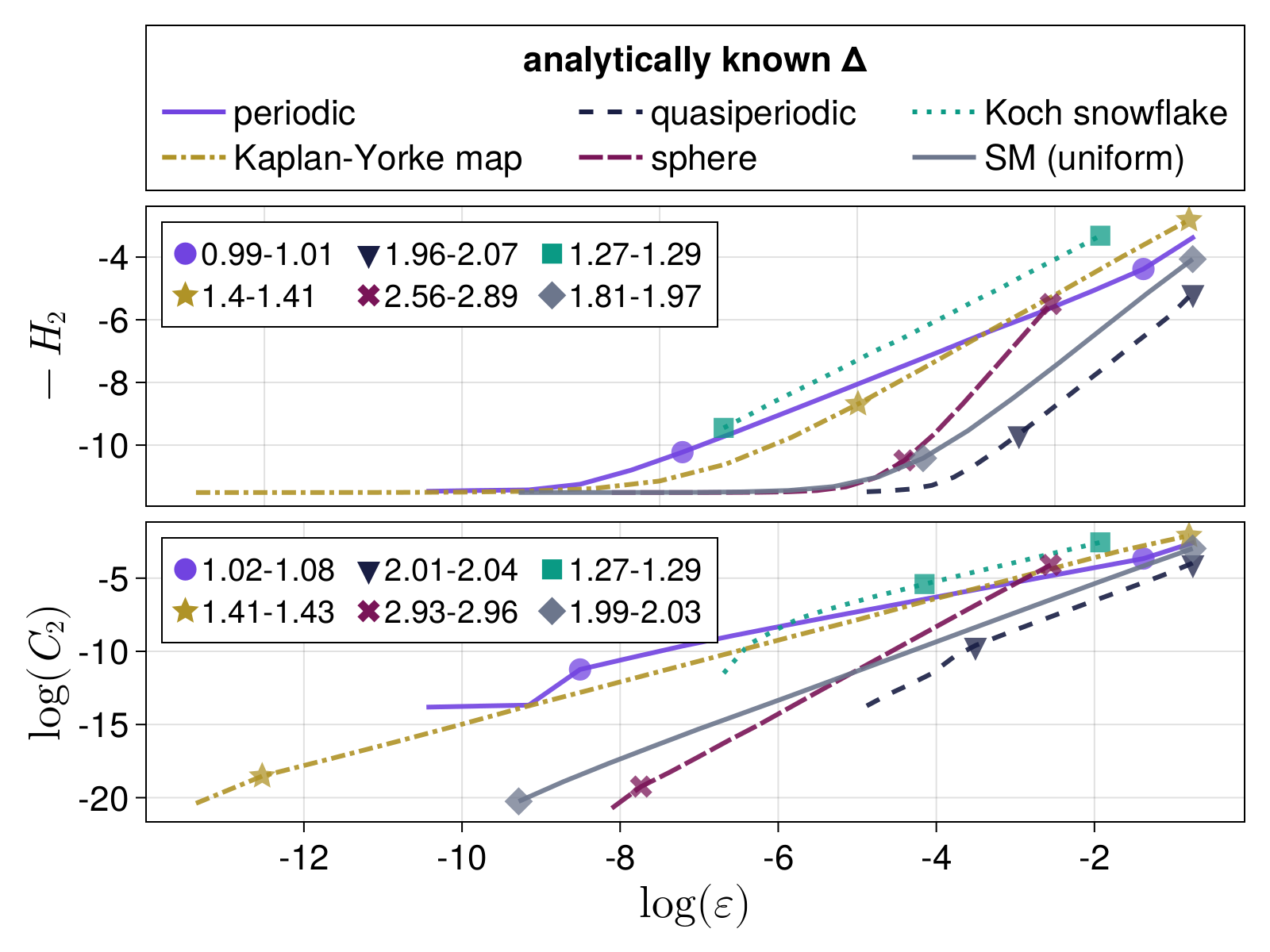}
    \caption{Fractal dimension estimates of $\Delta_2^{(H)}$ and $\Delta_2^{(C)}$  for sets with analytically known fractal dimension. The figure title maps colors to set names. The legends within each axis map colors to two numbers which mean the (5\%, 95\%) confidence intervals corresponding to the estimation of the fractal dimension (curve slope). The markers on the curves denote the start and end of the estimated linear region.}
    \label{fig:analytic}
\end{figure}

The best place to start is a simple sanity and accuracy test of the two main estimators for sets whose fractal dimension can be computed analytically in a straightforward manner. This is shown in Fig.~\ref{fig:analytic}. We use a periodic orbit from the R\"ossler system ($\Delta=1$), a quasiperiodic orbit of order 2 from the H\'enon-Heiles system ($\Delta=2$), the Koch snowflake ($\Delta = \log(4)/\log(3)\approx 1.262$), the Kaplan-Yorke map ($\Delta = 1 - \log(2)/\log(0.2)\approx 1.4306$), a uniform filling of the 3D sphere ($\Delta = 3$) and a chaotic trajectory of the Standard Map (SM) for very high $k=64$, which covers uniformly the state space and thus has $\Delta = 2$.

Some results that will be repeatedly seen throughout this section become clear. For \emph{noiseless data}, the correlation sum method is clearly better for three reasons.
First, its linear scaling region covers a wider range of scales, while $H_2$ saturates much more quickly to flatness for small $\varepsilon$. Notice that $\log(C_2)$ cannot saturate for small $\varepsilon$, but instead diverges to $-\infty$, but we can never reach this point due to the process that chooses the appropriate overall range of $\varepsilon$ (Sec.~\ref{sec:estimating_slope}).
Both curves would in principle saturate to flatness for very large $\varepsilon$, specifically exceeding the total size of $X$, but we again do not reach this threshold based on the choice of the range of $\varepsilon$.

Second, within the linear scaling region the curves of $\log(C_2)$ fluctuate less than those of $H_2$, resulting in a narrower confidence interval. Comparing confidence intervals is only meaningful if the same method is used to extract them. That is why specifically for Fig.~\ref{fig:analytic} we used the standard least squares fit for $\log(C_2)$ instead of the aforementioned correction of Ref.\cite{Sprott2001}.
Third, the actual numbers we obtain for $\Delta$ from the correlation sum method are closer to the analytically expected values, than those of the entropy-based method. Especially for sets that should have an integer fractal dimension the correlation sum method is much closer to the actual result.

\subsection{Different dynamical systems}
\label{sec:known_rule}
In the following we cross-compare with the value obtained from the Lyapunov (Kaplan-Yorke) $\Delta^{(L)}$ dimension which for all systems of interested used in this paper is found in Table~\ref{tab:kaplan}. We note that $\Delta^{(L)}$ is conjectured to equate to the $q=1$ version of dimensions, however here we use $q=2$. That is because the correlation sum algorithm does not apply to $q=1$, unless one uses a fixed mass approach, but we explain in Sec.~\ref{app:fixedmass} why we do not. We assume that the differences between the two estimators, when compared to $\Delta^{(L)}$, should not depend much when changing $q=1$ to $q=2$, hence the following results remain valid.

\begin{figure}[t]
    \centering
    \includegraphics[width=\columnwidth]{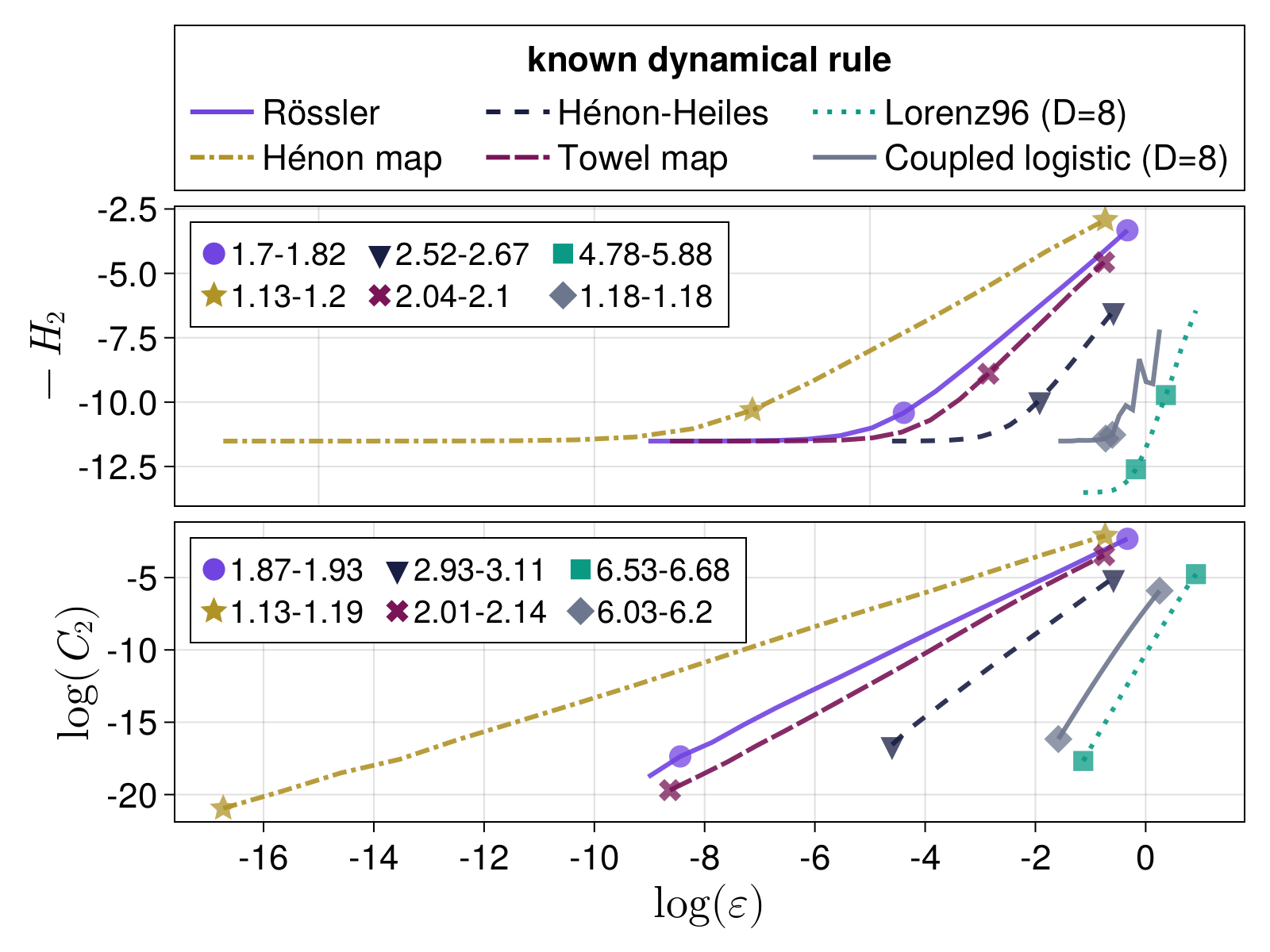}
    \caption{Fractal dimension estimates $\Delta_2^{(H)}$ and $\Delta_2^{(C)}$ for sets coming from different dynamical systems with known dynamical rule. CLM stands for Coupled Logistic Maps. We are not sure why $H_2$ of CLM yields a jagged curve instead of a clear linear slope, but the focus of this paper is not individual systems so we leave this for further study.}
    \label{fig:different_ds}
\end{figure}

In Fig.~\ref{fig:different_ds} we compare 3 discrete and 3 continuous dynamical systems of different input dimensionality (systems are defined in detail in App.~\ref{app:systems}).
We confirm the result of the previous section, i.e. the correlation sum method is more accurate than the entropy one because it is much closer to the values expected from the Lyapunov dimension (Kaplan-Yorke conjecture).
Furthermore, the entropy method seems to underestimate the fractal dimension more strongly as the state space dimensionality increases.
This is expected given the fact that the entropy method works via a histogram approximation of the natural density, and it is well known that the higher dimensional the data, the less accurate producing a histogram for them becomes (i.e., space is covered more sparsely by $N$ points as $D$ increases).

This figure also allows us to indeed confirm that the logarithmic correction to the correlation sum by Sprott and Rowlands\cite{Sprott2001} can be impactful, especially for high-dimensional sets where the convergence is the slowest. For example, the standard linear regression fit would give confidence intervals $(6.44, 6.6)$ for the 8-dimensional Lorenz96 model and $(5.78, 6)$ for the 8 coupled logistic maps. These values are closer to the entropy based estimates but further away from the $\Delta^{(L)}$ estimates of Table~\ref{tab:kaplan}.

\subsection{Data length}
\label{sec:length}
\begin{figure}
    \centering
    \begin{subfigure}{\columnwidth}
        \includegraphics[width=\columnwidth]{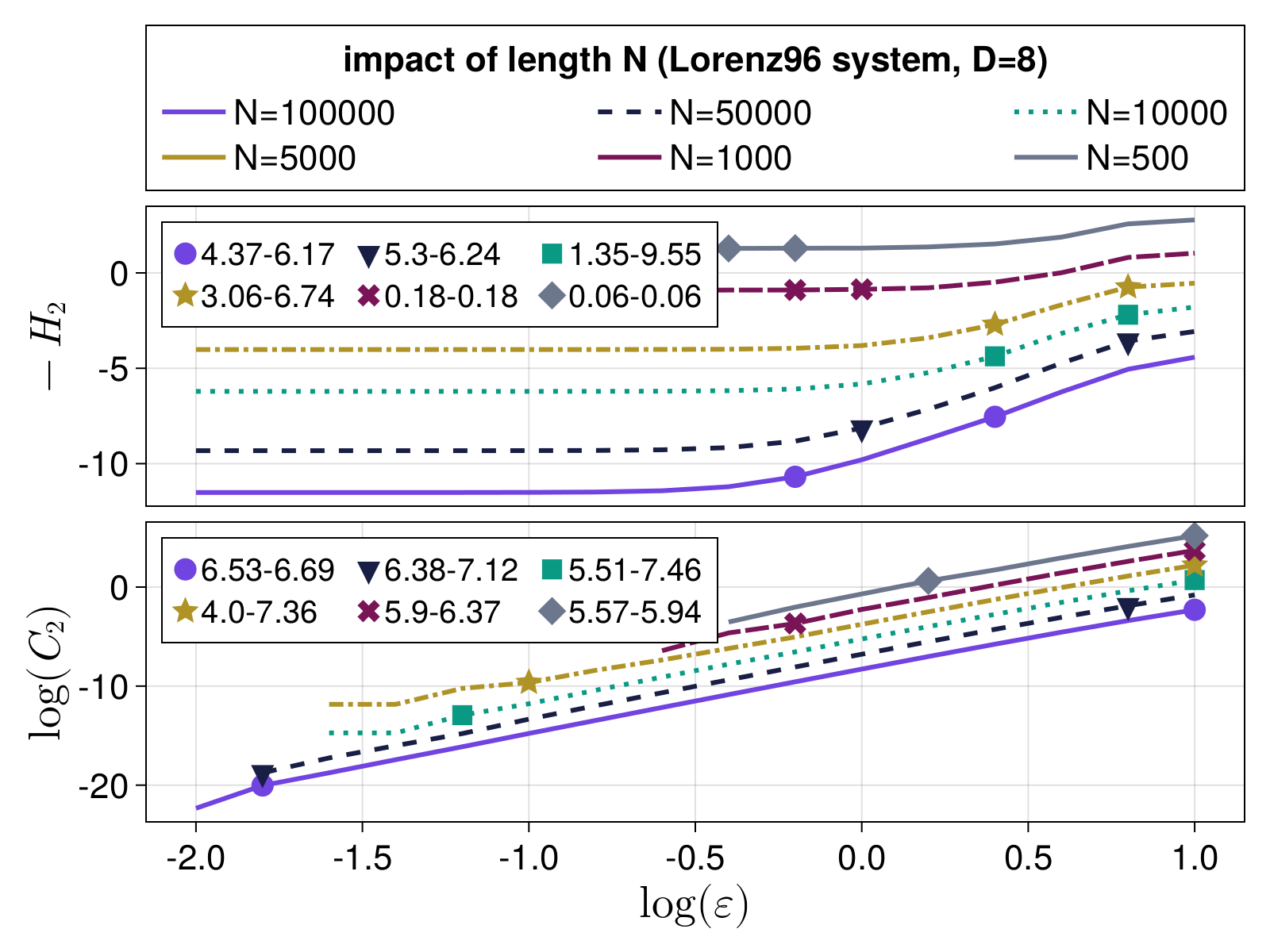}
    \end{subfigure}
    \begin{subfigure}{\columnwidth}
        \includegraphics[width=\columnwidth]{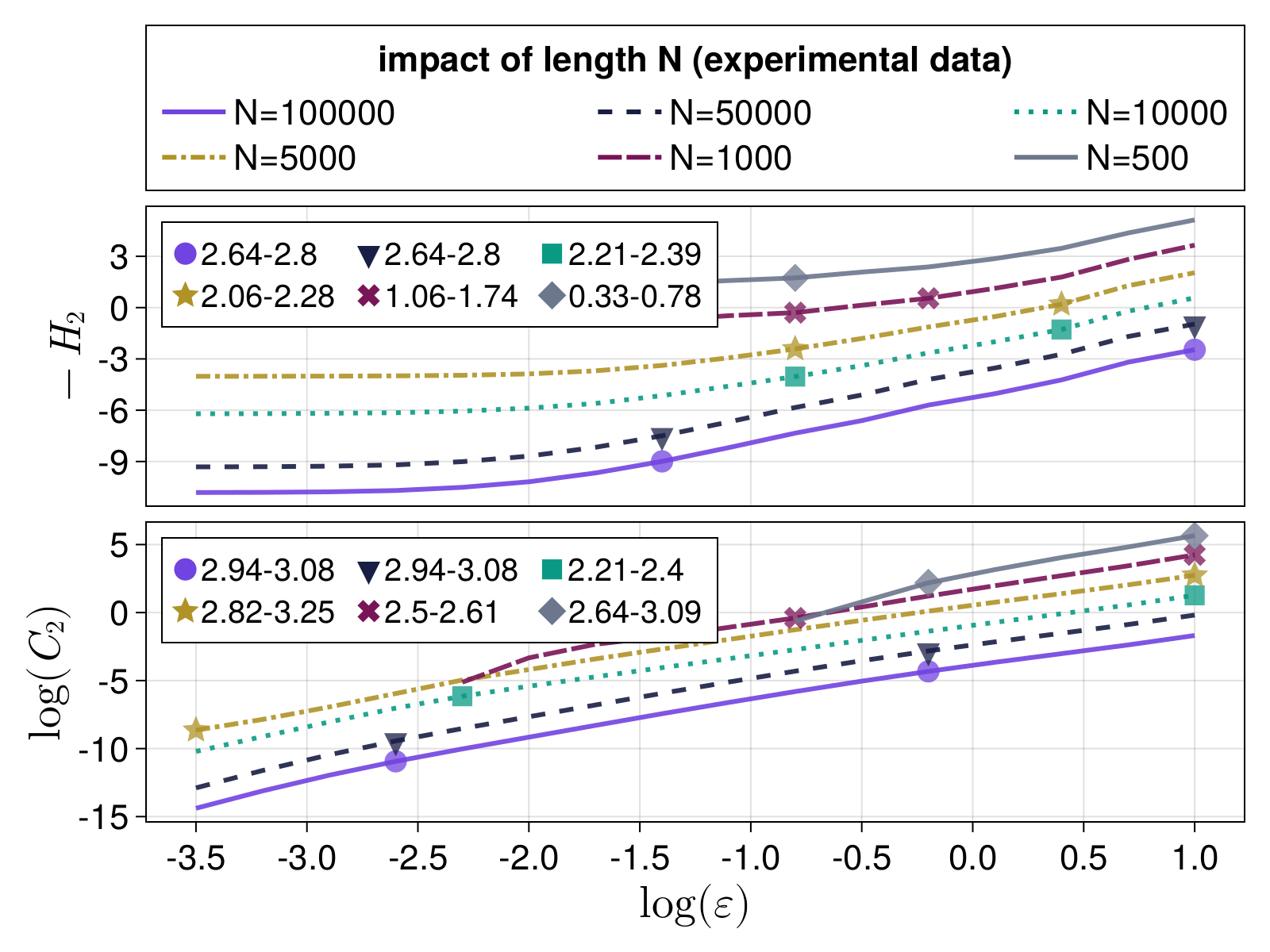}
    \end{subfigure}
    \caption{Dependence of estimators on data length $N$. The plot in the top panel shows results  for data from an 8-dimensional Lorenz96 system and the bottom plot for the experimental dataset ``electroch. 1'' from Fig.~\ref{fig:experimental}. The curves have been vertically offset for visual clarity. For these plots we have used the automatic $\varepsilon$ range estimation described in Sec. \ref{sec:estimating_slope} only for the largest datasets and use the same range for the rest. As a result $\log(C_2)$ cuts off to $-\infty$ as $N$ decreases and curves are truncated correspondingly.}
    \label{fig:length}
\end{figure}

Keeping the dimensionality constant but varying data length leads us to an interesting observation in Fig.~\ref{fig:length}.
The entropy-based method performs very poorly for small $N$ and clearly should never be used with small datasets. In contrast, the correlation dimension is much more robust and yields already useful lower bounds for $\Delta_2^{(C)}$ for very short time series like $N =500$. Eckmann and Ruelle~\cite{Eckmann1992}  discussed the data requirements for estimating the correlation dimension and pointed out that the minimal number $N_{\text{min}}$ of data points needed to estimate a dimension $\Delta_2^{(C)}$  quickly increases like $\log(N_\text{min}) \propto \Delta_2^{(C)}$. Assuming linear scaling for $\varepsilon = \rho E$, where $E$ denotes the diameter of the set (largest pointwise distance) and $\rho$ is a small number like 0.1, Eckmann and Ruelle state that for $N$ data points the estimation of $\Delta_2^{(C)} $ will not provide values larger than
\begin{equation}
    \Delta^{(C)}_{2 \ \text{max}}(N) = \frac{2\log N}{\log(1/\rho)} . 
    \label{eq:eckmann}
\end{equation}
For $\rho = 0.1$ the upper limits for $N=500$, 1000, 10000, 100000 are $ \Delta^{(C)}_{2 \ \text{max}}(N) =  5.4$, 6, 8, 10, respectively. In this context other choices of $\rho$ may also be justified and will decrease (for $\rho < 0.1$) or increase (for $\rho > 0.1$) the upper limit $\Delta^{(C)}_{2 \ \text{max}}(N) $. The method for detecting linear scaling regions used in this study, for example, starts searching at $\varepsilon$-values corresponding to $\rho \approx  1/e \approx  0.368 $ (see Sec.~\ref{sec:estimating_slope}). Therefore, depending on the data, in some cases linear scaling may occur for $\rho > 0.1$ resulting in larger upper limits $\Delta^{(C)}_{2 \ \text{max}}(N)$. 

The results for the 8D Lorenz96 model shown in Fig.~\ref{fig:length} are essentially consistent with the estimates of Eckmann and Ruelle. For $N=500$ the $(5\%,95\%)$ confidence interval of the slope is $(5.57, 5.94)$, i.e. slightly larger than  $ \Delta^{(C)}_{2 \ \text{max}}(N) = 5.4$ assuming $\rho = 0.1$ (a value of $\rho = 0.113 $, for example, would provide $ \Delta^{(C)}_{2 \text{max}}(N) = 5.7 $). With increasing $N$ the estimated slopes converge towards the value $\Delta^{(L)}= 6.91$ of the Kaplan-Yorke dimension of the 8D Lorenz96 system, a value that is already included in the interval (5.51, 7.46) obtained with $N=10000$, a number of points large enough to estimate correlation dimensions up to $ \Delta^{(C)}_{2 \ \text{max}}(N) = 8$ according to the Eckmann-Ruelle limit.

The bottom panel of Fig.~\ref{fig:length} shows log-log plots of $C_2$ for real-world experimental data (see Sec. \ref{sec:experimental} for a description). The dimension estimates $\Delta_2^{(C)}$ for different lengths $N$ are all about 3 and only the corresponding confidence intervals shrink for increasing $N$. This is also in agreement with the Eckmann-Ruelle bound, because dimensions  $\Delta_2^{(C)} \approx 3$ can in principle be achieved with $N=500$ or larger.
 
On the other hand, Tsonis et al.\cite{Tsonis1993}, using the results of Nerenberg and Essex\cite{Nerenberg1990}, argue that the minimum number of points $N_\mathrm{min}$ to estimate a dimension $x$ with 95\% confidence is scaling like $N_\mathrm{min}\sim 10^{2 + 0.4x}$. In our example, assuming true value $x \approx 7$, it would require at least $N_\mathrm{min} = 63095$ points, which is too high compared to what we can estimate from Fig.~\ref{fig:length}. However, the estimate presented by Tsonis et al. is surrounded by its fair share of ambiguity, because it involves deciding a-priori a scaling region extent, and it does not say whether the dimension $x$ in the expression should be the embedding dimension or the actual fractal dimension (that is unknown).
We will discuss this topic again in Sec. \ref{sec:inappropriate}.

If only small data sets are available, finite sample corrections  derived by  Grassberger \cite{Grassberger1988} could be used to improved estimates of $H_q$ or $C_q$.

\subsection{$q$-order dimensions (multi-fractality)}
Here we examine how well the estimators capture multi-fractal properties, i.e. the dependence of $\Delta_q$ on $q$, or the absence of multi-fractality, i.e., results that should be invariant to $q$.
This dependence on $q$ was discussed in the review by Theiler~\cite{Theiler1990}, but we think an even better reference is in the book by Rosenberg~\cite{Rosenberg2020}.
A known theoretical result is that $\Delta_q$ is a non-increasing function of $q$, $\Delta_{q_2} \le \Delta_{q_1}$ for $q_2 > q_1$, see e.g. Ref.\cite{Grassberger1983b}.

\begin{figure}
    \centering
    \includegraphics[width=\columnwidth]{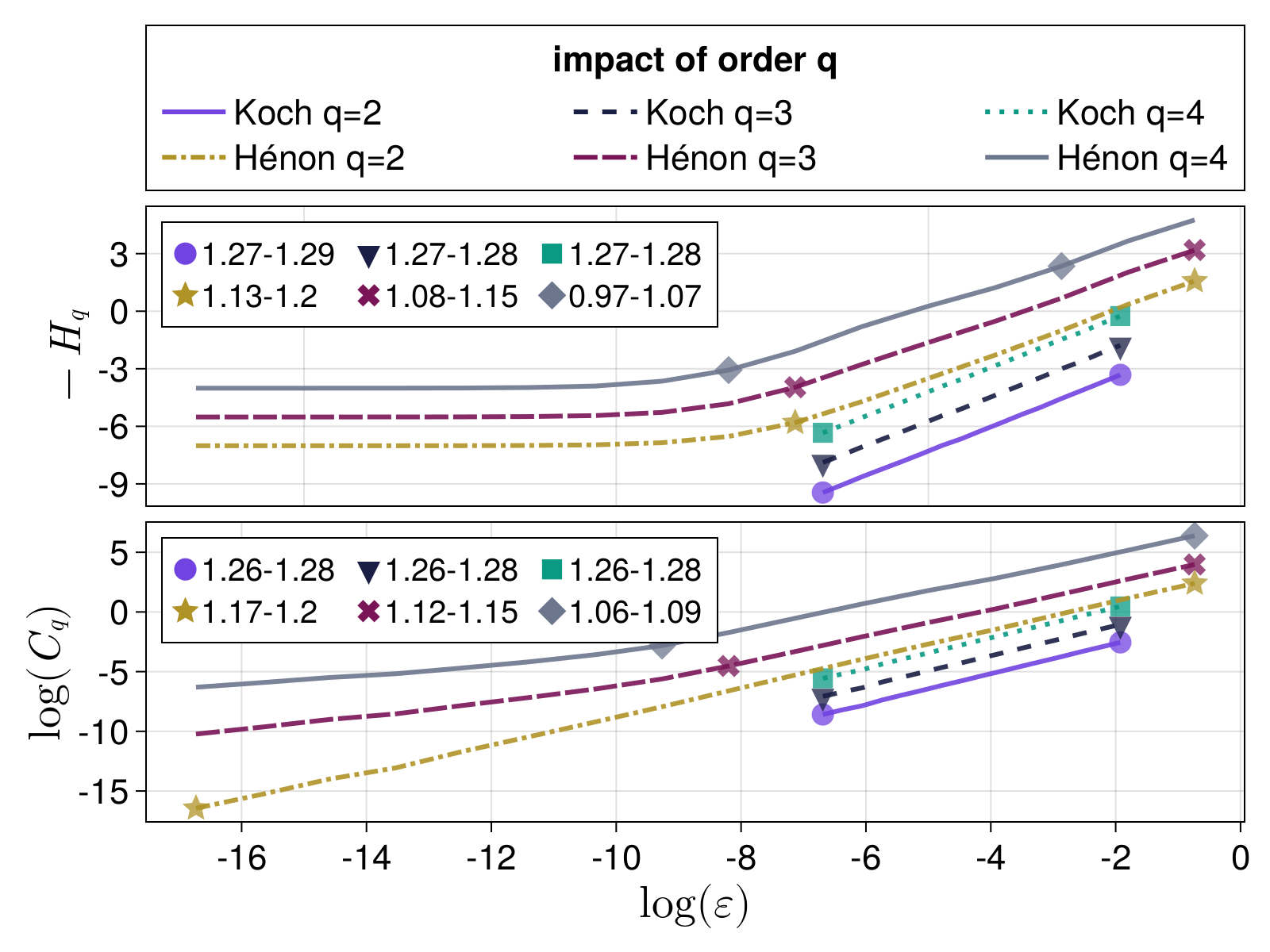}
    \caption{Impact of order $q$ on the fractal dimension. We used the standard least squares fit for $\log(C_q)$ as Ref.\cite{Sprott2001} does not discuss $q\ne2$. The curves have been vertically offset for visual clarity.}
    \label{fig:orderq}
\end{figure}

Here we will use two examples. The first is the Koch snowflake, which has an (approximately) uniform density and thus its FD should have no dependence on $q$ whatsoever.
The second is the H\'enon map, which has a strongly non-uniform natural measure, giving the expectation of a clear decrease of $\Delta_q$ with increasing $q$.
The results are shown in Fig.~\ref{fig:orderq}.

Both the entropy and correlation sum approaches perfectly capture the absence of multi-fractality, giving identical curves for all $q$ for the Koch snowflake.
For the H\'enon map estimates, both methods satisfy the criterion of a decreasing $\Delta_q$ with $q$.
But there is a problem.
For the correlation sum method and for $q\ne 2$, the function $\log(C_q)$ vs. $\log(\varepsilon)$ is no longer a straight line, but is composed of two linear regions with significantly different slope, making the results ambiguous. It is also not obvious why the slopes change at the given $\epsilon$ value, or why the slopes below that indicate a significantly lower dimension value. The slopes change at $\log\varepsilon \approx -10$ and for $q=3$ the left slope becomes $\approx 0.58$ while for $q=4$ the left slope becomes $\approx 0.39$ (right slopes are shown in figure legend). We observed this behavior of $\log(C_q)$ having two slopes in practically all example sets with non-uniform measure.

We could not find anything in the literature about this observation, and in fact, we could not find a single figure in the literature plotting $C_q$ versus $\varepsilon$ for $q\ne 2$, even though the correlation sum for $q\ne 2$ is provided in several publications\cite{Grassberger1983b,KantzSchreiber2003,Grassberger2007}, which report a value for $\Delta_{q\ne 2}^{(C)}$ (but it is unclear whether they have encountered the same problem as we have, or not). We have extensively tested our code and we are confident that the implementation of Eq.~\eqref{eq:correlationsum} is correct.

For multi-fractal analysis we are (typically) interested in quantifying the most fine properties of the fractal.
Perhaps then one should determine the slope of the linear region at the \emph{smallest} $\varepsilon$ values, instead of the slope of the linear region covering the largest range of $\varepsilon$ (but for very small $\varepsilon$ the statistics becomes worse for finite data sets). However, the slopes of the smallest $\varepsilon$ are clearly incorrect; the correct slopes are the ones of the largest $\varepsilon$ values (those also highlighted in Fig.~\ref{fig:orderq}).
In any case, using the slope of the largest $\varepsilon$ values gives the correct results, but this strong dependence of slope with $\varepsilon$ when $q\ne2$ is worrisome and indicates that more clarity regarding $C_{q\ne2}$ must be established in the literature.

\subsection{Dimension (delay embedding)}
\label{sec:main_delay}
\begin{figure}[t]
    \centering
    \includegraphics[width=\columnwidth]{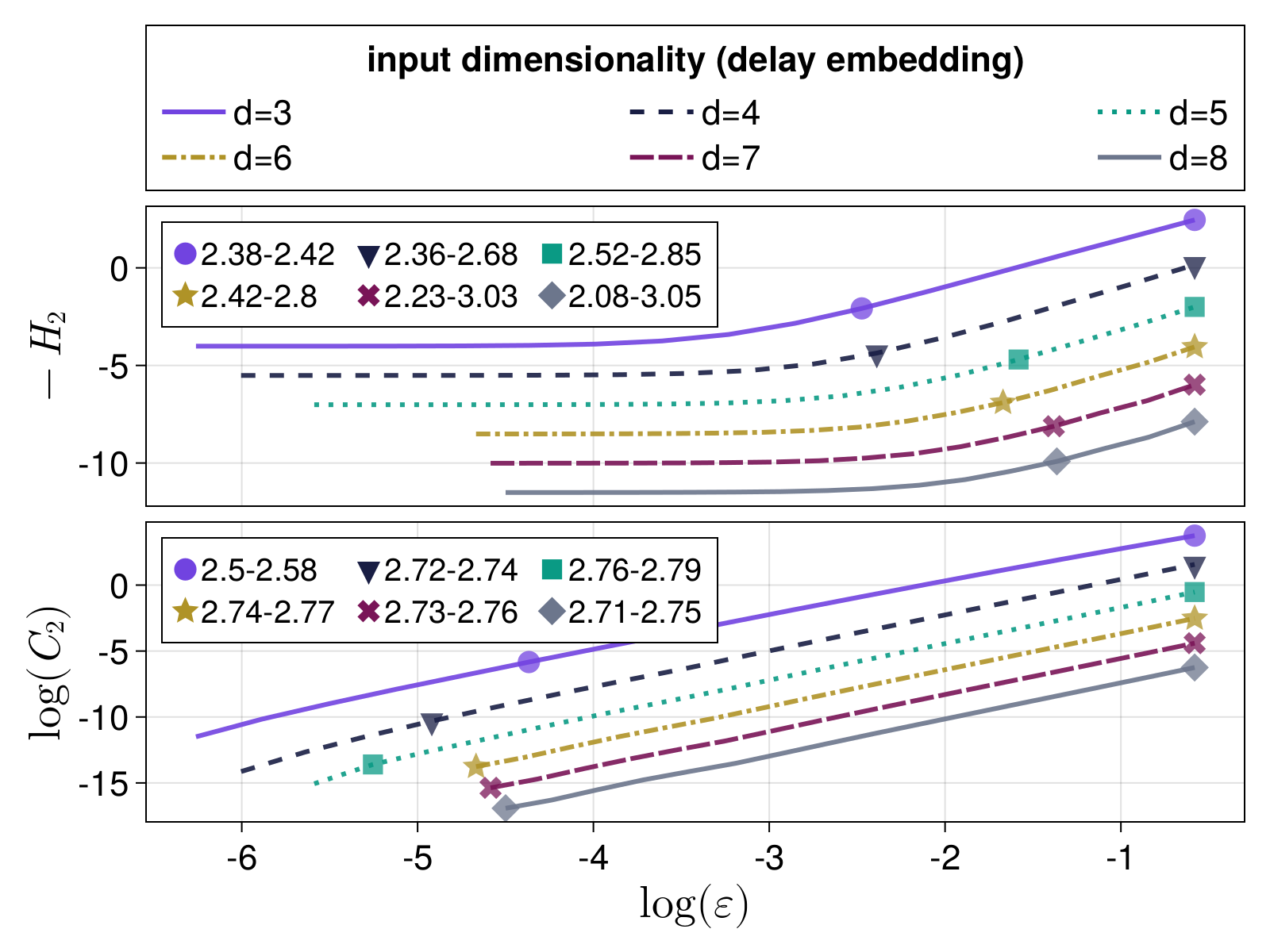}
    \caption{Effect of increasing input dimensionality of an input timeseries from a set of known fractal dimension utilizing delay embeddings of dimension $d$. The timeseries used is the first variable of a chaotic H\'enon-Heiles trajectory.}
    \label{fig:embed}
\end{figure}

This section examines the impact of varying state space dimensionality $D$ of input data, which is common case when delay-embedding timeseries~\cite{Packard1980, Takens1981, KantzSchreiber2003} (because there one increases $D$ and searches for convergence of $\Delta$).
In principle, provided the condition $d > \Delta_0^{(H)}$ is met, with $d \in \mathbb{N}$ the embedding dimension, then the reconstructed set has the same fractal dimension as the original set the timeseries was recorded from~\cite{Sauer_et_al_1991}.
In Fig.~\ref{fig:embed} we check how the methods fare with this statement, and whether their accuracy decreases with increasing input dimensionality, i.e. embedding dimension.

In Fig.~\ref{fig:embed} we used a chaotic timeseries from the H\'enon-Heiles system. This has $\Delta = 3$ and as such we expect convergence of the fractal dimension estimates to a value around 3 for $d \ge 4$. We see in Fig.~\ref{fig:embed} that this is indeed the case. $\Delta^{(C_2)}$ does not seem to drop in performance with increasing $d$, besides a very small decrease in the overall range of order of magnitudes the linear scaling region covers. Same results were obtained using a timeseries from the chaotic R\"ossler system with $\Delta \approx 2$ or a chaotic Lorenz96 system with $D=4$, which also has $\Delta\approx 3$. $\Delta_2^{(H)}$ seems to perform poorer with increasing $d$ by either significantly reducing the linear scaling region or being less accurate in the dimension estimates. However, how much poorer it performs depends on the dataset: for the H\'enon-Heiles it is still decent, while for Lorenz96 it performs much worse (not shown). Hence, we conclude that $\Delta_2^{(C_2)}$ performs much better as the state space dimensionality of data increases (while keeping data length and other aspects constant) versus $\Delta_2^{(H)}$, which is somewhat already evident from Fig.~\ref{fig:different_ds}.

\subsection{Noise}
\label{sec:noise}
\begin{figure}[t]
    \centering
    \includegraphics[width=\columnwidth]{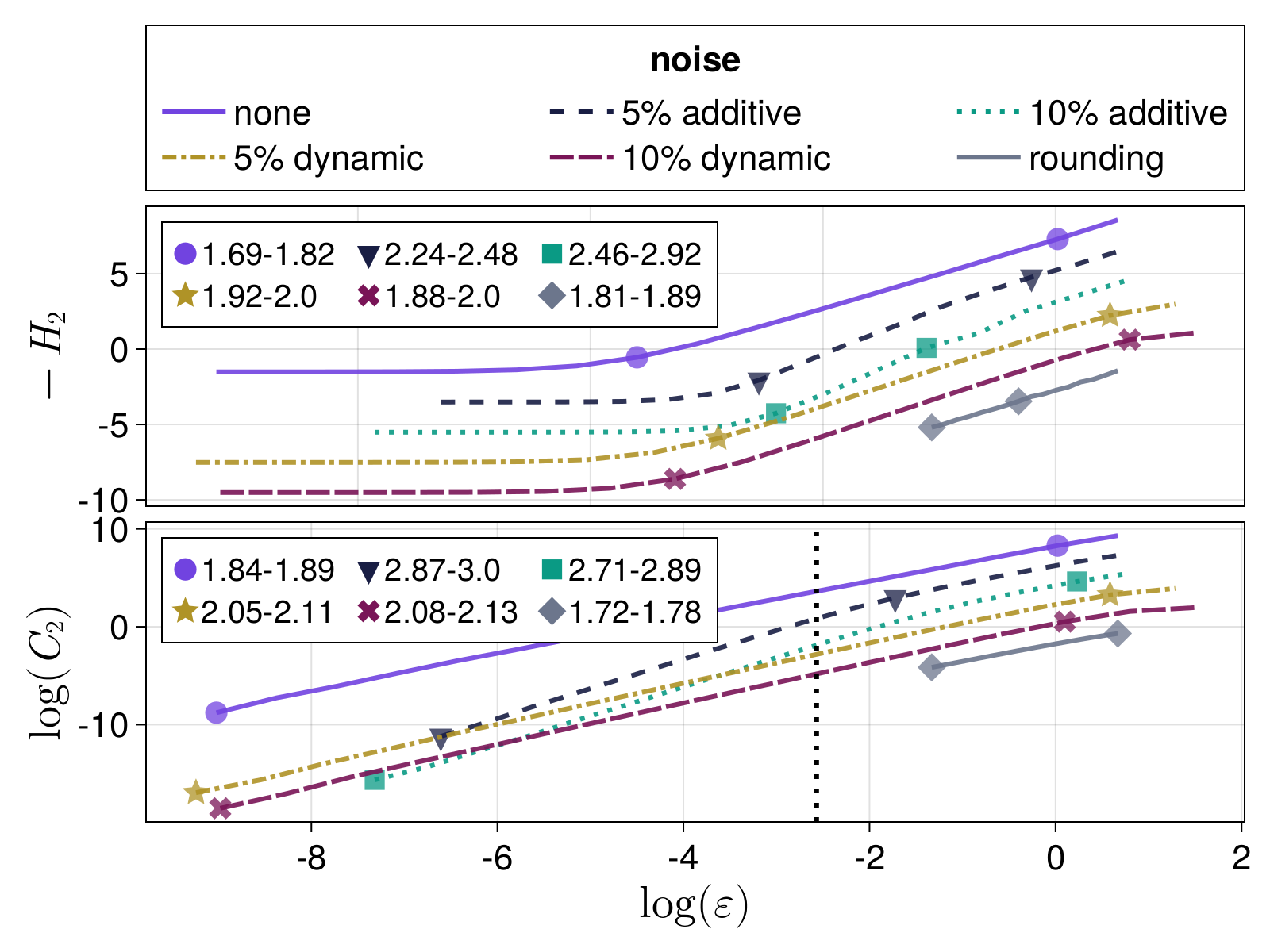}
    \caption{Impact of noise using the (chaotic) R\"ossler system as an example. The noise percentages approximately indicate ratio of the std. of the noise divided by the std. of the deterministic dynamics. See App.~\ref{app:boxassisted} for a discussion of the vertical dotted line.}
    \label{fig:noise}
\end{figure}

Real world data is always accompanied by noise, and therefore the impact of noise on the calculation is highly important for the choice of the method.
As it is well known, the presence of noise in the data makes estimating a fractal dimension harder, as the fractal dimension of the noise is equal to that of the state space, and hence almost always larger than the fractal dimension of the clean data.
Figure~\ref{fig:noise} shows results for various kinds, and amount, of noise added to the (normalized) chaotic R\"ossler attractor.
On purpose for this plot we have used $\zeta = 1$ (see Sec. \ref{sec:estimating_slope}), and the standard linear regression method for $\log(C_2)$, because the logarithmic correction of Ref.\cite{Sprott2001} overestimates $\Delta > 3$ for noisy data (the input dataset is three dimensional and thus cannot have $\Delta$ > 3).

We start with the case of additive noise. There it is known that there is some distance $\varepsilon_\sigma$ called the ``noise level'', below which the slope of $\log(C_2)$ changes from being the fractal dimension of the chaotic set to that of the noise \cite{Deshmukh_et_al_2020}, This fact can be used to actually estimate the noise level of the data~\cite{KantzSchreiber2003}. 
In Fig.~\ref{fig:noise} we see how quickly this really happens for $\log(C_2)$. Even for 5\% additive noise, the curve is already dominated with the noise slope (which has $\Delta \approx 3$ for 3-dimensional additive noise), and only a small segment of the curve at large $\varepsilon$ values where the slope becomes the deterministic value.
The entropy curve does not have this property, and remains having a single slope throughout (except the saturation part of course), while the slope value is an average between the purely deterministic $\Delta$ and that of the noise. On one hand, this may be considered a downside, because it doesn't allow estimation of noise level.
On the other hand, we should note the majority of the $\log(C_2)$ curve is already reflecting the noise slope. Thus, if we estimated the average slope of the $\log(C_2)$ curve, it would be much larger than that of $H_2$, i.e., it puts much more weight on the noise dimension than the deterministic data. These results regarding additive noise are typical and do not seem to strongly depend on the system considered. 
Note that in the case of delay reconstruction, the slope of the noise should increase with increasing embedding dimension. The slope corresponding to the FD of the deterministic set should remain constant for embedding dimensions larger than a minimum value required for successful reconstruction of the state space of the dynamical system generating the data.

For dynamic noise we turned the ODE of the R\"ossler system into a stochastic differential equation by adding a Wiener process term $\eta dW$ in the second equation. For small amount of dynamic noise (which here reflects a proportionality of $\eta$ with the expected size of the $x$ variable), the fractal dimension increases slightly as expected, but does not have any noticeable change in its numerical value up to 10\% noise. When one turns up the dynamic noise more, the dynamics collapse and there is no ``chaotic'' attractor anymore (not shown). Both entropy and correlation sum methods perform equally well versus this kind of noise, and there is no noise radius or change of slope discernible in the correlation sum case.
We also looked at low-resolution data, by rounding R\"oessler timeseries to 2 digits after the decimal. This obviously decreases the valid $\varepsilon$-ranges one can do the computation for (see Sec. \ref{sec:estimating_slope}).
For $\Delta_2^{(C)}$ this also significantly changes the result to a value smaller than ``correct'', while for $\Delta_2^{(H)}$ it has no impact. This means that $\Delta_2^{(H)}$ performs better for rounded data, which makes sense given the way it is computed (as long as points are in the same box, it doesn't matter how close they are to each other).

\subsection{Real world data}
\label{sec:experimental}
\begin{figure}[t]
    \centering
    \includegraphics[width=\columnwidth]{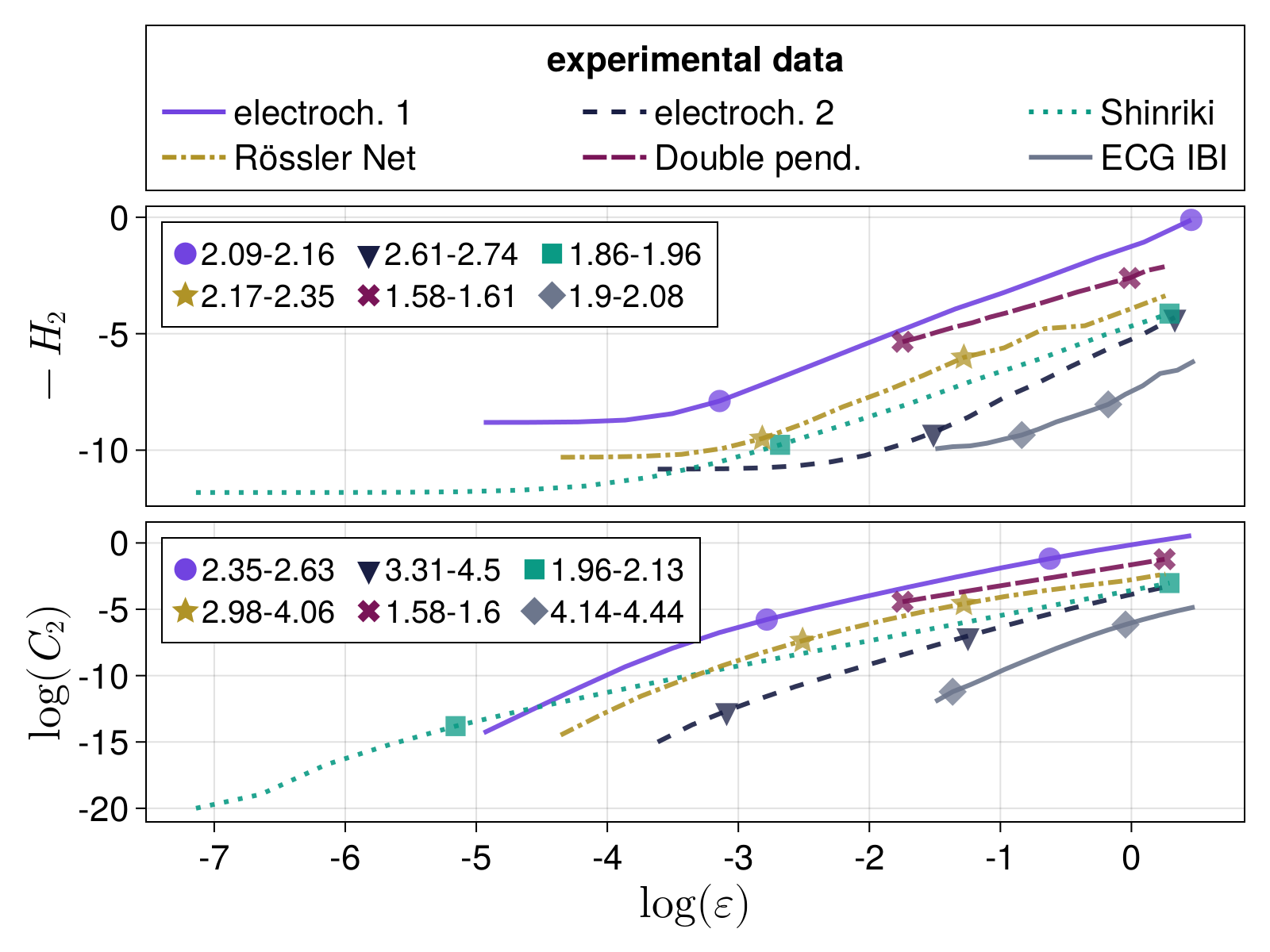}
    \caption{Fractal dimension estimates for experimental systems known to be underlined by low-dimensional nonlinear dynamics.}
    \label{fig:experimental}
\end{figure}

In Fig.~\ref{fig:experimental} we show fractal dimension estimates for real world experimental data. 
We focused specifically in experiments that are relatively clean (large signal to noise ratio) and where \emph{the underlying dynamics is well known to display low dimensional deterministic chaos}. This is important, because for this review we do not want to mix the scientific question of whether an observed system accommodates a low-dimensional deterministic representation, with the technical/computational question of whether an estimator would actually detect that.

In Sec.~\ref{sec:inappropriate} we further discuss what happens with real world data where neither of these two conditions apply.
We limited densely sampled experimental data to at most $N=50,000$, with sampling of about 10 samples per characteristic timescale.
Because of the observations of the previous subsection, we have used $\zeta = 1$ and the standard linear regression method for $\log(C_2)$ instead of the logarithmic correction of Ref.\cite{Sprott2001}.

The datasets are as follows: two electrochemical oscillator datasets (the second being more chaotic than the first)\cite{Kraemer2021}; timeseries from a circuit replicating the dynamics of the Shinriki oscillator \cite{Shinriki_et_al_1981}; the mean field of a network of 28 circuits following R\"ossler dynamics from Ref.~\cite{Vera-Avila2019}; data from a mechanical double pendulum from Ref.~\cite{asseman2018learning};
ECG recordings during a pacing experiment of healthy individual from Ref.~\cite{CEPS2023}.
All experimental timeseries were delay embedded using the recent automated method due to Kraemer et al.~\cite{Kraemer2021}, see App.~\ref{app:delayparameters} for the embedding parameters. The method yielded embedding space of 7 or less for all experimental timeseries, giving even more confidence that the data may display low-dimensional deterministic chaos.

For dataset ``R\"ossler Net'' the curve $(\log(\varepsilon), \log(C_2))$ continuously changes its slope instead of having one, or at most two, constant-slope segments. This makes deducing a single fractal dimension ambiguous. Slight curving can be observed also in ``electroch. 1'', ``electroch. 2'' and ``ECG IBI'' but it is a weak enough effect that two scaling regions can nevertheless be extracted (for ``ECG IBI'' Fig.~\ref{fig:experimental} reports the slope of the noise). On the other hand, the entropy-based approach does not suffer from this problem and, besides the expected saturation for small $\varepsilon$, seems to be described quite accurately by a single slope and thus a specific FD value.
The results shown in Fig.~\ref{fig:experimental} are from 5-6 dimensional embeddings yet both $H_2, C_2$ yield FDs that are less than the embedding dimensions (excluding the `R\"ossler Net'' case for $C_2$ where a FD cannot be estimated). 
Hence, we can assume that the FD of the underlying dynamics has to be somewhere between the low bounds of $\Delta^{(H)}_2$ and $\Delta^{(C)}_2$.

The ``constant slope curving'' of $C_2$ is something we have not seen before with synthetic data. 
From the discussion of Sec. \ref{sec:noise}, the problem may be because realistic noise may be neither white nor stationary, or because a too high of a noise level in the data. Indeed, we saw that for 5-10\% relative noise, the slopes of $C_2$ already reflect the noise FD. And in Fig.~\ref{fig:experimental}, $C_2$ yields consistently higher FD than $H_2$, even though we know that an underlying low-dimensional representation exists.
To extract this lower FD value from $C_2$ one therefore must focus on the larger scales $\varepsilon$ and try to find this consistent smaller slope by increasing embedding dimension (as is standard practice\cite{KantzSchreiber2003}). We show such an analysis in App.~\ref{app:embedding_real}. Nevertheless, it appears true that $C_2$ is more strongly affected by noise when compared to $H$.

\subsection{Extreme cases}
\label{sec:inappropriate}
\begin{figure}[t]
    \centering
    \includegraphics[width=\columnwidth]{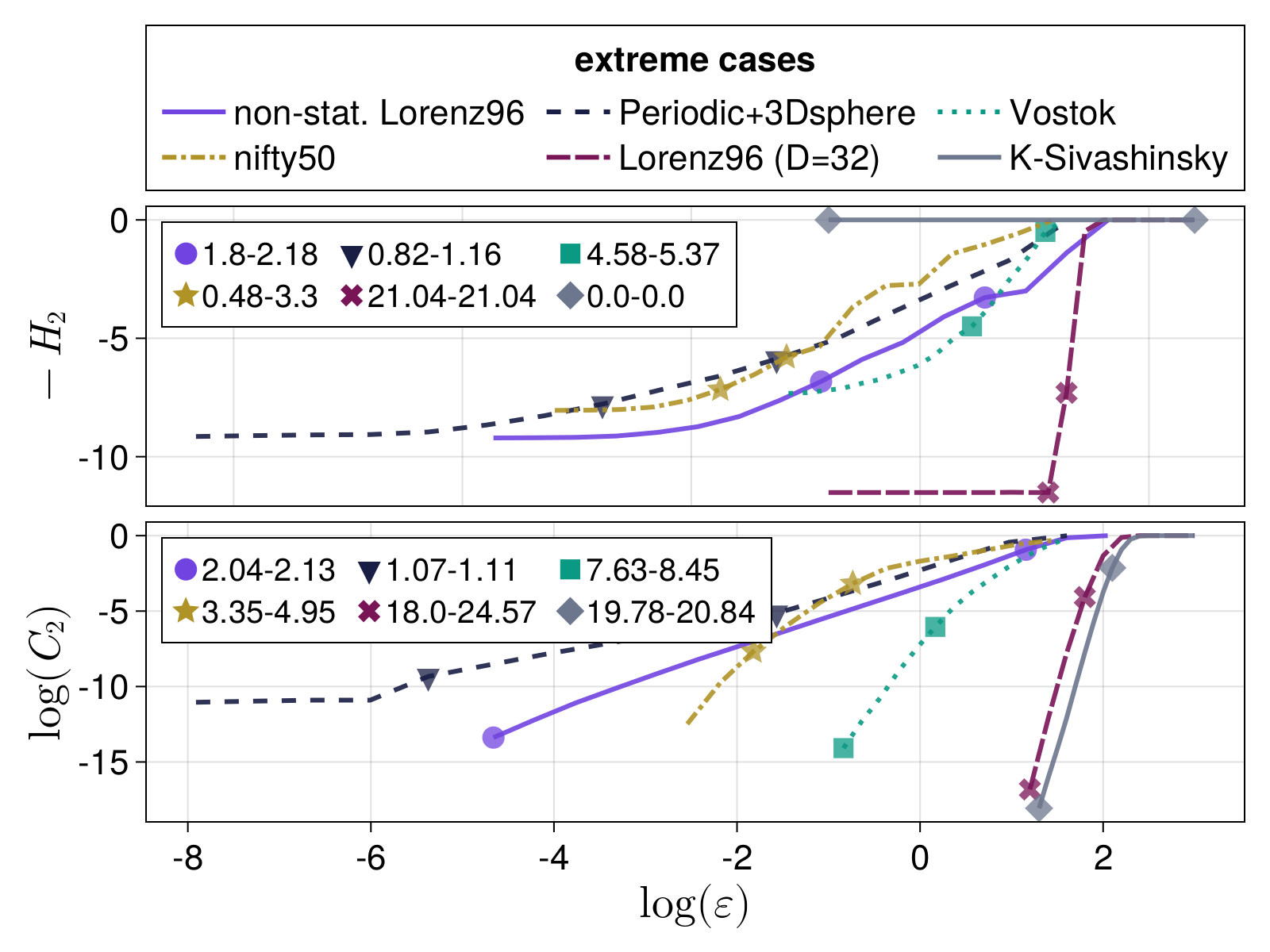}
    \caption{Fractal dimension estimates for extreme cases.}
    \label{fig:extreme}
\end{figure}

In this subsection we examine the result of applying the aforementioned methods to ill-conditioned data which may be non-deterministic or non-stationary, or to extremely high-dimensional data, where there exists this notion in the literature that the methods used so far are unlikely to succeed.
For the first dataset we used data from the Lorenz-96 model, with $D=6$, while having the $F$ parameter increase linearly during the time evolution from 1.0 (periodic motion) to 24.0 (chaotic motion with $\Delta \approx 5$). 
The second dataset is the concatenation of a periodic trajectory from the R\"ossler system with noise uniformly distributed on the 3D sphere.
The third dataset is paleoclimate temperature timeseries from the Vostok Ice core, embedded in 8-dimensional space, which is unlikely to be stationary or to accommodate a low-dimensional representation~\cite{Grassberger1986}.
The fourth dataset is a stock market timeseries for the ``nifty50'' index embedded in 6-dimensional space, which is definitely non-stationary and rather unlikely to be deterministic. The last two datasets are extremely high dimensional data of the Lorenz96 and the Kuramoto-Sivashinsky spatiotemporal system (the latter having 101 dimensions after discretization). The results of the dimension estimation are shown in Fig.~\ref{fig:extreme}.

Generally speaking, in Fig.~\ref{fig:extreme} the results of the first four datasets show that something is ``wrong''. There is a large miss-match between the estimates of the entropy and correlation sum methods and the curves do not seem to be composed of a single slope. Oddly, for the non-stationary Lorenz96 data, the correlation sum has a clear straight slope with fractal dimension somewhere between the extreme values.
The Vostok data in particular are plagued both by a continuous change in slope, especially in $\log(C_2)$, but also, the resulting FD values do not converge when increasing the embedding dimension (not shown). Additionally, the FD values obtained from $H_2$ or $C_2$ are very different.
Notice that in all cases our automated algorithm finds a value for $\Delta$ nevertheless. This only serves to highlight how careful one should be, and to always plot the curves of $H_2$, $\log(C_2)$.

Although in the literature there exist several tests for non-stationarity using e.g. permutation entropy or other methods, we will now describe a simple, fractal-dimension-based scheme. One can divide the data into $m$ equal parts in two ways: Making $m$ segments of length $N/m$  of successive points, or by choosing every $m$-th point, each time starting from point 1 to $m-1$.
For these subsets, the same fractal dimension estimation is done. If there is non-stationarity, the first kind of selection will show significantly different estimates across its sub-datasets, while the second will show approximately the same estimates.

Let us now consider the last two datasets of Fig.~\ref{fig:extreme} (32-dimensional Lorenz96 and Kuramoto-Sivashinsky). Surprisingly, $C_2$ shows a rather clear linear scaling region that has very high slope, but not  as high as the expected dimension values $\Delta^{(L)}$ (28 and 32, respectively). This is consistent with the upper bound 
Eq.~\eqref{eq:eckmann} of Eckmann and Ruelle with $ \Delta^{(C)}_{2 \ \text{max}}(100000) \approx  10$ for $\rho =0.1$ or  $ \Delta^{(C)}_{2 \ \text{max}}(100000) \approx  23$ for $\rho = 1/e$, because even in cases where linear scaling occurs already  for relatively large values $\varepsilon = \rho E$ it cannot be expected to obtain slopes of size 28 or 32 with $N=100000$ data points, only.
Furthermore, Fig.~\ref{fig:extreme} shows that for these data 
the range of scales covered by the linear region is very small: only one $e$. Increasing the amount of data also increases the linear scaling region
and the resulting confidence intervals shrink as we get more data points in the linear region (not shown here).

As discussed already in Sec.~\ref{sec:length} these examples show again that with high dimensional data one needs much longer timeseries to properly cover the (typically high-dimensional) chaotic set, and this affects any kind of estimate of dynamical properties; it is not a problem specific to the correlation dimension. Additionally, one needs very low signal-to-noise ratio, because due to the coverage of a very small range of scaling factors $\varepsilon$, even a small amount of noise may ruin the estimation. But, if one does have such a clean high dimensional dataset, $\Delta_2^{(C)}$ may still provide a useful estimator at least for a lower bound of the correlation dimension of that data set, where the value obtained should always be compared with  $ \Delta^{(C)}_{2 \ \text{max}}(N)$ for the available amount of data $N$ and dimension estimates of surrogate data (see Sec.~\ref{sec:surrogates}) to avoid wrong conclusions.

\subsection{Estimation of slopes and sizes $\varepsilon$}
\label{sec:estimating_slope}
To estimate the value of $\Delta^{(H_q)}$ or $\Delta^{(C_q)}$ we need to find the slope of $-H_q$ or $\log(C_q)$ versus $\log(\varepsilon)$.
This matter is typically resolved in a context-specific manner, where each plot is carefully examined and the ``linear region'' is decided by the practitioner by eye.
This approach cannot work in an objective comparison.
Here we formulate an entirely objective and sensible (but not flawless) automated process that is separated into two parts: the choice of which sizes $\varepsilon$ to calculate  $-H_q$ or $\log(C_q)$ for, and how to estimate a linear scaling region from the respective curves. 
Once the linear scaling region is identified, actually obtaining the fractal dimension is a simple least squares fit\footnote{Note that small deviations from a straight line for densely sampled values of $\varepsilon$ may occur due to \emph{lacunarity} of fractal sets\cite{badii_intrinsic_1984,smith_lacunarity_1986,arneodo_fractal_1987,smith_intrinsic_1988,theiler_lacunarity_1988}. While we can also observe this for very densely sampled $\varepsilon$, this effect is so miniscule that we consider it irrelevant for almost all data sets in practice.
These osciallations are not suitable for the quantification of lacunarity in fractal sets, but other measures exist for this purpose\cite{gefen_geometric_1983,allain_characterizing_1991}.}

The range of $\varepsilon$ is always decided with generating formula $\varepsilon = e^x$ where $x$ are $k$ linearly spaced values from $\log(\varepsilon_\text{min}) + \psi$ to $\log(\varepsilon_\text{max}) - \zeta$. I.e., the values of $\varepsilon$ are exponentially ranged in base $e$. $\varepsilon_\text{min}$ is the smallest inter-point distance existing in the set and $\varepsilon_\text{max}$ the average of the lengths along each of the variables of the set, and $\psi,\zeta$ constants.
Unless stated otherwise, we have used $k=16, \psi=1, \zeta=1$ in Sec. \ref{sec:main}.
In essence, we are limiting $\varepsilon$ to be one order of magnitude (in base $e$) larger than the smallest inter-point distance and one order of magnitude smaller than $\varepsilon_\text{max}$. 
If the resulting $\varepsilon$ range does not cover at least two orders of magnitude (common case in high dimensional data), we use $\zeta = 0$ instead.
This choice of $\varepsilon_\text{max}$ brings very good performance in the box-assisted algorithm for the correlation sum (\ref{app:boxassisted}) but it is not so small as to make the computation meaningless for realistic and/or noisy data.
Notice that the automated fractal dimension estimates can be sensitive on $k, \zeta, \psi$. In practice we would recommend to produce several estimates by varying these parameters and obtain the median of $\Delta$.

\begin{figure}[t]
    \centering
    \includegraphics[width=\columnwidth]{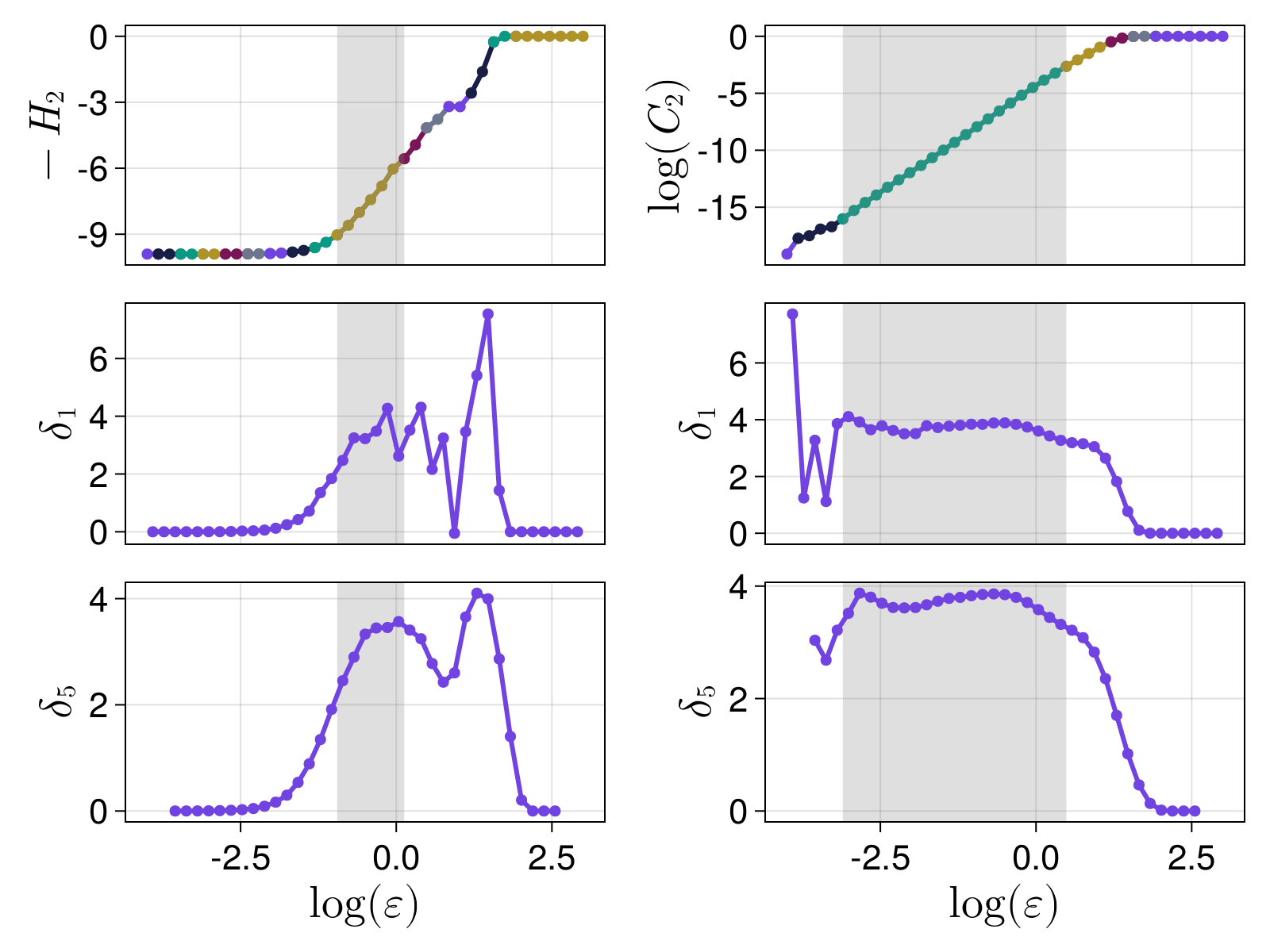}
    \caption{Demonstration of the algorithm estimating fractal dimensions from the curves of $H_2, \log(C_2)$. Shaded interval are the estimated linear scaling regions. See text for the definition of $\delta_1, \delta_5$ (differences of the top panels).}
    \label{fig:demonstration}
\end{figure}

To estimate the linear region we proceed as follows.
We scan the local slopes of each one of the $k-1$ segments of the curve $y$ vs $\log(\varepsilon)$ starting from the leftmost one (here $y = -H_q$ or $\log(C_q)$).
If the local slope of the preceding segment is approximately equal to that of the next one with relative tolerance tol, i.e. $|s_{i-1} - s_i| \le \mathrm{tol} \cdot \mathrm{max}(s_{i-1}, s_i)$, then these two segments belong to the same linear region.
We move to the next segment and compare it in the same way with the first segment belonging to the same linear region. 
When we find a miss-match, we start a new linear region.
This way we have segmented the curve $y$ vs $\log(\varepsilon)$ into approximately linear regions. We then choose the linear region which spans the largest amount of the $\log(\varepsilon)$ axis, and label it ``the'' linear region.
We finally perform a least squares fit there and report the 5-95\% confidence interval of the fitted slope.
In Fig.~\ref{fig:demonstration} we visually demonstrate the process. We also compare it with another standard way fractal dimension related plots are presented: the successive local slopes $\delta_1$ of each point of the curves and the same slopes  $\delta_5$ but fitted in a 5-long data window. Our linear regions approach is equivalent with finding the largest plateau in the local slopes plots.

Notice that there is a clear pitfall here. This algorithm will deduce a linear region no matter what.
In many scenarios this region might be meaningless, being too small to be of actual value, or could even be the scaling of the noise in noisy data, as shown in Sec. \ref{sec:noise}.
So after all, careful consideration of the result is always necessary.

Besides the algorithm presented here, also worth mentioning is recent work by Deshmukh et al.~\cite{Deshmukh2021} that offers an alternative way to estimate a slope. All possible slopes that could be estimated from the curve (by choosing all possible segments of length more than a specified minimum), are estimated. These are weighted by their length and by their inverse error and compose a distribution. The mean of the distribution is presented as the slope, while the quantiles of the distribution can be used as confidence intervals. For the work presented here, we believe our approach is more fitting, because how large a scaling region is is also part of the accuracy of a FD estimator. Furthermore, we still wanted to display problems in the presence of two scaling regions (as in e.g., Sec. \ref{sec:noise}), while the approach of Deshmukh et al. transforms the ``looking at the curve $y(x)$ for two scaling regions'' problem into ``looking at the distribution of all possible slopes for two peaks'', which, while easier to resolve, still would add more information content in our already extensive article. Nevertheless, the approach Deshmukh et al. is most likely better suited to use in practice, when estimating a fractal dimension from experimental data, as the authors have extended their method to also provide convergence criteria of fractal dimension estimates in Ref.~\cite{Deshmukh2023}.


\section{Extreme value theory analysis}
\label{sec:mainevt}
In this section we thoroughly analyse the power and shortcomings of the extreme value theory based FD $\Delta^{(E)}$ introduced in Sec. \ref{sec:evt_def} for computing FDs under a similar lens as the comparison of the previous section. Unless stated otherwise we use $N=10^5$ as the data length, standardize all input sets before any computations, and use $p=0.99$ as the quantile probability of Sec. \ref{sec:evt_def} (due to the discussion in Sec.~\ref{sec:evt_quantile}). Since $\Delta^{(E)}$  is obtained via an arithmetic mean, it can be compared to $\Delta^{(H)}_2$ and $\Delta^{(C)}_2$ which is what is used by most plots in Sec.~\ref{sec:main}.

\subsection{Exemplary sets}
\label{sec:evt_exemplary}
\begin{figure}[t]
    \centering
    \includegraphics[width=\columnwidth]{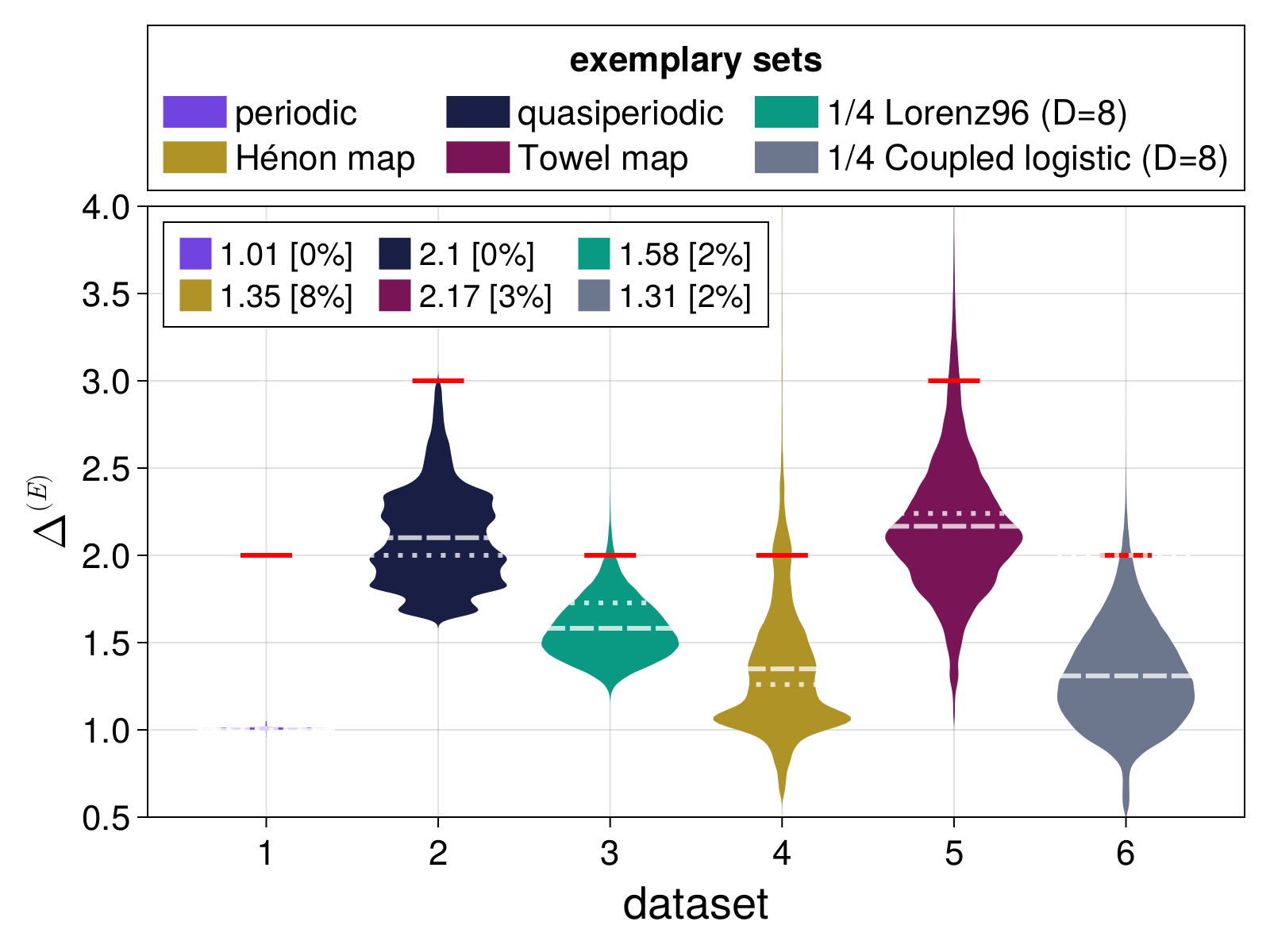}
    \caption{Fractal dimension estimates for exemplary systems using the extreme value theory method. Shown are the distributions of $\Delta^{(E)}_i$ for each system. The inner legends display the mean of the distribution and in brackets the percentage of the distribution outside a strict cutoff value (the state space dimensionality). For the two 8-dimensional systems (Lorenz96, coupled logistic maps), we divide dimension values by 4 for visual purposes. More details in Sec. \ref{sec:evt_exemplary}.}
    \label{fig:evt_exemplary}
\end{figure}

We start with Fig.~\ref{fig:evt_exemplary} which shows $\Delta^{(E)}$ computed for exemplary sets. The figure should be compared with Figs.~\ref{fig:analytic} and \ref{fig:different_ds}. The figure style is typical for the rest of this section and shows the dimension estimates as distributions, with dashed white lines indicating the mean, and dotted white lines indicating the ``expected'' value, for which we use $\Delta^{(L)}$ if possible, otherwise $\Delta^{(C)}_2$. Small horizontal red lines cap the strict limit that the dimension estimates should not exceed, i.e., the state space dimensionality. The inner legend in the plot displays information about the distribution: the mean, and in parenthesis the percentage of values that exceed the dimensionality limit cap (the red line).

All in all the $\Delta^{(E)}$ estimates seem to match well those obtained by $\Delta_2^{(C)}$ with two notable exceptions: the method performs ``poorly'' for a quasiperiodic ($\Delta = 2$) trajectory of the H\'enon-Heiles system, for the chaotic attractor of the H\'enon map and for the Lorenz-96 chaotic attractor. Here ``poorly'' means significant inaccuracy in the first decimal digit. It is not clear exactly why the EVT method is not particularly accurate for these systems, however we can speculate the respective reasons. For quasiperiodic trajectories, the sampling time may be near-commensurate with one of the two periods, making some points very rarely visited in the finite set, even if they would be uniformly visited in the limit $N\to\infty$. As the method assigns higher dimension value to a point according to its visitation frequency, these rarely-visited points on the quasiperiodic orbit get higher dimension values than they should (within the EVT framework, the lower the visitation frequency of a state space point, the higher its local FD). For the H\'enon map, the only thing to note is that the attractor natural density is extremely singular and the assumption of Eq.~\eqref{eq:evt_additional_assumption} likely does not hold.
For the Lorenz96 and the coupled logistic maps, the thing to note is that the attractors have a high ``expected'' (here Lyapunov) FD, which means that for given amount of data and high underlying FD, the EVT underestimates the FD more than $C_2$.

\subsection{Quantile probability}
\label{sec:evt_quantile}
Unlike the entropy and correlation sum methods for computing FD presented so far, the EVT one is \emph{parametric}\footnote{While one could argue that estimating the slope of e.g., the entropy versus a size requires a choice of sizes, we counter-argue that the choice of sizes comes objectively and naturally from the dataset itself, as we illustrate in Sec. \ref{sec:estimating_slope}.}: it requires the choice of an ``extreme'' probability value $p$ for which to extract the quantile of $g_i$ when calculating the exceedances $E_i$ in Eq.~\eqref{eq:exceedances}. 
Therefore, before performing any further evaluation of the method, we must examine how it depends on its parameter $p$.
We have found no formal mathematical definition of an ``extreme'' in the literature of this EVT methodology, or how to practically compute an ``optimal'' value for $p$, or whether an optimal value exists at all. Ref.~\cite{Faranda2017} provides some methodology for checking whether the chosen $p$ is inappropriate that we evaluate in Sec.~\ref{sec:evt_significance}.

\begin{figure}[t]
    \centering
    \includegraphics[width=\columnwidth]{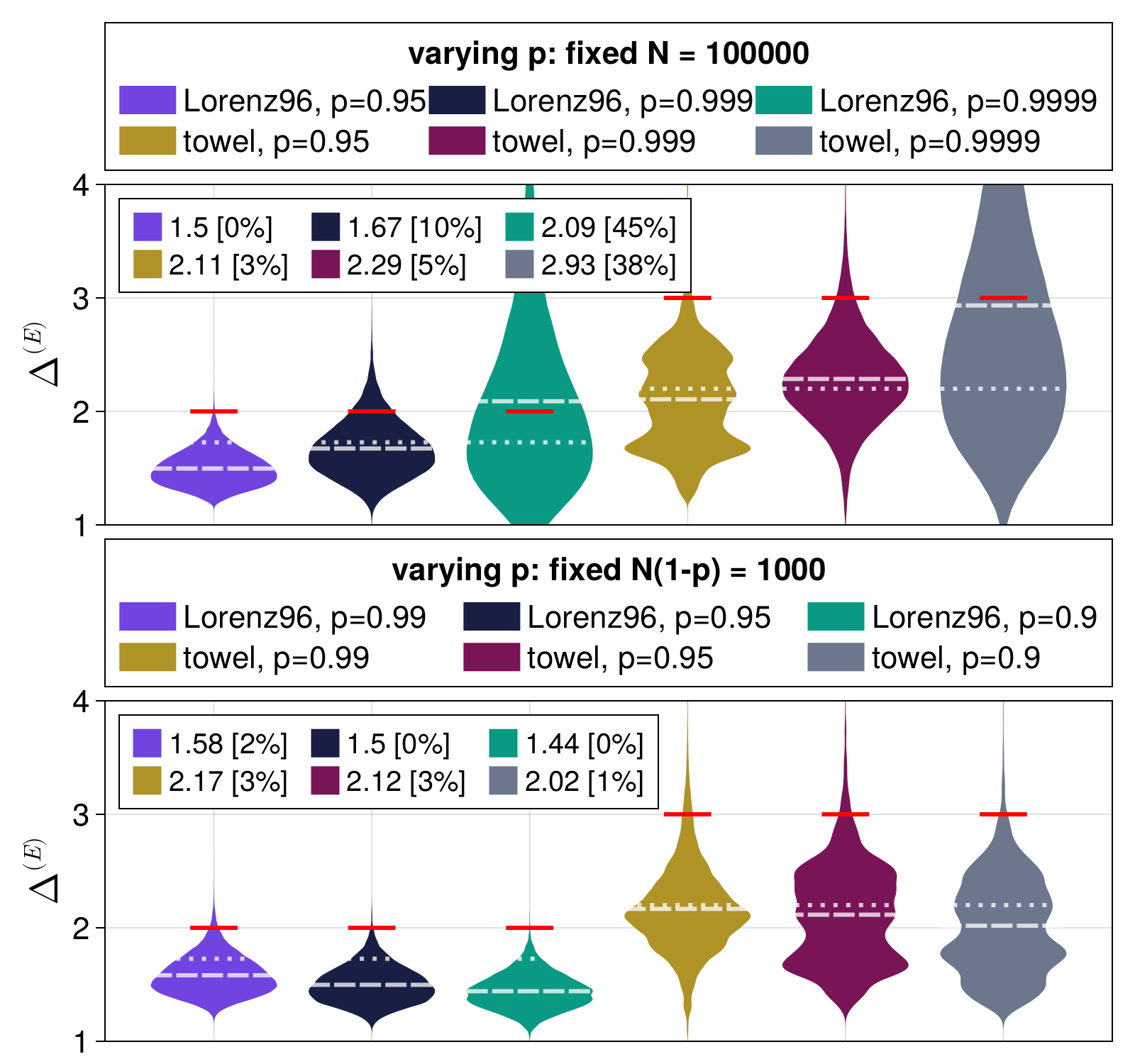}
    \caption{Impact of choice for extreme probability $p$ when estimating a quantile for Eq.~\eqref{eq:evt_distances}. Top: under fixed data length, bottom: under fixed length of exceedances. $D=8$ is used for Lorenz96 and $\Delta^{(E)}$ for it is divided by 4 as in other plots.}
    \label{fig:evt_quantile}
\end{figure}

In this subsection we examine the impact of choices of $p$. This choice is somewhat linked with the data length $N$, as the local dimension estimation for each state space point is done based on $N(1-p)$ points. In Fig.~\ref{fig:evt_quantile} we therefore vary $p$ with fixed $N$ but also co-vary $N,p$ with fixed $N(1-p)$.

The results show that increasing $p$ increases $\Delta^{(E)}$. It also appears also that not only the mean of the distribution of $\Delta^{(E)}_i$ depends on $p$, but the shape of the distribution as well. On one hand, it is somewhat re-assuring than once choosing a fixed $N(1-p)$ the results do not vary as wildly as when $N$ is fixed, provided that $N(1-p)$ is large enough. But we also noticed that under fixed $N(1-p)$ the estimated dimension seems to monotonically decrease, further and further away from the expected value, when decreasing $p$.

On the other hand, in most real world applications it is $N$ that is fixed and hence the results of the top panel of Fig.~\ref{fig:evt_quantile} are what is of most interest. Besides, even if it was possible to co-vary $N, p$ in a realistic application, we still cannot provide instructions of what the best choice should be for $p$: while Fig.~\ref{fig:evt_quantile} reveals the dependence on $p$, it doesn't lead to any obvious conclusions on what $p$ should be.
A saving grace here is that while there is a clear dependence on $p$, the mean value $\Delta^{(E)}$ does not change significantly (i.e., differences span less than one integer, which is anyways the accuracy we are interested in in practice), provided that $p$ remains in a range so that $N(1-p) \ge 100 - 1000$.

\subsection{Quantifying significance}
\label{sec:evt_significance}

The entropy or correlation sum approaches provide a relatively straightforward way to check for significance of results: there should a single slope covering several orders of magnitude $\varepsilon$. The larger the range of magnitudes, the more significant the results. Such a simple visual significant check does not exist for the EVT approach. In this section we examine possible ways to test for the significance of the EVT results based on what has been suggested in the literature or with alternative means we devised while composing this review.

In practical applications as in Ref.~\cite{Faranda2017} the authors identify a range of $p$ values that are appropriate using a statistical hypothesis test of whether $E_i$ follows an Exponential distribution (EXPD). Examples of such statistical tests are the Anderson-Darling~\cite{Anderson1952} or the Kolmogorov-Smirnov~\cite{Smirnov1948}. Here we used the Kolmogorov-Smirnov exact one sample test. The test proceeds as follows: for a given $p$, each of the $E_i$ are first fitted to a EXPD. Then, the fitted EXPD is used in a statistical test for the null hypothesis: ``the data coming from the given GPD''.
The test yields a p-value. Typically, if the p-value is very small, e.g., p<0.05, the null hypothesis can be rejected, which may mean unsuitable data all-together, or not enough data (e.g., quantile probability $p$ was chosen too high). In practice, one hopes that the majority of the p-values (each for each $E_i$) is significantly larger than some low threshold of e.g., 0.05.

Formally speaking however, a p-value greater than some threshold does not mean we can accept the null hypothesis; only that we fail to reject it. Any other distribution may have generated the data equally well. Hence, the convincing power of this line of argumentation (checking for large p-values) is weak from a statistical inference point of view. An alternative test mentioned in the literature is to check how stable the distributions for $\sigma$ of the fitted EXPDs are when varying $p$; but we didn't find this argument convincing (stability of parameters does not mean significant fit), so we ignored it.

A third way that one may judge the significance of the results is to directly estimate the error of the EX{D} fit for each exceedances vector $E_i$. To do this, we use a form of a normalized mean squarer error given by:
\begin{equation}
    \mathrm{NRMSE} = \sqrt{\frac{\sum_{j=1}^{k}(P_j - G_j)^2}{\sum_{j=1}^{k}(P_j - U)^2}}.
    \label{eq:nrmse}
\end{equation}
Here $P_j$ is the empirical probability density of the measured $E_i$ values at their $j$-th bin (i.e., the $j$-th bin's histogram height). $G_j$ is the fitted EXPD estimated at the bin's mid point. $U$ is the same as $G_j$ but assuming a uniform distribution fitted to data instead of a EXPD (hence, $U$ does not depend on $j$). The denominator normalizes the measure, so that the error of the EXPD fit is measured with respect to the error of the uniform distribution (and hence, it is meaningful to compare NRMSEs across different $E_i$). If NRMSE > 1, the uniform distribution is a better distribution model.
We have used this error measure in the past in varying scenarios, and noticed that values < 0.5 correspond to a  visually relatively correct fit~\cite{Isensee2019, Datseris2022} (without any mathematical guarantee of course).

\begin{figure}[t]
    \centering
    \includegraphics[width=\columnwidth]{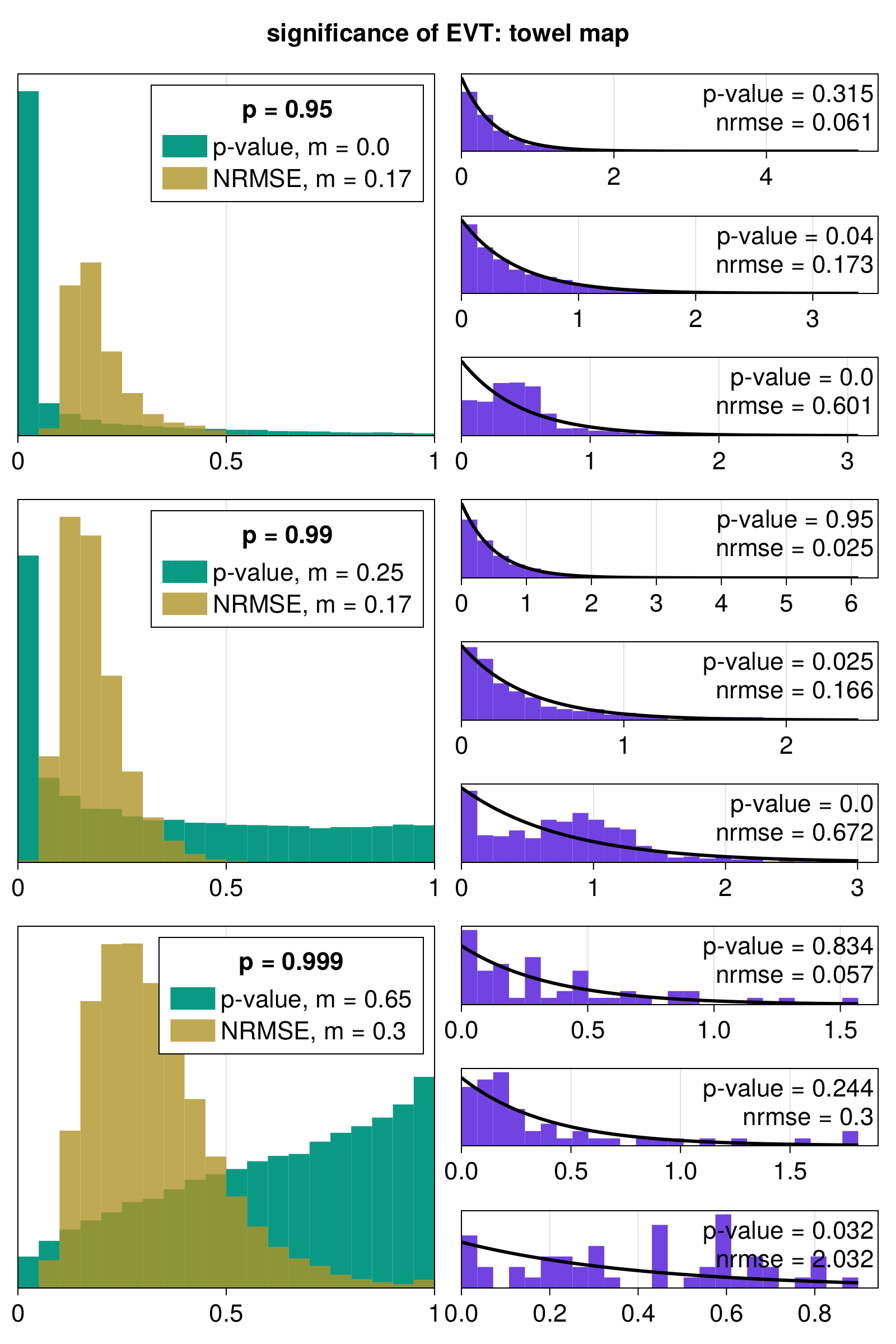}
    \caption{Distributions of p-values of the statistical hypothesis test of whether the data are coming from a EXPD, and the relative error of the EXPD fits normalized by the error of a uniform distribution. Inner legends show the distribution \emph{medians}. The figure is divided into three rows for three quantile probabilities $p$. Next to the distributions of p-values and NRMSE there are three exemplary plots of the distributions of the exceedances $E_i$ and with black line the fitted EXPD (the text indicates the corresponding p-value and NRMSE).
    The bottom panel has the NRMSE distribution clampled in [0, 1].
    The input dataset $X$ is a chaotic trajectory of the towel map. See discussion in Sec.~\ref{sec:evt_significance} for more.}
    \label{fig:evt_significance}
\end{figure}

In Fig.~\ref{fig:evt_significance} we show distributions of p-values and NRMSEs for a chaotic trajectory of the towel map. We provide more similar such plots in App.~\ref{app:more_significance}, which establish that our observations do not depend at all on the input set $X$. 

The results are very surprising. It appears that the p-value based test is unhelpful and/or misleading. For example, it shows that quantile $p=0.95$ is a bad choice, because the overwhelming majority of p-values are $\le$ 0.05, eluding to a rejection of the hypothesis ``the data come from a EXPD''. Yet, if we look at the actual fitted data to the right of the top panel, we do not observe ``bad quality'' fits at all. This is further established by the distribution of NRMSE values which has all of its mass in low values. We are not sure why the hypothesis test behaves this way in this scenario. 

On the other extreme of $p=0.999$ the p-value based test is again misleading. This is the case where most of the p-values are > 0.05, hence, it would be the most trustworthy in terms of the p-value test. Yet clearly, this is the case where the actual fits are the worst, by far. We have performed extensive numerical tests and are confident that the code implementations yielding the p-value are correct\footnote{We have repeated the numerical experiments with the approximate one sample Kolmogorov-Smirnov and the Cramer Von Mises tests and got practically identical results. We also tested obtaining p-values from data directly sampled from a constructed EXPD, and found the p-values uniformly distributed as is theoretically expected~\cite{Altman2017}.}.

This leads us to conclude that the NRMSE based test is much more trustworthy. If a practitioner wants to transform the NRMSE data into a Boolean decision ``is this $p$ okay?'', we would argue to check if the majority (i.e., 99\%) of the mass of the NRMSE distribution is less than 0.5.
Even so, the NRMSE test may only provide a range for $p$ where the EXPD fits are of sufficiently high quality. 
It cannot instruct how to pick a $p$ from that range. Thankfully, from what we have seen in Sec.~\ref{sec:evt_quantile}, the fluctuations of $\Delta^{(E)}$ with $p$ are relatively small if $p$ is in an appropriate range.

What made the discussion of this subsection difficult is that we have not found any information regarding the $p$-values of this test in the literature, despite the plethora of real world applications (see App.~\ref{app:evt}). 
While it has been mentioned that the $p$ chosen in these real world applications ``satisfy'' this p-value test~\cite{Faranda2017}, the actual p-value distributions were not shown.

In fact, we haven't found any discussions in general regarding the significance of the EVT results. 
The main argument the EVT FD literature has used in favor of significance of results is that the $\Delta^{(E)}_i$ distributions did not change much within a range of appropriate $p$.
But by itself this argument is not convincing of the \emph{validity} of the results, only of their \emph{stability}.
We will discuss these aspects again in Sec.~\ref{sec:evt_extreme} and in the conclusions.
For the rest of the manuscript we will be using $p=0.98-0.99$, as these values seem to yield correct results for synthetic $X$ with lengths $10^4-10^5$.

\subsection{Comparison with pointwise dimension}
\label{sec:evt_pointwise}

\begin{figure}[t]
    \centering
    \includegraphics[width=\columnwidth]{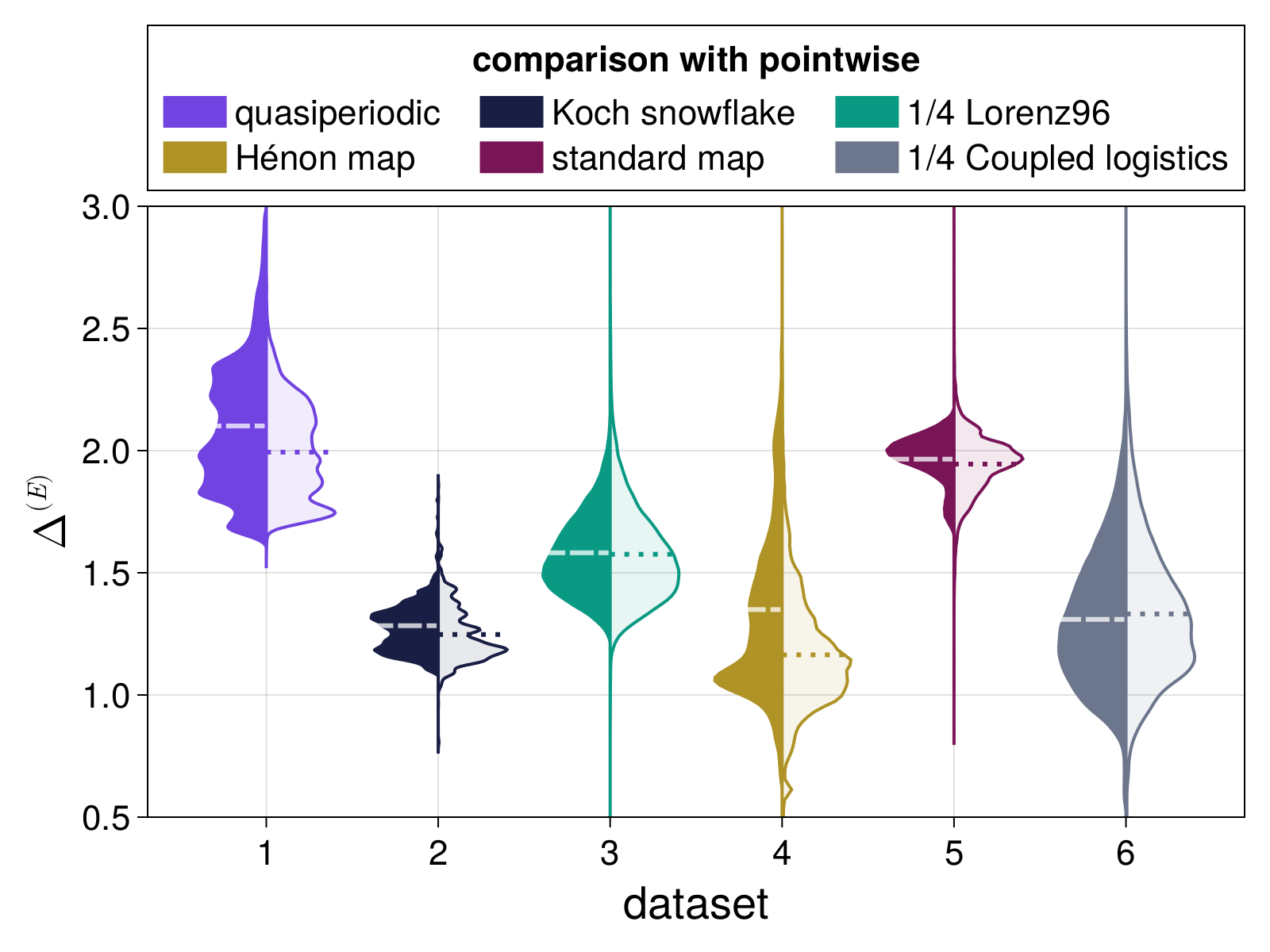}
    \caption{Comparison of local fractal dimension estimates using the extreme value theory (EVT) method and the pointwise dimension. Left and shaded distributions are EVT, right and transparent are pointwise. Dashed white lines are means of EVT, dotted colored lines are means of pointwise.}
    \label{fig:evt_pointwise}
\end{figure}

In Fig.~\ref{fig:evt_pointwise} we again compute $\Delta^{(E)}$ for exemplary sets as in Sec.~\ref{sec:evt_exemplary}, but now we compare it with the pointwise dimension, i.e., the scaling of the inner sum of Eq.~\ref{eq:correlationsum} versus $\varepsilon$. Our goal with this comparison is to see how well either method captures the ``spread'' of dimension values. We expect that for rather uniform fractal sets the distribution should be narrow, and wide for sets with highly non-uniform natural measure. Note that the average of the pointwise dimensions of the correlation sum \emph{does not coincide} with $\Delta^{(C)}_2$. As is also made clear in Ref.~\cite{Theiler1990}, $\Delta^{(C)}_2$ gives a more accurate result for the FD of the whole attractor because it utilizes more points to estimate the scaling behavior.

The most important result here is that for the chosen $p=0.99$ the two methods yield very similar results, hence establishing the overall accuracy of the EVT method for synthetic data. The pointwise dimension estimates however are more accurate for the H\'enon map attractor and a quasiperiodic orbit, which we already discussed in Sec.~\ref{sec:evt_exemplary}. Hence, $C_q$ is slightly more accurate for noiseless deterministic sets.

\subsection{Length, dimension, sampling time}
\label{sec:evt_aspects}

\begin{figure}[t]
    \centering
    \includegraphics[width=\columnwidth]{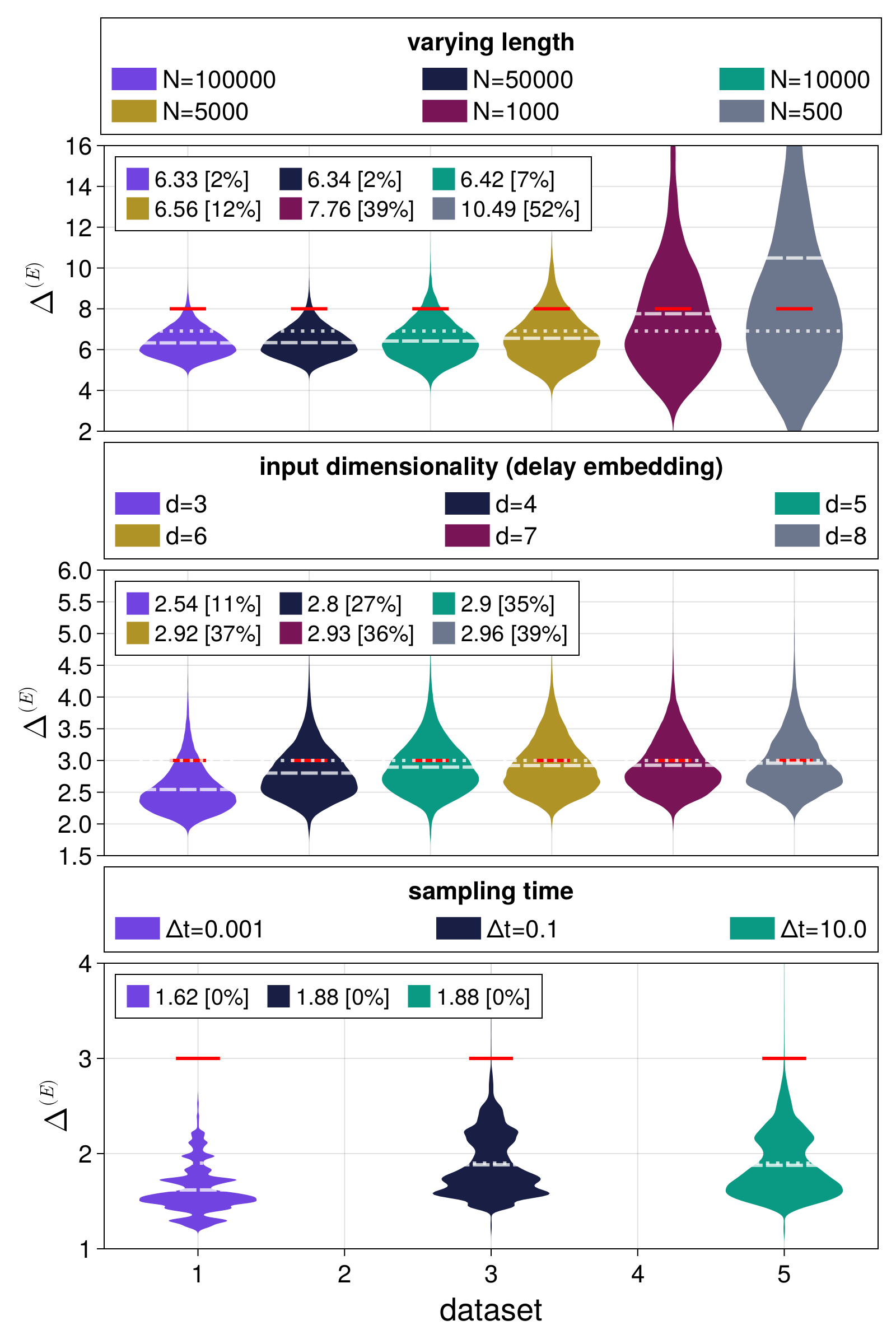}
    \caption{Fractal dimension estimates using the extreme value theory (EVT) when varying data aspects: length (top panel, using a chaotic Lorenz96 $D=8$ trajectory), dimensionality (middle panel, using delay embeddings of a chaotic H\'enon-Heiles timeseries), or sampling time (bottom panel, using a chaotic R\"ossler trajectory).}
    \label{fig:evt_dataaspects}
\end{figure}

We now test the EVT approach while varying various aspects of the input data: length, state space dimensionality (using delay embeddings of increasing dimension as in Sec. \ref{sec:main_delay}) and sampling time. The reason to judge the quality of the EVT approach versus sampling time is because, unlike the correlation sum approach which explicitly takes into account dense time sampling via the Theiler window $w$, the EVT approach is typically presented as agnostic to the sampling time. The analysis is presented in Fig.~\ref{fig:evt_dataaspects}.

Regarding data length, EVT scales well with decreasing $N$ up to a threshold. When $N$ becomes too low, so that $N(1-p)$ becomes less than 50, the results significantly loose accuracy, making the the estimated $\Delta^{(E)}$ increase rapidly. Interestingly, with decreasing $N$ the EVT \emph{overestimates} the FD instead of underestimating it like $H_2$ or $C_2$.
Note that one cannot simply fix this problem by reducing $p$ because, as illustrated in Sec.~\ref{sec:evt_quantile}, this has its own downside of decreasing the estimated dimension. Still, we may conclude that for the data considered here, with $N_\mathrm{max} > 10^3$ the EVT performs well, which is a better scaling with $N$ than $H_2$, but worse than $C_2$. Like with $C_2$ this data length $N_\mathrm{max}$ should scale with the FD, however there is no analytic treatment as to how (while for $C_2$ analytic bounds are discussed in Sec.~\ref{sec:length}). Given that real world data are typically small in length, one has to be particularly aware about this point.

As far as input dimensionality is concerned, we observe similar results as with $\Delta_2^{(C)}$ in Fig.~\ref{fig:embed}: $\Delta^{(E)}$ seems quite robust when increasing dimensionality.

When it comes to sampling time, the correlation sum approach utilizes the Theiler window, which in practice shortens the distance calculations from $N$ to $N-w$. This has negligible impact on data length but significant positive impact on the FD estimate in cases where data are sampled densely in time~\cite{DatserisParlitz}. Similar results are obtained for $\Delta^{(E)}$: for very small sampling times, the FD is biased towards lower values. Hence, it would make sense to include a Theiler window in Eq.~\eqref{eq:evt_distances}, i.e., only include $j$ with absolute distance from index $i$ greater than some $w$ like in Eq.~\eqref{eq:correlationsum}.
We note that Ref.~\cite{Buschow2018} also considered the impact of ``temporal neighbors'' and reached the same negative bias conclusion.

\subsection{Noise} \label{sec:evt_noise}
\begin{figure}[t]
    \centering
    \includegraphics[width=\columnwidth]{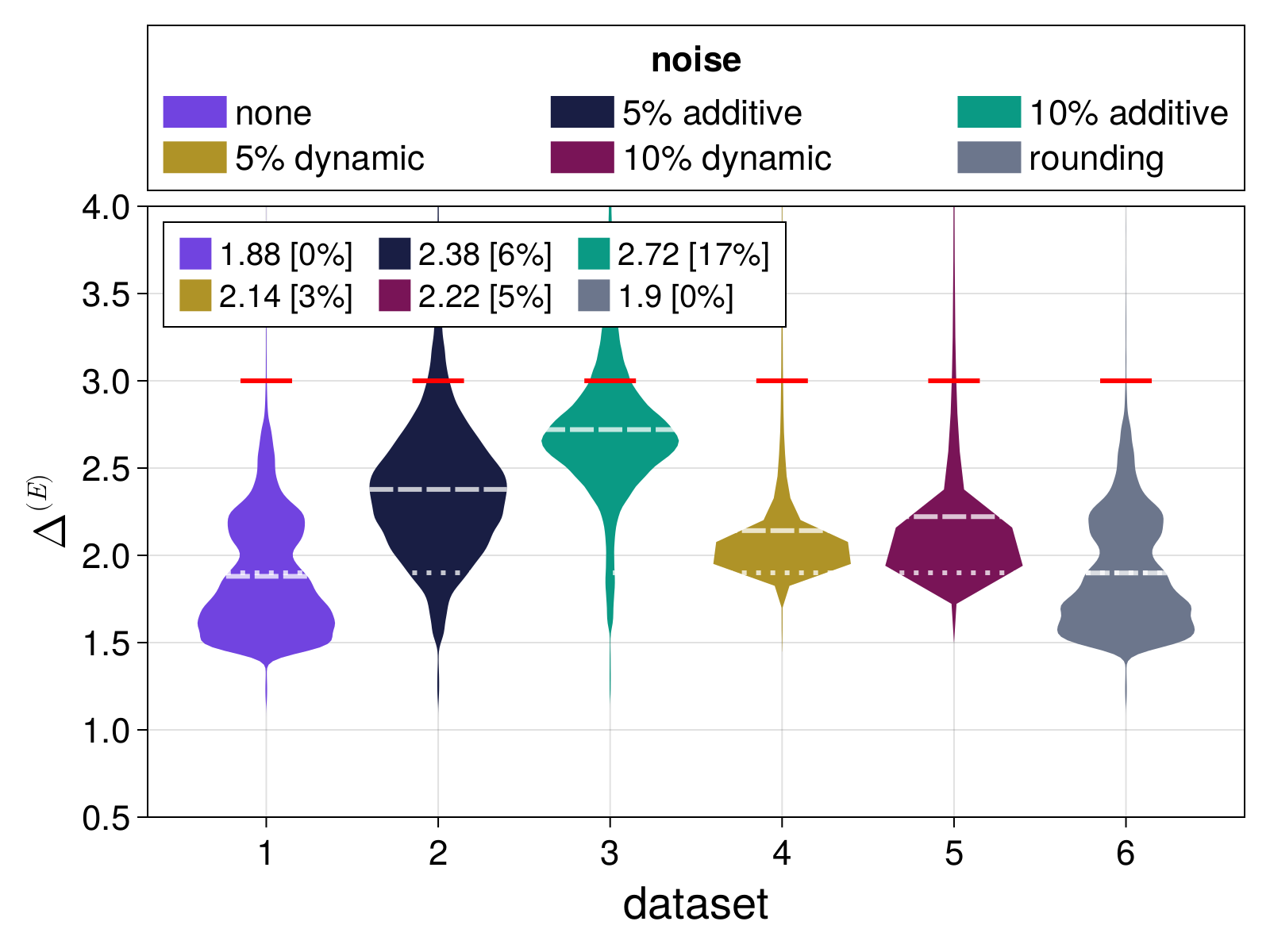}
    \caption{Impact of noise in the fractal dimension estimates of extreme value theory.}
    \label{fig:evt_noise}
\end{figure}

We repeat here the analysis of Sec. \ref{sec:noise} in Fig.~\ref{fig:evt_noise} using the R\"ossler system combined with various forms of noise. For additive noise $\Delta^{(E)}$ behaves much more similarly to $\Delta_2^{(H)}$ because the mean FD value of EVT is an average without focus at a specific scale. This means that the FD value in the presence of additive noise is between 1.9 (deterministic) and 3 (state space dimensionality). Additionally, the FD values of $\Delta^{(E)}$ are the smallest (and hence, closest to the deterministic FD value) out of the three ($E, C_2, H_2$). $\Delta^{(E)}$ also seems to be completely unaffected by rounding.

These observations make sense if one considers how $\Delta^{(E)}$ is computed (Sec. \ref{sec:evt_def}). The logarithms of all inter-point distances are taken into account for the computation of the quantile of $g_i$. Rounding will have negligible effect on the distribution of inter-point distances, and additive noise will have diminished effect due to being ``averaged out'' in some sense when computing the quantile. These properties make $\Delta^{(E)}$ preferable in the presence of noise, unless one wants to identify the noise radius, in which case $C_2$ is more suited.

Lastly, we mention that in the case of dynamic noise, $\Delta^{(E)}$ provides slightly higher values than $H_2$ or $C_2$ (but of course we don't know whether any of the three is the more ``correct'' number). It is worth noting nevertheless that the distribution of $\Delta^{(E)}$ for dynamic noise is much narrower than for other types of noise, which we found unexpected. We examined no further however, as the results with dynamic noise depend strongly on the system used and the exact form the noise was added, and hence can't lead to any general statements.

\subsection{Real world data}
\begin{figure}[t]
    \centering
    \includegraphics[width=\columnwidth]{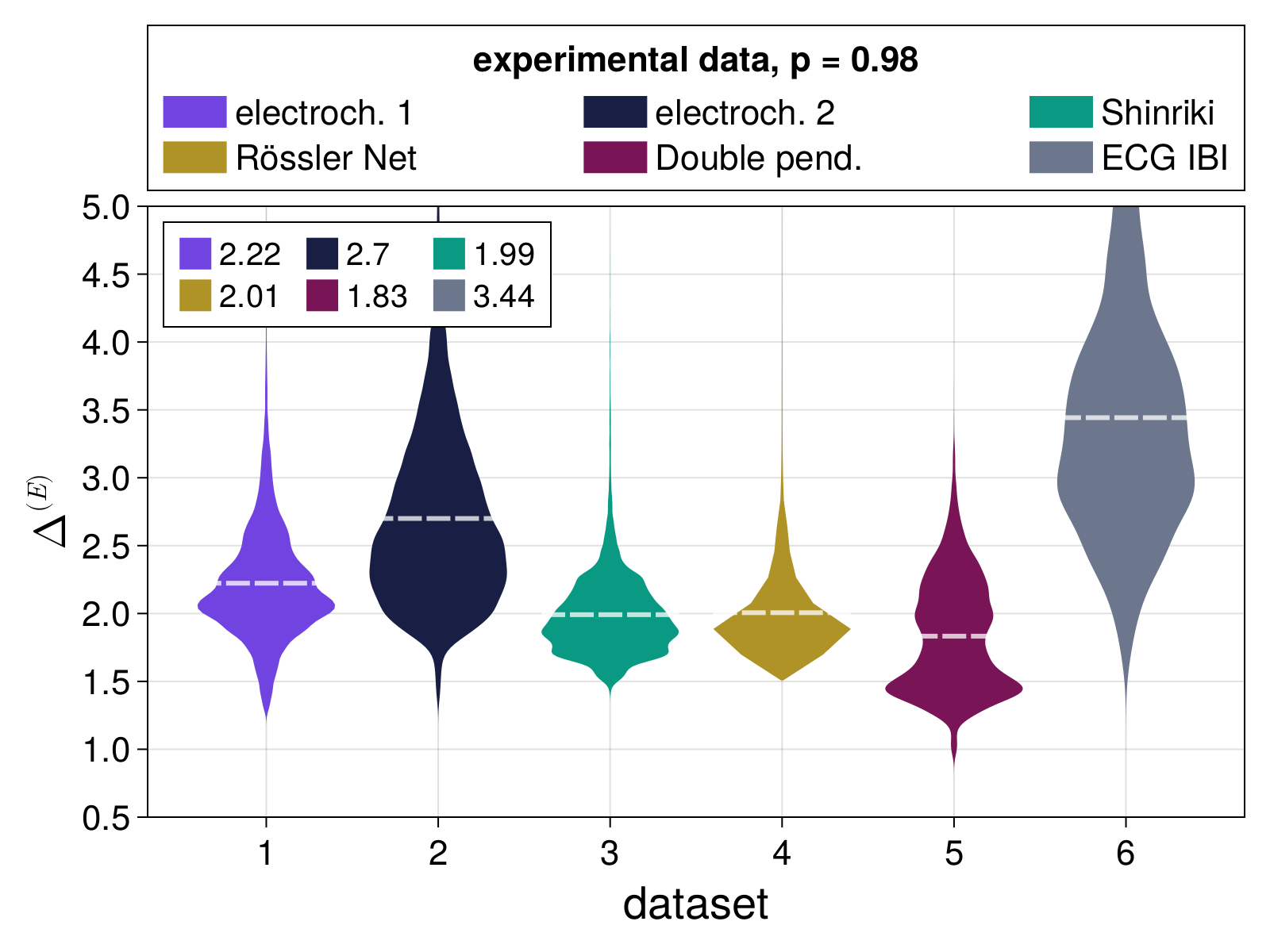}
    \caption{Fractal dimension estimates of extreme value theory for experimental data known to accommodate a low-dimensional deterministic representation.}
    \label{fig:evt_experimental}
\end{figure}

In Fig.~\ref{fig:evt_experimental} we use the same datasets as in Sec.~\ref{sec:experimental}. Because the datasets are smaller in size than the typical length we used before ($N=10^5$), we used $p = 0.98$ instead of $p = 0.99$. This value for $p$ also satisfies the ``NRMSE test'' we described in Sec.~\ref{sec:evt_significance}, in the sense of most NRMSE values being less than 0.5.

Besides the large extent of some of the distributions of $\Delta^{(E)}_i$, we do not notice any downside or incorrectness in the mean FD values: they are comparable with those coming from $H_2$ (and we can't know whether $H_2$ or EVT are more correct in their estimate). Given the results of the preceding Sec.~\ref{sec:evt_noise} this is expected due to the better response EVT has to noise (versus $C_2$).

\subsection{Extreme cases}
\label{sec:evt_extreme}
\begin{figure}[t]
    \centering
    \includegraphics[width=\columnwidth]{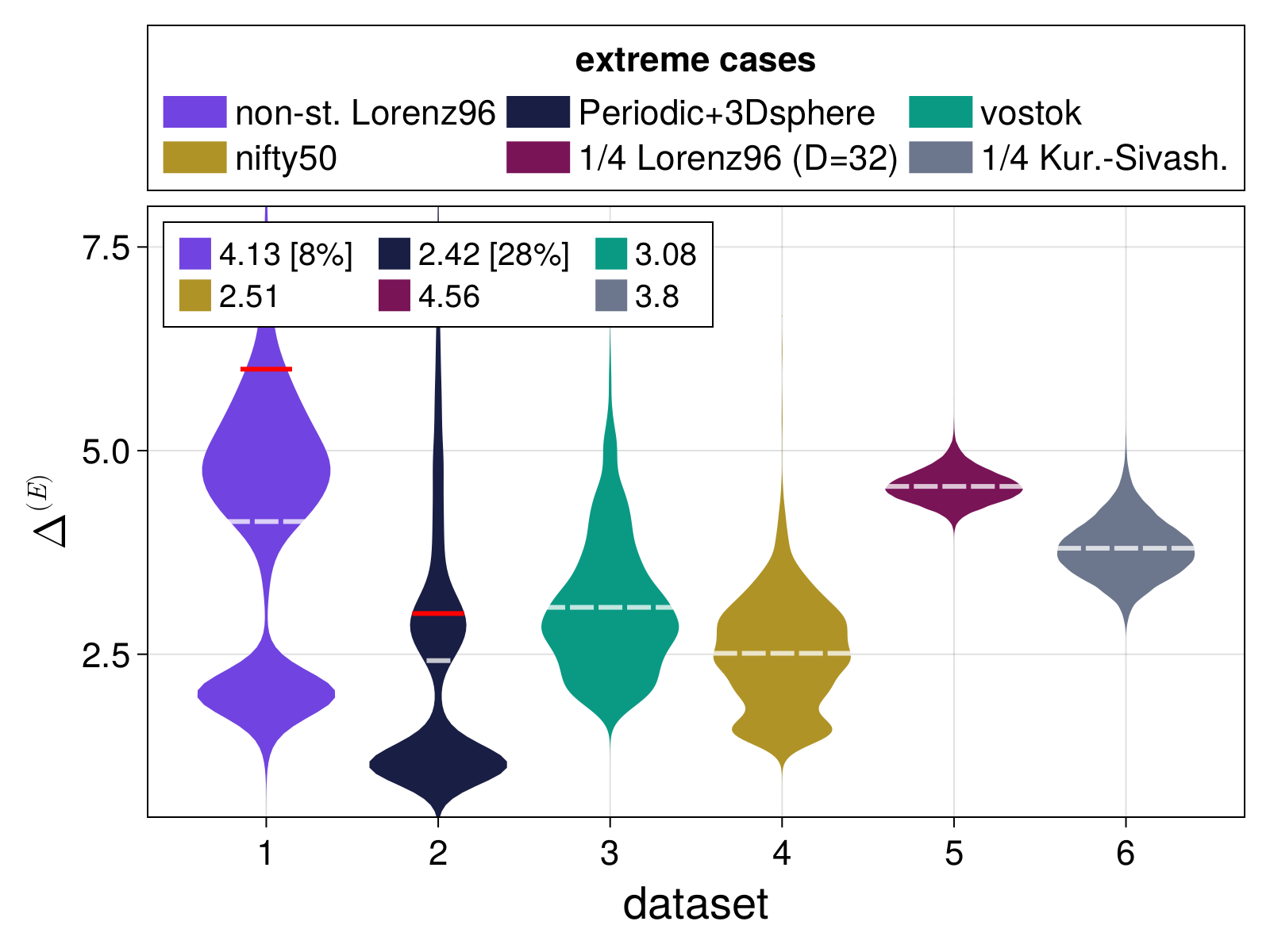}
    \caption{Application of extreme value theory fractal dimension estimates on extreme cases. Values for Lorenz96 and Kuramoto-Sivashinsky have been divided by 4 for visual purposes. We used $p=0.95$ for the ``nifty50'' and ``vostok'' data, because they are much smaller (otherwise $p=0.98$). However, mean FD values depend only weakly on $p$.}
    \label{fig:evt_extreme}
\end{figure}

In Fig.~\ref{fig:evt_extreme} we apply $\Delta^{(E)}$ in various extreme cases as in Sec. \ref{sec:inappropriate}.
$\Delta^{(E)}$ does a good job distinguishing that two sets of very different dimension have been artificially merged with each other in the first two cases of Fig.~\ref{fig:evt_extreme}, because it yields bi-modal distributions for $\Delta^{(E)}$. 
However, this is mainly due to the way that the sets were created. When compared to the case of adding noise to the data, we didn't see bimodal distributions with peaks at 2 and 3 dimension values, i.e., $\Delta^{(E)}$ can't be used to identify a noise radius unlike $\Delta^{(C)}_2$. 
Nevertheless, $\Delta^{(E)}$ could be a good tool detecting non-stationarity in an observed set, if that non-stationarity significantly changed the FD value over time.

Unfortunately, for the sets that do not accommodate a low-valued FD description, like the Vostok and nifty50, a straight-forward application of $\Delta^{(E)}$ gives a rather ``clear'' picture that a low-dimensional FD value describes the data. This is especially obvious in the nifty50 set (which we remind is a stock market timeseries), where $\Delta^{(E)}$ gives the lowest dimension values and a narrow distribution. Surrogate testing did not help here either: generated random-Fourier surrogates~\cite{Theiler1992} had consistently higher $\Delta^{(E)}$ than the original data, enhancing the wrong conclusion that the estimated $\Delta^{(E)}$ of the data is valid. Additionally, performing the significance tests of Sec.~\ref{sec:evt_significance} made things worse: the resulting plots (shown in Fig.~\ref{fig:evt_significance3}) look very similar to those obtained from systems with a legitimate low-dimensional representation, again re-enforces the wrong conclusion.
Lastly, we attempted to perform the standard analysis of checking whether the FD result converges with increasing embedding dimension of the timeseries. In App.~\ref{app:embedding_real} we find that, in contrast to the $H_2, C_2$ methods, the EVT method shows a convergence of the ``nifty50'' timeseries FD. Already in embedding dimension $d = 5$ it shows a constant mean $\Delta^{(E)}$ for any $d \ge 5$, which, again, re-enforces the wrong conclusion. Same results were obtained for the ``Vostok'' timeseries.

It appears that four different significance testing methods (p-values, NRMSEs, surrogate timeseries, convergence with increasing embedding dimension) all yielded with confidence that inappropriate data like ``vostok'' or ``nifty50'' have a small FD, which is incorrect.
These results have major implications for both previous and future applications of the EVT method in real world data. 
We haven't found a way to test whether the FD values yielded from the EVT method actually represent a low-dimensional deterministic system or not, i.e., there is no way to \emph{falsify} the method.
Hence, extreme care must be taken when applying the EVT method to arbitrary real world datasets, and whether the data accommodate a deterministic representation must be confirmed by other means (e.g., a self-consistent physical theory or using the correlation sum with increasing embedding dimension as is standard practice in nonlinear timeseries analysis~\cite{KantzSchreiber2003}).

For extremely-high dimensional, but deterministically chaotic data, $\Delta^{(E)}$ does an excellent job in identifying a very high FD (note the FD values are divided by 4 in the figure), that is also very close to the expected value $\Delta^{(L)}$. This was expected, as $C_2$ also does an excellent job identifying a very high dimension for clean data, and in general EVT and $C_2$ perform very similarly overall when it comes to deterministic noiseless data. For the Kuramoto-Sivashinsky example we see that EVT estimates higher FD than $C_2$ (while typically we noticed that for high dimensional data it under-estimates the FD when compared to $C_2$). However, in this example we are not sure the results are because EVT is indeed more accurate, or because EVT over-estimates the FD due to having too small $N$ when compared to the expected FD value (see discussion in Sec.~\ref{sec:evt_aspects}).

\section{A note on surrogate timeseries}
\label{sec:surrogates}
Surrogate timeseries have been recommended by Theiler et al. in the early 90s~\cite{Theiler1992} as a mean to test for nonlinearity in noisy timeseries.
It was suggested there that a discriminatory statistic for the test can be the FD computed via the correlation sum.
That requires a bit of care.
If one uses the algorithm we described here, i.e., using Eq.~\eqref{eq:correlationsum} and then deducing the slope of the largest linear scaling region, then the user risks estimating the fractal dimension of the noise (already existing in the original data) instead of the ``underlying deterministic nonlinear dynamics data'' (if any exist), thus invalidating the hypothesis testing approach in the first place. Other discriminatory statistics should be used instead, see~\cite{Lancaster2018} for example.

If a fractal dimension is chosen as a discriminatory statistic nevertheless, we propose the following alternatives for computing it: (i) use Takens' estimator (Sec. \ref{app:takens}) with the empirically good estimate $\epsilon_\text{max} = 0.25 \text{std}(x)$ with $x$ the original timeseries.
Because Takens' estimator performs a maximum likelihood estimation instead of a linear fit, and thus considers \emph{all} regions of the correlation sum up to some $\varepsilon_\text{max}$, it produces an ``average'' fractal dimension of the noise and the underlying data. It should be clear, however, that the Takens estimator is used in this context (just) as a discriminating statistics and its results must not be interpreted as meaningful dimension estimates.
Or (ii) use the EVT version $\Delta^{(E)}$ which also similarly produces an ``average'' of fractal dimensions of noise and deterministic set.


\section{Conclusions}
\label{sec:conclusions}
In this paper we have analyzed many different practically relevant fractal dimension (FD)  estimators we found in the literature. From all these estimators, we focused on an extensive quantitative comparison between the entropy-from-histogram method $H_q$ (Sec. \ref{sec:entropy} and App.~\ref{app:histogram}), the (potentially box-assisted) correlation-sum method $C_2$ (Sec. \ref{sec:correlationsum}) without any logarithmic corrections, and the extreme value theory method (EVT, Sec. \ref{sec:evt_def}). Based on our review, these three estimators are the ones most worthy using in practice (App.~\ref{app:all}).

To keep this paper within sensible limits, we have used a relatively small set of possible dynamical systems and real-world data. We cross-checked our results with different systems (not shown), and we are confident that the results presented are robust. Nevertheless, it is impossible to guarantee their universality for all possible input datasets. To amend this, the reader can repeat our extensive analysis for any other dynamical system or input dataset of choice by only changing a couple of lines of code in the provided code base, see App. \ref{app:software}.

Our conclusions are as follows (see also Fig.~\ref{fig:summary}).
When comparing $H_2$ with $C_2$, we found that for \emph{synthetic (i.e., noiseless) data} $C_2$ is clearly superior to $H_2$, retaining much better accuracy for decreasing $N$ (amount of points) increasing $D$ (state space dimension), or decreasing $\varepsilon$ (size scale). $C_2$ can also be used to detect the ``noise radius'', as e.g. illustrated by Kantz and Schreiber~\cite{KantzSchreiber2003}. For real data this can become a downside, making the FD estimation using $C_2$ ambiguous due to either an almost continuous change of slope or the majority of the slope reflecting the slope of the noise (Sec.~\ref{sec:experimental}). This can be partly alleviated by examining the behavior of $C_2$ with increasing embedding dimension $d$.
In Sec. \ref{sec:length} we saw that, provided that a deterministic chaotic attractor is known to generate the data, it is still quite sensible to estimate a FD of relatively small datasets using $C_2$, as the estimated FD remains very accurate even for relatively small data lengths. 
On the other hand, $H_q$ is very sensitive to data length and underestimates the FD strongly even for moderately small data (which is often the case in experiments).
We had difficulties making sense of correlation sum for order $q\ne 2$, even though it has been mentioned several times in the literature. $C_q$ for $q\ne2$ gives correct results only when one considers the slope at the largest $\varepsilon$ values and it is unclear why the slopes (i.e., FD values) at small $\varepsilon$ values are incorrect. As such, we suggest that when treating multifractality, the entropy method should be preferred, or, if one has too few data points (where $H_q$ performs poorly), then using $C_q$ and recording the slope of $\log(C_q)$ at largest $\varepsilon$ is the best alternative.

We then compared $C_2$ with EVT. In deterministic datasets we found a very high degree of agreement between $C_2$ and EVT, confirming EVT's accuracy. EVT performed equally well for high dimensional data but worse than $C_2$ for decreasing $N$. Still, EVT performs better with decreasing $N$, increasing $D$, or increasing underlying FD when compared to $H_2$. When it comes to noise, EVT appears to have similar results with $H_2$, i.e., the results are an average of the deterministic (smaller) FD and that of the noise. EVT reports the smallest FD values when contaminated by noise, and hence, EVT is affected less strongly by noise than $C_2$ or $H_2$. One more advantage of EVT versus $C_2, H_2$ is that it foregoes the identification, fitting, and extraction of slope from a scaling region, and hence does not face the same limitations when the data can cover only very small range of magnitudes of $\varepsilon$. This advantage is balanced by the disadvantage of EVT being much more of a black box method than $C_2, H_2$.

It appears that the EVT is a promising method that combines benefits from both $H_2, C_2$: it scales well with decreasing $N$ or increasing $D$ and is more tolerant to noise than $C_2$. However, we also observed that it has a huge downside: \emph{it cannot be falsified}. Or, at least, at the moment there does not exist a method (visual or statistical) in the literature that can confidently falsify it, nor that it can quantify its significance meaningfully. All four methods we utilized and discussed in Sec.~\ref{sec:evt_significance} (p-values, root-mean-squared errors, surrogate tests, increasing embedding dimension) failed to indicate that e.g., stock market timeseries are an inappropriate input (despite being non-stationary and not satisfying practically any of the assumptions underlying the EVT method). Instead, all ways to test for significance gave big confidence that stock market timeseries are described by a very small FD of $\approx 2.5$.

Surprisingly, we found practically no discussion in the literature about falsifiability, despite the plethora of real-world applications in very varied input datasets (Sec.~\ref{sec:evt_def}). 
Nevertheless, we believe that falsifiability is important, because with it one resolves controversies like the fractal dimension of ``global climate attractors''~Ref.\cite{Tsonis1993}.
The lack of falsifiability has major implications for both previous and future real world applications using the extreme value theory approach: the argument for low-dimensional determinism in the data cannot come from the EVT method itself, at least not at the moment.

It is clear, as it was before this work, that estimating a FD is not an easy task and hence focusing on only a single number can mislead.
The best practice we feel is to calculate several versions of $\Delta$, from different methods and with varying the parameters of each method (including the range of $\varepsilon$ or the quantile probability $p$) and produce a median of the results. Besides, given the software implementation we provide here, calculating all FD variants only necessitates a couple lines of code, see App.~\ref{app:software}. Furthermore, plotting of the appropriate quantities $-H_q, \log(C_q)$ versus $\varepsilon$ is a must, and can hint whether the methods are applied at inappropriate data. In addition, expecting more than one decimal point of accuracy is unrealistic in most practical applications.

As an outlook, we believe there is still some future research to be done regarding estimating FDs. Besides the Lyapunov dimension, which is not easily applicable to observed data, every other estimator disregards the time-ordering information in observed data (i.e., that the sequence of points follows the flow in state space instead of being randomly drawn samples on the attractor). Perhaps here is a way to make a more powerful estimator by using this discarded information of the time-ordering of the points in the dataset. For example, this time-ordering information has been used in e.g. estimating the transfer operator~\cite{Diego2019, ComplexityMeasures.jl}, which can also yield the natural density, and perhaps this operator can be utilized to create a FD estimator of higher accuracy or with better scaling with the number of points $N$. 

Regarding the EVT approach we believe future research can improve in two fronts: 1) developing a mathematically-rigorous framework for choosing $p$; 2) developing statistical tests for correctness and significance of the method's results that can successfully falsify the method with inappropriate data. With new statistical indicators, the method may be applied with more confidence to data of unknown dynamical origin.


\appendix 

\section{Algorithms for estimating a fractal dimension}
\label{app:all}

\subsection{Optimized histograms for arbitrary $\varepsilon$}
\label{app:histogram}
Here we describe an optimized method to calculate histograms, which to our knowledge neither has been published before, nor we could find a faster method that works for arbitrary $\varepsilon$.
The process has memory allocation scaling of $D\cdot N$ and performance scaling of $D\cdot N \log(D\cdot N)$, neither of which depends on $\varepsilon$. The processes is as follows.
Every point in the dataset is first mapped to its corresponding bin via the operation $\mathbf{b}_i = \lfloor (\mathbf{x}_i - \mathbf{x}_\text{min})/\varepsilon \rfloor$, where $\mathbf{x}_\text{min}$ is a vector containing the minimum value of the dataset along each dimension and $\lfloor \cdot \rfloor$ is the floor operation.
The resulting $\mathbf{b}$ ($N$ in total) are then sorted with a quick sorting algorithm, which results in all equal $\mathbf{b}_i$ being successive to each other. The sorting is lexicographic, i.e., sorting by first dimension, then by second, and so on.
We then count the successive occurrences of equal $\mathbf{b}_i \equiv \mathbf{b}_{i+1}$, which gives the amount of points present in the corresponding bin and move on to the next bin (there are no actual bins being created in memory, a bin is conceptually defined as the group of successively equal $\mathbf{b}_i$).
Dividing by the total amount of points gives the probabilities that can then be plugged into Eq.~\eqref{eq:renyi_entropy} to yield the entropy $H_q$.

\subsection{The Box-Counting Algorithm by Molteno}
This box-counting algorithm for the generalized dimension was introduced by Molteno \cite{Molteno1993} as an improvement for the method introduced by Grassberger\cite{Grassberger1990}.
It claims to be of order $D\cdot N$.
The algorithm partitions the data into boxes and counts the number of points in each box to retrieve the probabilities $p_i$ necessary to calculate the generalized dimension $\Delta_q^{(H)}$.
The faster runtime of the algorithm is due to the use of integer manipulations for the division of the data points into boxes.

To perform these manipulations, all data values are converted to unsigned integers.
These are calculated by finding the maxima $\textbf{x}_\mathrm{max}$ and minima $\textbf{x}_\mathrm{min}$ in each dimension to then identify the dimension that covers the largest range $X$.
\begin{equation}
    \textbf{x}_\textrm{Int}(\textbf{x}) = \left\lfloor2^{N_\textrm{bits}-1}\left(\frac{\textbf{x} - \textbf{x}_\textrm{min}}{X}\right)\right\rfloor,
\end{equation}
where $N_\textrm{bits}$ is the bit size of the unsigned integer type that was used and $\lfloor\,\cdot\,\rfloor$ is the floor function.

The first box contains all indices to the array that stores the data points, where a box refers to an array of indices. 
The boxes are subsequently partitioned, until the mean number of points per filled box falls below a threshold $k_0$, recommended by Molteno to be 10.

The $k$th step of partitioning divides the previous box into $2^D$ new boxes with the dimension of the data $D$. 
For a data point $\textbf{x}$, the index $i$ of its new box is calculated as 
\begin{equation}\label{eq:bitshift}
    i = \sum_{j=1}^D ((x_j \, \texttt{>>} \, (N_\textrm{bits} - k)) \, \texttt{and}  \, 1) \cdot 2^i.
\end{equation} 
Here \texttt{>>} is the logical shift operation, shifting the bit representation of the first value to the right by the second value, \texttt{and} is the bit-wise ``and'' operation. 
Thereby $(x_j \, \texttt{>>} \, (N_\textrm{bits}-k)) \, \texttt{and} \, 1$ translates to checking whether the bit at position $N_\textrm{bits}-k+1$ is one. 
In a cycle of partition, all boxes that contain more than one data point are split up and empty ones are deleted. 
To retrieve the probabilities $p_i$, the number of data points contained in each box is counted and divided by the total number of data points.

The algorithm provides exponentially scaled sets of values, that suit the approximation of the limit in Eq.~\eqref{eq:generalized_dim} and it only needs to compute the operation \eqref{eq:bitshift} $D \cdot N$ times per partitioning process. 
The main disadvantage is the static box size choice. 
It works well for low-dimensional data sets, but for larger dimensions the number of boxes increases exponentially with dimension, thereby increasing the number of points necessary to calculate the dimension exponentially.
This drastically limits the  $\varepsilon$ range to which the algorithm is applied over, and for some sets makes the computation of low accuracy or even straight-out impossible.
Given how fast our histogram algorithm already is (all $H_q$ curves in this paper took on average 0.1 seconds to compute), we saw no reason to use the Molteno method.
Also, the Molteno method does not allow the user to decide $\epsilon$ values, making it less flexible.

\subsection{Correlation sum}
\label{app:vanilla}
The original correlation sum can be calculated as given in (\ref{eq:correlationsum}).
For $q = 2$, the second sum can be changed to only include indices higher than the current one ($i$), 
\begin{equation}
    C_2(\varepsilon) = \frac{2}{(N-w)(N-w-1)}\sum_{i=1}^{N}\sum_{j=1+w+i}^{N} B(||X_i - X_j|| < \varepsilon)
\end{equation}
which makes the computation twice as fast. In all cases, the calculation time scales with $D\cdot N^2$ and therefore this method is exceptionally slow for high $N$.
Note that the version of the $q$-order correlation sum given in Eqn.~\eqref{eq:correlationsum} differs from the version provided by Kantz and Schreiber ~\cite{KantzSchreiber2003} in the exponent of $1/(q-1)$. 
If the $q$-order correlation given by Kantz and Schreiber is called $\bar{C_q}(\varepsilon)$, it scales as $\bar{C_q}(\varepsilon) \sim \varepsilon^{(q-1)\Delta^{(C)}_q}$, then $\bar{C_q}(\varepsilon)^{1/(q-1)} = C_q(\varepsilon) \sim \varepsilon^{\Delta^{(C)}_q}$ and both versions are equal.
Besides the exponent, the formulation of $C_q$ also differs in the indices of the outer of the sums which range from 1 to $N$, requiring a slightly changed normalisation.

\subsection{The Box and Prism assisted Correlation sum}
\label{app:boxassisted}
Theiler \cite{Theiler1987} proposed an improvement over the calculation of the correlation dimension by Grassberger and Procaccia\cite{GrassbergerProcaccia1983} that divides the data into boxes before calculating the distances between points. 
Thereby the number of distance calculations is reduced and the scaling becomes faster than $N^2$ (how much, it depends on the box size $r_0$).
After division into boxes the formula given in \eqref{eq:correlationsum} is used to calculate the correlation sum, therefore an extension to the $q$-order correlation sum is possible.

The integer representations of the boxes $b(\textbf{x})$ with side-length $r_0$ can be calculated by 
\begin{equation}
    b(\textbf{x}) = \left\lfloor\frac{\textbf{x} - \textbf{x}_\textrm{min}}{r_0}\right\rfloor
\end{equation}
for each point $\textbf{x}$ and the minimum of each dimension $\textbf{x}_\textrm{min}$.
This method is based on Appendix\eqref{app:histogram}, as Theiler does not specify a method for the calculation of the boxes.
The permutations needed to sort all representations in lexicographic order are calculated with a quick sort algorithm. 
The representations have the same ordering as the points they were calculated with, therefore sorting the points into boxes reduces to the iteration through the permutations and searching for changing integer representations.
If the integer representation changes, all previous permutations are stored in an array. 

For a point inside a box, there may be points outside the box that are within the given distance of the point.
Therefore the neighboring boxes w.r.t. the current box also have to be found and checked. 
The distances between the points in the box and its neighbors are calculated as given in Eq.~\eqref{eq:correlationsum}.
The first sum runs over all points of the initial box and the second sum uses all points in the box and adjacent boxes. 
For $q = 2$ only boxes with equal or larger indices are included in the neighbor search and the optimization, presented in the previous chapter, is used.

A point of criticism for this algorithm is its poor runtime for higher dimensions.
The advantage of distributing the points into boxes beforehand diminishes as the number of boxes increases considerably with dimension.
To reduce the amount of boxes, Theiler proposed a prism algorithm, where only the first $P$ dimensions are used to distribute the data into boxes.
These new boxes, where $P$ sides are of sidelength $r_0$ and the other $D-P$ sides cover the whole range, are called prisms.
The best choice given by Theiler is $P = 0.5\log_2N$ and should be used when $D$ exceeds $0.75\log_2N$. 
A downside of this prism approach is that for any $P<D$, some point pairs that should have been discarded may be included due to having small distances in the first $P$ dimensions but a larger distance in at least one of the remaining dimensions.

According to Theiler, the size of the boxes $r_0$ can be computed as 
\begin{equation}
    r_0 = R(2/N)^{1/\Delta_2^{(C)}},
\end{equation}
where $R$ is the size of the chaotic attractor and the dimension $\Delta_2^{(C)}$ is estimated by computation of the correlation sum for $\sqrt{N}$ points. 
Since these points are chosen randomly, the estimated dimension and in consequence the box size and even the final dimension can vary strongly.
A downside is that for some sets the proposed box size can drop below the minimum interpoint distance of the set.
Furthermore, our tests showed an irregular output value of $r_0$ with differences as high as two orders of magnitude for two box size estimates on the H\'enon map.

Bueno-Orovio and Pérez-García\cite{BuenoOrovio2007} proposed a different, and more stable, algorithm for the calculation of the optimal box size $r_0$.
They optimized the expected calculation times for the optimal number of filled boxes to
\begin{equation*}
    \eta_\textrm{opt} \approx N^{2/3}\left[\frac{3^{\Delta_2^{(C)}} - 1}{3^{P} - 1}\right]^{1/3}.
\end{equation*} 
$P$ is the number of dimensions used for the boxing, $D$ for boxes and $P$ for prisms.
Introducing the effective length of the chaotic attractor $\ell = r_\ell\eta_\ell^{1/\Delta_2^{(C)}}$ and solving $\eta_\textrm{opt} = (\ell/r_0)^{\Delta_2^{(C)}}$ for the box size $r_0$ yields 
\begin{equation}\label{eq:r_buenoorovio}
    r_0 = \frac{\ell}{\eta_{\textrm{opt}^{1/\Delta_2^{(C)}}}}.
\end{equation}
$\ell$ is the effective length, $r_\ell$ is the box size used to calculate the effective length with $r_\ell = R/10$ and $\eta_\ell$ is the number of filled boxes in case of the effective length.
$\eta_\ell$ is calculated by distributing $N/10$ points into boxes of size $r_\ell$.
The dimension $\Delta_2^{(C)}$ used for the calculation of $r_0$, $\eta_\textrm{opt}$ and $\ell$ is again estimated by computation of the correlation sum for $\sqrt{N}$ points.

The box size estimator still varies in its choice of box size but shows fluctuations of smaller amplitude than the Theiler estimator.
The choice of a box size smaller than the smallest interpoint distance only occurs for high dimensional data sets with a low number of points.
Furthermore, Bueno-Orovio and Pérez-García chose a prism dimension of always $P = 2$ (for $D>2$).
However a prism dimension of 2 can result in box size estimates smaller than the minimal interpoint distance for high dimensional datasets with comparably low size.
In this paper we used $P=2$ but we have different $r_0$, see below.

The main benefit of a small $r_0$ is that it makes the computations much faster. 
Unfortunately, both suggestions, and especially that of Ref.\cite{BuenoOrovio2007}, often fail in practice. 
They give much too small $r_0$ values and for data with any amount of noise whatsoever this value is well below the noise radius. 
This is displayed in Fig.~\ref{fig:noise}. The vertical dotted line shows the $r_0$ estimated by Eq.~\eqref{eq:r_buenoorovio} (the calculation of $C_2$ would be limited up to this $r_0$). 
In fact $r_0$ is so small, that even for only 5\% additive noise the computation would only show the noise dimension and no hint of the deterministic (and smaller) slope of $\log(C_2)$ vs. $\log(\varepsilon)$. 
That is why in this paper we decided to use the box-assisted version for better performance, but with $r_0 = \varepsilon_\mathrm{max}/e^2$ as discussed in Sec. \ref{sec:estimating_slope}. 
The performance is not too bad, e.g. for the typical data lengths $N=10^5$ considered here computing the entire $C_2$ curve takes about a minute on an average computer. 
Notice that even with the optimized-for-performance $r_0$ of Ref.\cite{BuenoOrovio2007}, the entropy-method is still massively faster (in fact it is even faster than just computing $r_0$).

\subsection{Fixed-Mass correlation dimension}
\label{app:fixedmass}

\begin{figure}[t]
    \centering
    \includegraphics[width=\columnwidth]{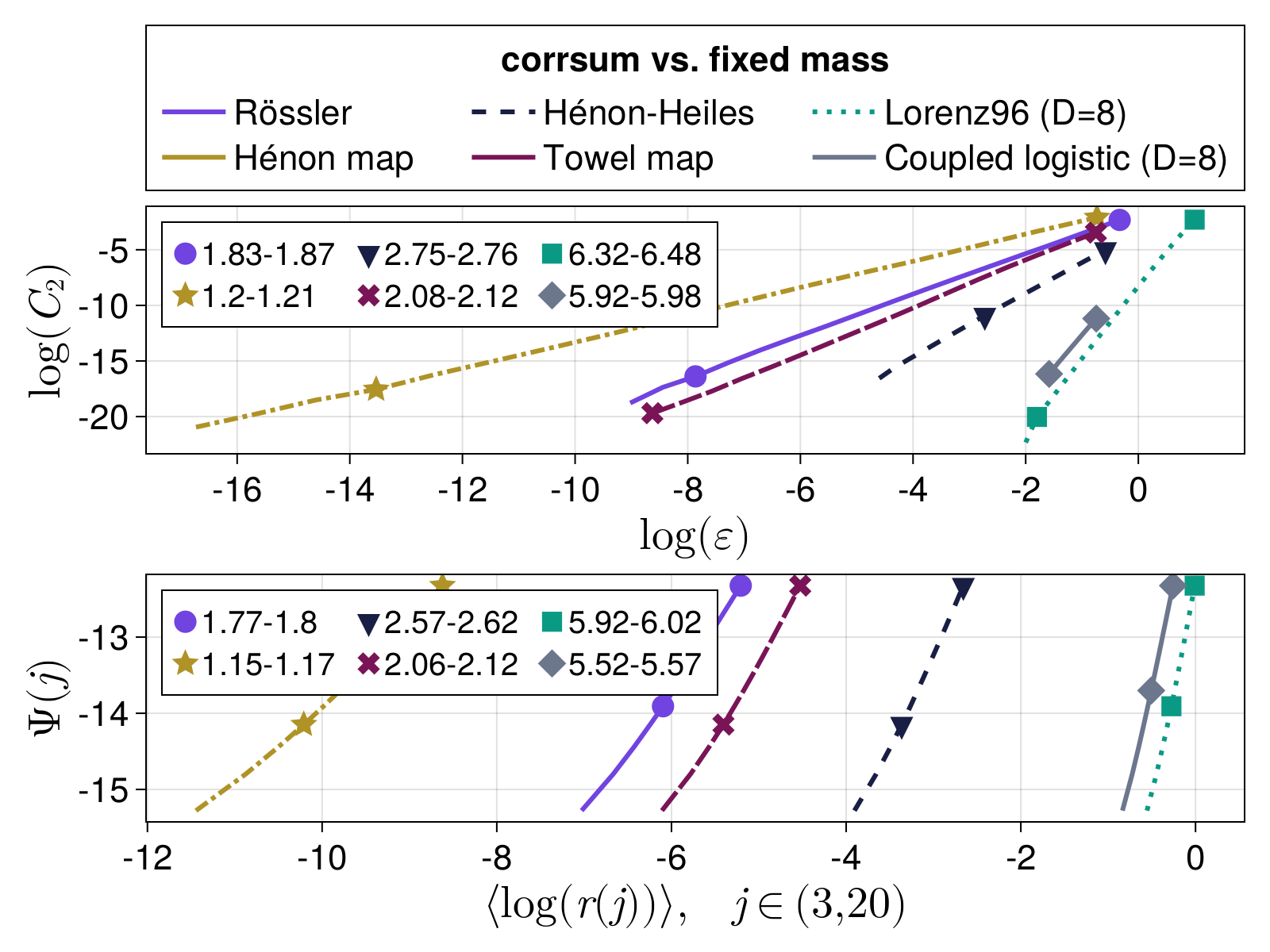}
    \caption{Fractal dimension estimates for sets coming from different dynamical systems with known dynamical rule to compare the correlation sum to the fixed mass formulation.}
    \label{fig:fixedmass}
\end{figure}

The correlation dimension stems from the assumption, that the probabilities $p_i$ scale as $\sum_i\left(p_i^q/\delta_i^{-(q-1)\Delta^{(M)}_q}\right) \sim 1$ for $\epsilon \to 0$ with $\delta_i$ being the diameter of the partition. 
In most methods that compute the dimension, the diameter is fixed and the probabilities are defined by the set.
Termonia and Alexandrowicz~\cite{Termonia1983,Asvestas1999,Grassberger1985} proposed a method using a fixed mass algorithm which does the opposite: the probabilities are fixed by defining $p_i = n/N$, where $n$ is a chosen number of points and $N$ the total number of points in the set.
The diameter is chosen to include $n$ points.
Following the explanation by Grassberger~\cite{Grassberger1988} the scaling can be rewritten to assume the form of Eq.~\eqref{eq:gen_fixmass}
\begin{equation}\label{eq:gen_fixmass}
   \overline{(r^{(j)})^{-\tau}} = \frac{1}{M}\sum_{i=1}^{M}\left(r^{(j)}_i\right)^{-(q-1)\Delta^{(M)}_q} \sim N^{q-1}\frac{\Gamma(j+1-q)}{\Gamma(j)}.
\end{equation}
$\overline{(r^{(j)})^{-\tau}}$ is the mean of all radii that contain $j$ points, $\tau = (q - 1)\Delta_q$, $M$ is the number of points of the set considered for the calculation, $\Gamma$ is the gamma function.
Eq.~\eqref{eq:gen_fixmass} is not solvable for $\Delta_q$ in general.
Eq.~\eqref{eq:fixmass} can be obtained from Eq.~\eqref{eq:gen_fixmass} from the limit $q \to 1$ and applying L'Hôspital's rule. 
\begin{equation}\label{eq:fixmass}
    \Delta_1 \overline{\log r^{(j)}} \sim \Psi(j) - \log N.
\end{equation}
$\Psi$ is the digamma function, $\Psi(x) = \mathrm{d} \log \Gamma(x)/\mathrm{d} x$.
With this general form that is not solvable this algorithm is restricted to $q = 1$.
We compared this version with the traditional correlation sum with $q=2$ and we found very similar accuracy in the estimated FD, provided that maximum $j$ was large enough. However, we also noticed that the fixed mass results spanned much less orders of magnitude in the estimated  $\overline{\log r^{(j)}}$, even for very high $j$, see Fig.~\ref{fig:fixedmass}. Furthermore, it is not clear to us how to estimate up to what maximum $j$ should the calculations be performed, i.e., it is not something that could be straightforwardly extracted from data as in Sec. \ref{sec:estimating_slope}. Regarding computational performance, the fixed mass algorithm is implemented using KDTrees~\cite{NearestNeighbors.jl}, giving very high performance for low-dimensional dataset, scaling poorly for very high dimensional ones. Hence, we decided to use only the traditional correlation sum version in the main comparison of Sec. \ref{sec:main}.

\subsection{Takens' Estimator}
\label{app:takens}
The estimator introduced by Takens\cite{Takens1985} aims to estimate the correlation dimension using the method of maximum likelihood estimation (MLE)\cite{mallet1986}. From $m$ interpoint-distances $\rho_j<\varepsilon_\text{max}$, one can estimate the correlation dimension as 
\begin{equation}\label{eq:takens-estimator}
    \Delta^{(T)} = -\frac{m-1}{\sum_{j=1}^m \log\frac{\rho_j}{\varepsilon_\text{max}}}.
\end{equation}
The derivation of this formula starts with the assumption that $\exists\,\varepsilon_\text{max}>0$, so that $\forall\,0\leq\varepsilon\leq\varepsilon_\text{max}$, the correlation sum 
\begin{equation}\label{eq:exactcorrelationsum}
    C_2(\varepsilon) = c\;\varepsilon^{\Delta_2^{(T)}}\quad\text{holds exactly},
\end{equation}
without any higher order terms, for some proportionality constant $c$.

The transformed variable $y_i=-\log(\rho_i/\varepsilon_\text{max})$ is distributed exponentially with parameter $\Delta^{(T)}$.
The log-likelihood-function of this distribution is given by
\begin{equation*}
    l(\Delta^{(T)};\lbrace y_i\rbrace) = \log L(\Delta^{(T)};\lbrace y_i\rbrace) = m\log\Delta^{(T)}-\Delta^{(T)}\sum_{j=1}^my_j.
\end{equation*}
$l$ is then maximised with respect to $\Delta^{(T)}$ to obtain the most likely value of the correlation dimension.
As Takens noted correctly, just taking the maximum of $l$ is biased\cite{Takens1985,borovkova1999}, which can be easily corrected by writing
\begin{equation*}
    \Delta^{(T)} = \underset{\textrm{correction}}{\frac{m-1}{m}}\underset{\textrm{original}}{\frac{m}{\sum_{j=1}^my_j}} = \frac{m-1}{\sum_{j=1}^my_j}.
\end{equation*}

For a Gaussian distributed random variable, the log-likelihood function is a parabola, that at $1\sigma$ has fallen by 0.5 from its maximum and at $2\sigma$ by 2.
By invariance\cite{zehna1966invariance}, this is also the case for a non-Gaussian random variable, letting us easily estimate the variance of $\Delta^{(T)}$.

\begin{figure}
    \centering
    \includegraphics[width=0.5\textwidth]{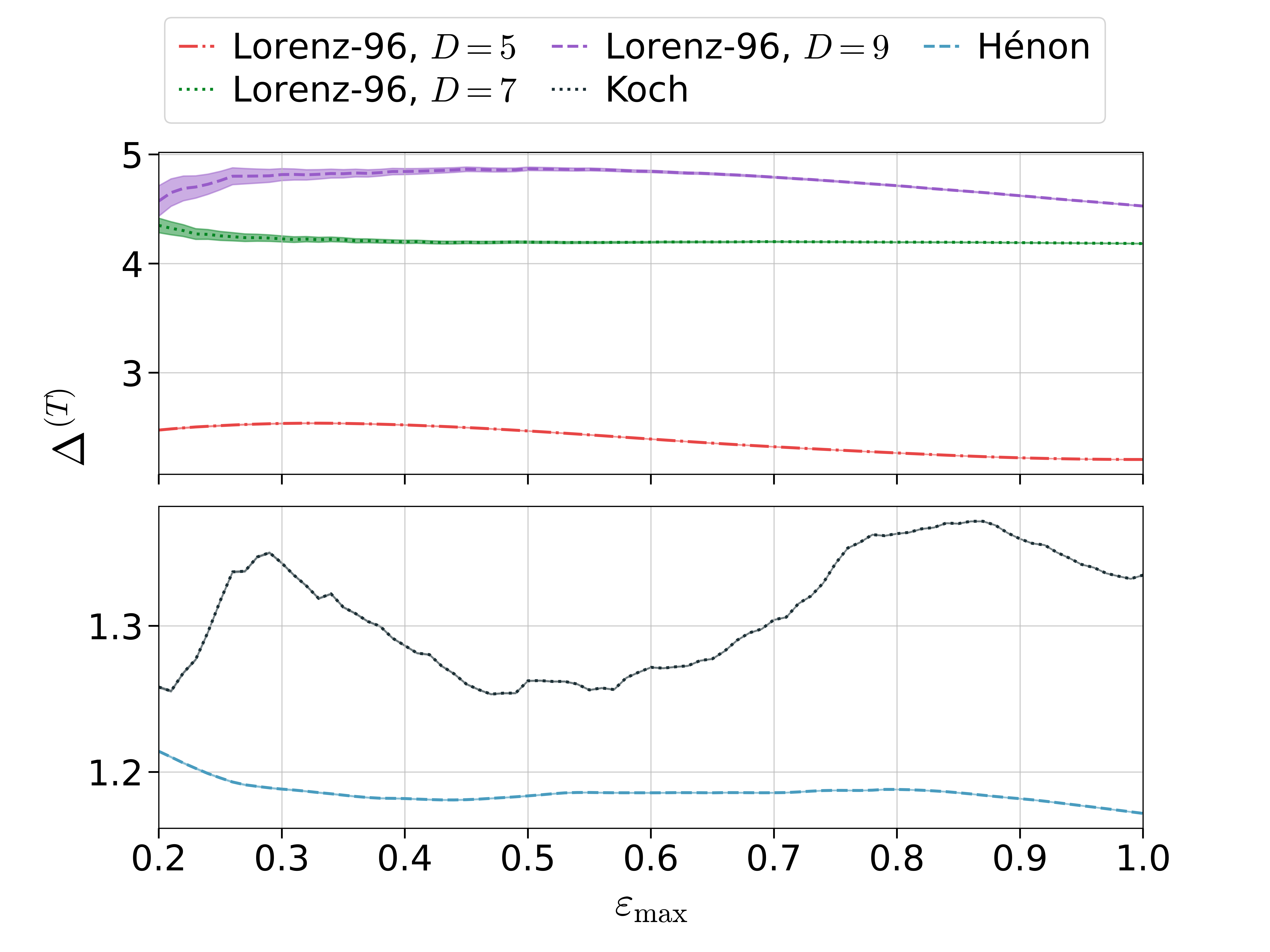}
    \caption{Dependency of $\Delta^{(T)}$ for different systems on the parameter $\varepsilon_\text{max}$. The shaded 5\%-95\% confidence intervals around the curves are not visible in most cases. Clearly, the variation of $\Delta^{(T)}$ over different values of $\varepsilon_\text{max}$ exceeds the confidence intervals.}
    \label{fig:takens_all}
\end{figure}

When testing the algorithm and its dependency on $\varepsilon_\textrm{max}$ (Fig.~\ref{fig:takens_all}) on different dynamical systems, we found that the variation of $\Delta^{(T)}$ exceeds the confidence intervals at any fixed $\varepsilon_\textrm{max}$ for low dimensional systems.

These variations occur because for the estimation, it is assumed that Eq.~\eqref{eq:exactcorrelationsum} holds. 
Thus, the estimated $\Delta^{(T)}$ and its confidence intervals are of no use as long as the validity of the assumption~\eqref{eq:exactcorrelationsum} is not known.

While Takens' estimator does not provide a significant advantage in the precise estimation of the fractal dimension, compared to correlation-sum based methods, it can be useful in the case of surrogate timeseries as described in Sec. \ref{sec:surrogates}.

\subsection{Judd's estimator}
Judd's ``improved'' estimator for the correlation dimension~\cite{Judd1992,Judd1994} also uses maximum likelyhood estimation.
To account for deviations from uniform scaling, it allows a polynomial $a$ of degree $t$, so that the assumed correlation sum is
\begin{equation}
    C_2(\varepsilon) \approx \varepsilon^{\Delta_2^{(J)}}(a_0+a_1\varepsilon+\ldots+a_t\varepsilon^t).
\end{equation}
It is stated that a degree $t\leq2$ is "usually sufficient".

\begin{figure}
    \centering
    \includegraphics[width=\linewidth]{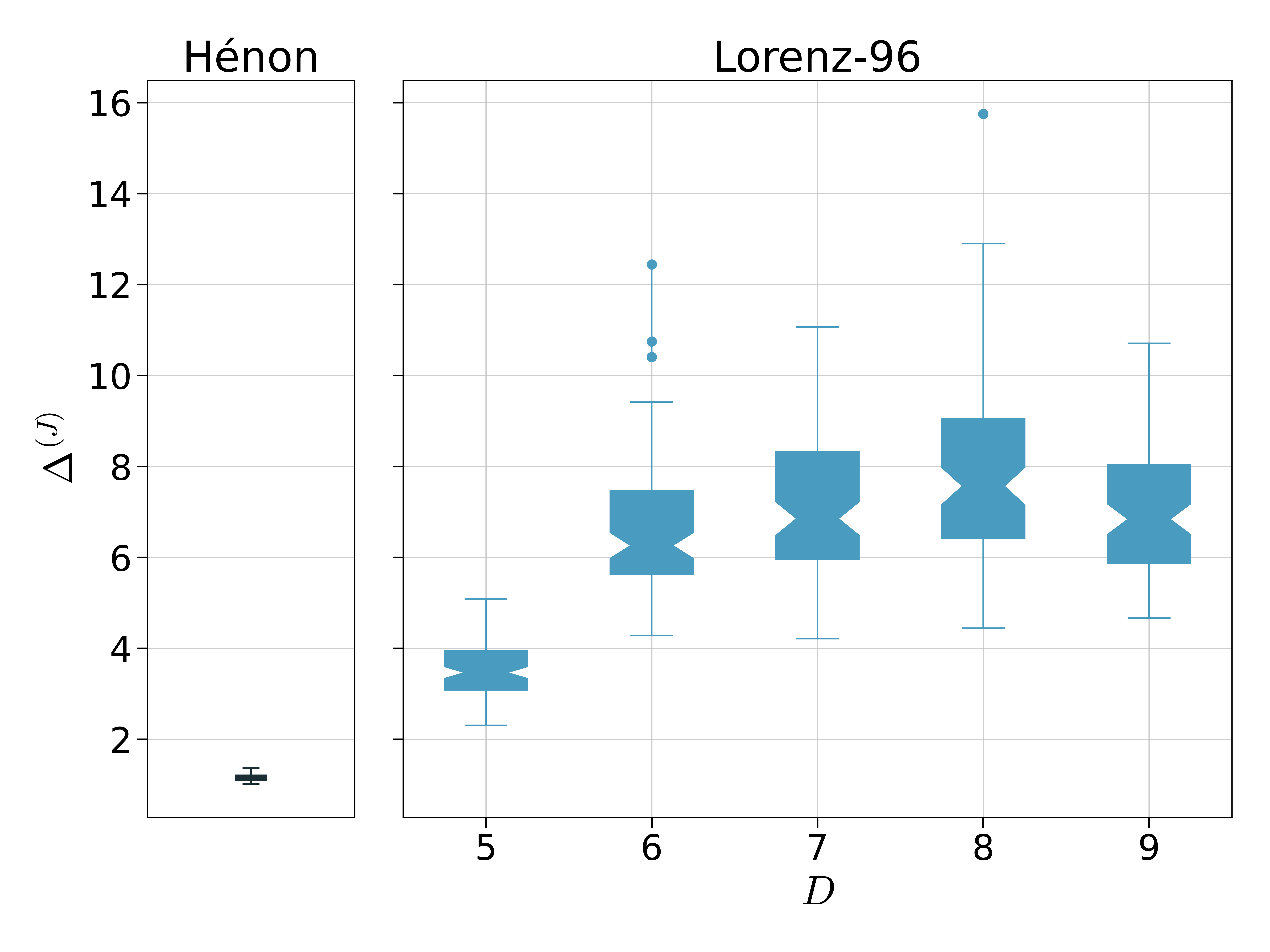}
    \caption{Estimates of $\Delta_2^{(J)}$ for different dynamical systems. For each box, 100 different samples have been drawn from a trajectory. Each individual sample contained 100 points. The degree of the polynomial was $\text{deg}(a)=1$.}
    \label{fig:judd_systems}
\end{figure}

The estimator performs a binned maximum likelihood estimation of $\Delta_2^{(J)}$ alongside the coefficients of the polynomial.
Judd therefore introduces a logarithmic binning, where the bins are defined by $B_0=[\varepsilon_0,\infty)$, $\varepsilon_i=\lambda^i\varepsilon_0$ and $B_i=[\varepsilon_i,\varepsilon_{i-1})$ for $i>0$ and $\lambda < 1$.
The parameter $\varepsilon_0$ is called the \emph{cutoff} and $w=\log(1/\lambda)$ the bin width.
Now, the probability of observing a distance in bin $B_i$ becomes $p_i=P_i-P_{i+1}$, with 
\begin{equation*}
    P_i = \left(\frac{\varepsilon_i}{\varepsilon_0}\right)^{\Delta_2^{(J)}}\left[a_0+a_1\left(\frac{\varepsilon_i}{\varepsilon_0}\right)+\ldots+a_t\left(\frac{\varepsilon_i}{\varepsilon_0}\right)^t\right]
\end{equation*}
If the bin contents $b_i$ are distributed multinomially, the negative log-likelihood function for the bin contents $\lbrace b_i\rbrace$ is
\begin{equation}\label{eq:log-likelihood-judd}
    l(\lbrace b_i\rbrace;\Delta_2^{(J)}, a) = -\sum_ib_i\log p_i + C,
\end{equation}
which must be minimized under the constraints
\begin{equation}\label{eq:judd-constraints}
    \sum_ip_i=1\qquad \text{and}\quad p_i>0.
\end{equation}
Eq.~\eqref{eq:log-likelihood-judd} does not necessarily only have one minimum because it depends on $t+2$ parameters and must be minimized numerically.

The optimal bin width for the estimator $w$ minimizes
\begin{equation}\label{eq:opt_w}
    \log w + \log \begin{pmatrix} n\\ b_1\ldots b_m \end{pmatrix} +  \log \begin{pmatrix} n+m+1\\ m \end{pmatrix},
\end{equation}
where $m$ is the index of the last bin for which the bin content $b_m\neq 0$ and $n$ is the number of considered interpoint distances.
Because the computation time of Eq.~\eqref{eq:opt_w} becomes unreasonably large for large numbers of distances, the estimator is restricted to small sample sizes.

Once the optimal bin width is found, the cutoff $\varepsilon_0$ is chosen as the right edge of the fullest bin of the histogram.
All bins to the right of this bin are joined to form $B_0 = [\varepsilon_0,\infty)$.

The minimization of Eq.~\eqref{eq:log-likelihood-judd} is subject to two difficulties that are already noted by Judd.
First, an optimizer cannot understand the idea that the exponential $\varepsilon^{\Delta^{(J)}}$ is the essence of the model, while the polynomial is only a device to correct for deviations from the scaling law.
Second, the optimizer is highly sensitive to the initial condition of the optimization, which could be reduced by a tailored optimizer\footnote{The optimization is split into to parts, where first $\Delta_2^{(J)}$ is optimized for a fixed $a$, and afterwards the coefficients of $a$ are optimized. 
This process is repeated until a convergence of the entire parameter vector, $(\Delta_2^{(J)}, a)^T$ is observed.}, but still one can observe a very broad distribution of $\Delta_2^{(J)}$ values over different samples of a long trajectory, especially for higher-dimensional systems, as is shown in Fig.~\ref{fig:judd_systems}.

Due to these problems, we decided not to include the estimator in the main comparison.

\subsection{Dimension from Lyapunov exponents}
In Sec. \ref{sec:lyapunov} we described the Lyapunov dimension $\Delta^{(L)}$ due to Kaplan and Yorke. We do not have anything to add here regarding $\Delta^{(L)}$, but we want to mention Ref.~\cite{Chlouverakis2005} by Chlouverakis and Sprott.
They suggest that instead of a linear interpolation to the sum of $\lambda_i$, a polynomial interpolation should be done instead.
However, as we found no theoretical foundation for this proposal, we decided to skip it (and we also did not notice any significant improvement with the numeric results).

\subsection{Mean return times}
According to the Poincar\'e recurrence theorem, any trajectory within an ergodic set\footnote{For the purposes of this paper we assume that all attractors are ergodic.} will return arbitrarily often and arbitrarily close to any neighborhood in the ergodic set\cite{Datseris2022}.
We represent this neighborhood as a hypersphere of radius $\varepsilon$, centered at some point $\mathbf{x}_0$ in the ergodic set, and define as $\gamma$ the mean return time to this hypersphere.
Then one expects that $\log(\gamma) \approx - \Delta^{(\gamma)} \log(\varepsilon)$ with $\Delta^{(\gamma)}$ the fractal dimension, as estimated from the mean return times.
A more formal discussion of this fact, and explicit connection with the natural measure of the ergodic set and the fractal dimension obtained via the generalized entropy \eqref{eq:generalized_dim} is given by Theiler~\cite{Theiler1990}.
Earliest reference we found using return times to estimate fractal dimensions was Ref.~\cite{Jensen1985}.

Unfortunately, the method using mean return times is not recommended at all. A fundamental limitation is that knowledge of the dynamic rule $f$ is necessary, otherwise the results of the method for measured data are too inaccurate to be considered seriously. Even for known rule $f$, the method converges slowly (numerically). Furthermore, it provides an estimate of the local dimension around the point of return, similarly with the point-wise dimension.
Thus, it has to be further averaged over several state-space points, requiring several orders of magnitude more computation time than the correlation sum method or the Lyapunov dimension method.
Lastly, we found its numeric output (not shown, see online repository) to be quite far from the results of the correlation dimension and thus we do not consider it accurate enough.

\subsection{Extreme value theory}
\label{app:evt}

We introduced the algorithm of estimating a FD via extreme value theory (EVT) in Sec.~\ref{sec:evt_def}. Here we will expand more on how the algorithm has been used in the literature and highlight potential ambiguities we have noticed in its real-world applications.

Recall from the discussion of Sec.~\ref{sec:evt_def} that the exponential distribution (EXPD)  the extremes of the $g(\varepsilon)$ function (Sec.~\ref{sec:evt_def}) follow is in fact a simplification of a Generalized Pareto Distribution (GPD). The cumulative function of GPD is
\begin{equation}
    F(x) = 1 - \left(1 +\frac{\xi x}{\sigma}\right)^{-1/\xi}  
\end{equation}
and is valid for $x \ge 0$ if $\xi \ge 0$ and for $0 \le x \le - \sigma/\xi$ if $\xi < 0$. It reduces to EXPD for $\xi = 0$. The first ambiguity we found in the literature concerns the recently published work of Pons et al.~\cite{Flavio2023}, as well as real-world applications that explicitly fit a GPD to data (i.e., allowing $\xi \ne 0$ instead of enforcing $\xi = 0$). It is correct that the exceedances $E$ in a dataset may follow a GPD, but if one wants to make the claim that the $\sigma$ parameter is connected with the fractal dimension, then one must assume that the data follow the reduced EXPD instead. That is because of Eq.~\eqref{eq:evt_equivalence}. If one uses the GPD cumulative function instead of the EXPD one, one would get the expression $\exp(-E\Delta_i)$ to be equated to $\left(1 + \xi E/\sigma_i\right)^{-1/\xi}$, from which it is impossible to claim $\Delta_i = 1/\sigma_i$ like in Eq.~\eqref{eq:evt_equivalence}. Hence, when using GPD fits instead of EXPD one cannot simply equate the (local) fractal dimension with the $\sigma$ parameter of the fit, in contrast to what has been done in the literature\cite{Flavio2023}.

The second ambiguity we encountered is the lack of application, and even discussion, of delay coordinates embeddings. Many real-world applications, e.g., Ref.~\cite{Faranda2017}, analyze a single dynamic variable (such as sea level pressure in the case of climate applications). This dynamic variable is a spatiotemporal field, and hence provides a multi-dimensional input dataset. Despite the high input dimensionality, this single variable is likely coupled to many other dynamic variables in a coupled dynamical system (which is especially true in the case of climate dynamics). The theory of delay embeddings is supposed to re-construct the missing dynamic variables and as a result provide a more correct representation of the dynamical flow. Delay embedding is missing from almost all applications of of EVT we reviewed and cited, even though they all use the timeseries of only one dynamic variable. Given that delay embedding is a well established analysis step~\cite{Kantz1993} that is completely separate from estimating fractal dimensions, we are not sure why there is a lack of discussion of it.

The third ambiguity we want to highlight is the report of relatively small values for the fractal dimensions of spatio-temporal, and highly complex, real-world data. For example, Ref.~\cite{Gualandi2020} reports dimensions of $\Delta \approx 3.5$ for the dynamics of slow earthquakes in the Cascadia region, Ref.~\cite{Giamalaki2021} reports $\Delta \approx 15$ for spatiotemporal atmospheric flow (of daily resolution; hence, large scale turbulence is considered), Ref.~\cite{Hochman2021} report a difference of at most $\delta \Delta_i \approx 2$ from the states of largest and smallest local fractal dimension of the 500-hPa geopotential height (Z500) dynamic variable for the weather of the Europian-Atlantic region. Especially in the last case this would mean that, out of the potentially millions of available degrees of freedom in this (discretized) spatiotemporal system, there is a difference of at most two additional degrees accessed by the state space flow between the least and most stable regions in state space. Given that in this review we noticed much higher differences of local dimensions in much lower dimensional systems (see e.g., Figs.~\ref{fig:evt_exemplary} or \ref{fig:evt_extreme}), we find this reported small number difficult to grasp. In general, given the discussion on falsifiability of Secs.~\ref{sec:evt_significance} and \ref{sec:evt_extreme}, as well as limitations that come from the length of input data that we discussed in Sec.~\ref{sec:length}, we believe that the absolute value of the fractal dimensions reported in these publications should be taken with a grain of salt, and not be equated with the available degrees of freedom in the state space. Whether or not the \emph{relative} values of the local dimensions $\Delta_i^{(E)}$(in the sense that relatively higher value means higher local state instability for the real system) can be used to draw conclusions or not depends on the confidence one has on the stability of the distributions of $\Delta_i^{(E)}$, see Sec.~\ref{sec:evt_significance}.

\subsection{Persistent homology}
Persistent homology methods are based on topological timeseries analysis and applications in dynamical systems. These techniques have been relatively recently applied to estimate fractal dimensions, and a quantitative review was recently published by Jaquette and Schweinhart~\cite{Jaquette2020}.

The methodology is based on tracking how $d$-dimensional holes form or disappear as the point cloud that composes $X$ is ``inflated'' or ``thickened''. This means that each point in $X$ is taken as a sphere with radius initially 0, and this radius is increased as the point cloud is ``inflated''. Estimators based on 0-dimensional persistent homology using minimal spanning trees were proposed already in the early 1990's by van der Weygaert et al. \cite{Weygaert1992} and Mart\'inez et al. \cite{Martinez1993} who stress that this approach provides an estimate of a (generalized) fractal dimension and also works with relative small data sets. For more information about implementations of the method, see Ref.~\cite{Jaquette2020}.

The review of Ref.~\cite{Jaquette2020} compared fluctuations in the output values and the distance of the output values themselves from the reference ``true'' values of the test sets applied to. The results showed that the persistent homology method has similar performance for $d=0$, and dramatically worse performance for $d > 0$, when compared to the correlation sum. Unfortunately, for $d=0$ the method performs poorly when noise is present, i.e., it does not distinguish two slopes (of the noise and of the deterministic set) and instead shows that of the noise, while it does find two slopes for $d=1$. Additionally the method output depends strongly on its meta-parameter $\alpha$, whose value cannot be deduced from input data. For these reasons, and because the methodology itself is more complicated to both explain and implement than the correlation sum of Sec. \ref{sec:correlationsum}, we deem the method worse than the correlation sum and only considered the correlation sum for a more in-depth comparison in Sec. \ref{sec:main}.

\section{Software implementations and code base}
\label{app:software}
The work done in this paper, as well as the figures produced, are available as a fully reproducible code base, which can be found on GitHub~\cite{GitHub}. It is written in the Julia language~\cite{Julia-2017}, and is using the software: DynamicalSystems.jl~\cite{Datseris2018}, DifferentialEquations.jl~\cite{DifferentialEquations.jl-2017}, BenchmarkTools.jl~\cite{BenchmarkTools.jl-2016}, ComplexityMeasures.jl~\cite{ComplexityMeasures.jl}, LsqFit.jl, and DrWatson~\cite{DrWatson2020}. Figures were produced with Makie~\cite{Makie}.
All methods, with the exception of Judd's algorithm and the persistent homology method, are implemented, documented, and tested extensively, in the FractalDimensions.jl~\cite{FractalDimensions.jl} submodule of DynamicalSystems.jl.
The implementations follow best practices in scientific code~\cite{goodscientificcode} and are highly optimized, utilizing multi-threading whenever possible.
The following code is a simple example of calculating $\Delta^{(C)}_2$, $\Delta^{(H)}_1$ and $\Delta^{(E)}_i$ with DynamicalSystems.jl and the Julia language:

\begin{lstlisting}[language=Python]
using DynamicalSystems
# some input data (chaotic Roessler system):
ds = Systems.roessler()
X, t = trajectory(ds, 10000; Dt=0.1, Ttr=100)
# Estimate lengths as in Sect. III I:
es = estimate_boxsizes(X)
# Estimate correlation sum, App. A 4:
Cs = boxed_correlationsum(X, es; q = 2)
# find and fit largest linear region:
DeltaC = slopefit(log.(es), log.(Cs))[1]
DeltaC # the correlation dimension
# Estimate entropies and fit dimension
DeltaH = generalized_dim(X, es; q = 1)
# Estimate local dimensions via EVT
p = 0.99 # quantile probability
DeltaEi = extremevaltheory_dims(X, p)
\end{lstlisting}

\section{Dynamic rules of known systems}
\label{app:systems}
All dynamical systems used for generating data are listed in  Table.~\ref{tab:all_systems_rules}.

\begin{table*}[t]
    \centering
    \begin{tabularx}{\textwidth}{llll}\toprule
        System & Dynamical rule &  Initial conditions & Parameters \\  \midrule
        H\'enon map & \parbox{.4\textwidth}{\begin{equation*}
            x_{n+1} = 1 - ax^2_n+y_n,\quad y_{n+1} = bx_n
            \end{equation*}} & $(0.08, 0.12)$ & $a = 1.4$, $b=0.3$ \\\midrule
        Kaplan-Yorke map & \parbox{.4\textwidth}{\begin{equation*}
        x_{n+1} = 2x_n \% 1, \quad y_{n+1} = \lambda y_n + \cos(4\pi x_n)
        \end{equation*}} & $(0.15, 0.2)$ & $\lambda = 0.2$ \\\midrule
        Towel map & \parbox{.4\textwidth}{\begin{subequations}
            \begin{align*}
            x_{n+1} &= 3.8 x_n (1-x_n) -0.05 (y_n +0.35) (1-2z_n),\\
            y_{n+1} &= 0.1 \left( \left( y_n +0.35 \right)\left( 1+2z_n\right) -1 \right)
            \left( 1 -1.9 x_n \right),\\
            z_{n+1} &= 3.78 z_n (1-z_n) + b y_n
            \end{align*}
            \end{subequations}} & $(0.085, -0.121, 0.075)$ & \\\midrule
        H\'enon-Heiles & \parbox{.4\textwidth}{\begin{equation*}
            \dot{x} = p_x,\;
            \dot{y} = p_y ,\;
            \dot{p}_x = -x -2 xy ,\;
            \dot{p}_y = -y - (x^2 - y^2)
            \end{equation*}} & $(0,-0.25,0.42081,0)$ & \\\midrule
        Coupled logistic maps & \parbox{.4\textwidth}{\begin{subequations}\begin{align*}
            u_{n+1}^{(i)} &= 4 v_n^{(i)}(1- v_n^{(i)}),\\
            v_n^{(i)} &= u_n^{(i)} + k\left(u_n^{(i-1)} - 2u_n^{(i)} + u_n^{(i+1)}\right)
            \end{align*}\end{subequations}} & $(0.1, \dots, 0.9)$ & $k = 0.1$\\\midrule
        R\"ossler & \parbox{.4\textwidth}{\begin{equation*}
            \dot{x} = -y-z,\quad
            \dot{y} = x+ay \quad
            \dot{z} = b + z(x-c)
            \end{equation*}} & $(0.1, -0.2, 0.1)$ & $a=b=0.2, c=5.7$\\ 
            &  & periodic parameters: & $a=b=0.2, c=3$\\\midrule
        Lorenz-96 & \parbox{.4\textwidth}{\begin{equation*}
            \dot{x_i} = (x_{i+1}-x_{i-2})x_{i-1} - x_i + F
            \end{equation*}} & $(j\times 0.1 \,\mathrm{for}\, j \in 0\ldots D-1)$ & $F = 24$\\
        \bottomrule
    \end{tabularx}
    \caption{Description of various systems considered in this paper.}
    \label{tab:all_systems_rules}
\end{table*}

\section{Delay embedding parameters for experimental data}
\label{app:delayparameters}

The method of Kraemer et al.~\cite{Kraemer2021} finds optimal delay times that may not be equispaced. The amount of delay times found is equal to the embedding dimension. The delay times found are listed in the below list for each system (delay times are always integers, in units of the sampling time). For Vostok and ``nifty50'' data we used traditional techniques of optimizing delay time and embedding dimension individually via Cao's method and minimum of mutual information, because the method of Ref.~\cite{Kraemer2021} (correctly) yields that the data do not accommodate a proper embedding.

\begin{itemize}
    \item ``electroch. 1'': (0, 26, 13, 5, 20)
    \item ``electroch. 2'': (0, 25, 16, 148, 138, 87, 60, 105)
    \item ``Shinriki'': (0, 19, 38, 57)
    \item ``nifty50'': (0, 43, 86, 129, 172, 215)
    \item ``vostok'': (0, 50, 100, 150, 200, 250, 300)
    \item ``double pendulum'': (0, 51, 25, 39, 12)
    \item ``Roessler'': (0, 6, 3, 14)
    \item ``EEG IBI'':  (0, 13, 26, 39, 7)
\end{itemize}

\section{More plots for significance of EVT}
\label{app:more_significance}

\begin{figure}[t]
    \centering
    \includegraphics[width=\columnwidth]{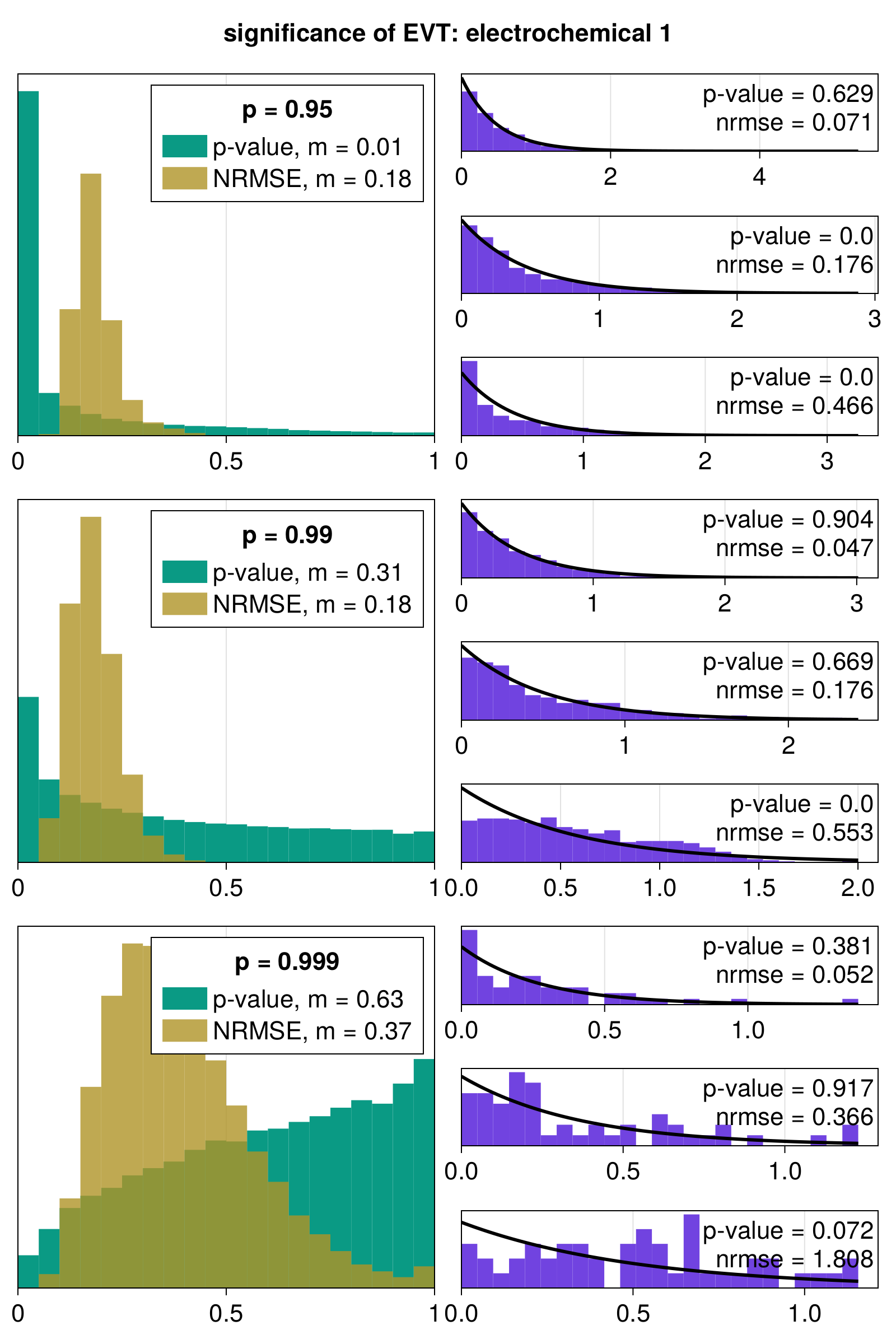}
    \caption{Same as Fig.~\ref{fig:evt_significance} but now for the experimental dataset ``electrochemical 1''.}
    \label{fig:evt_significance2}
\end{figure}

\begin{figure}[t]
    \centering
    \includegraphics[width=\columnwidth]{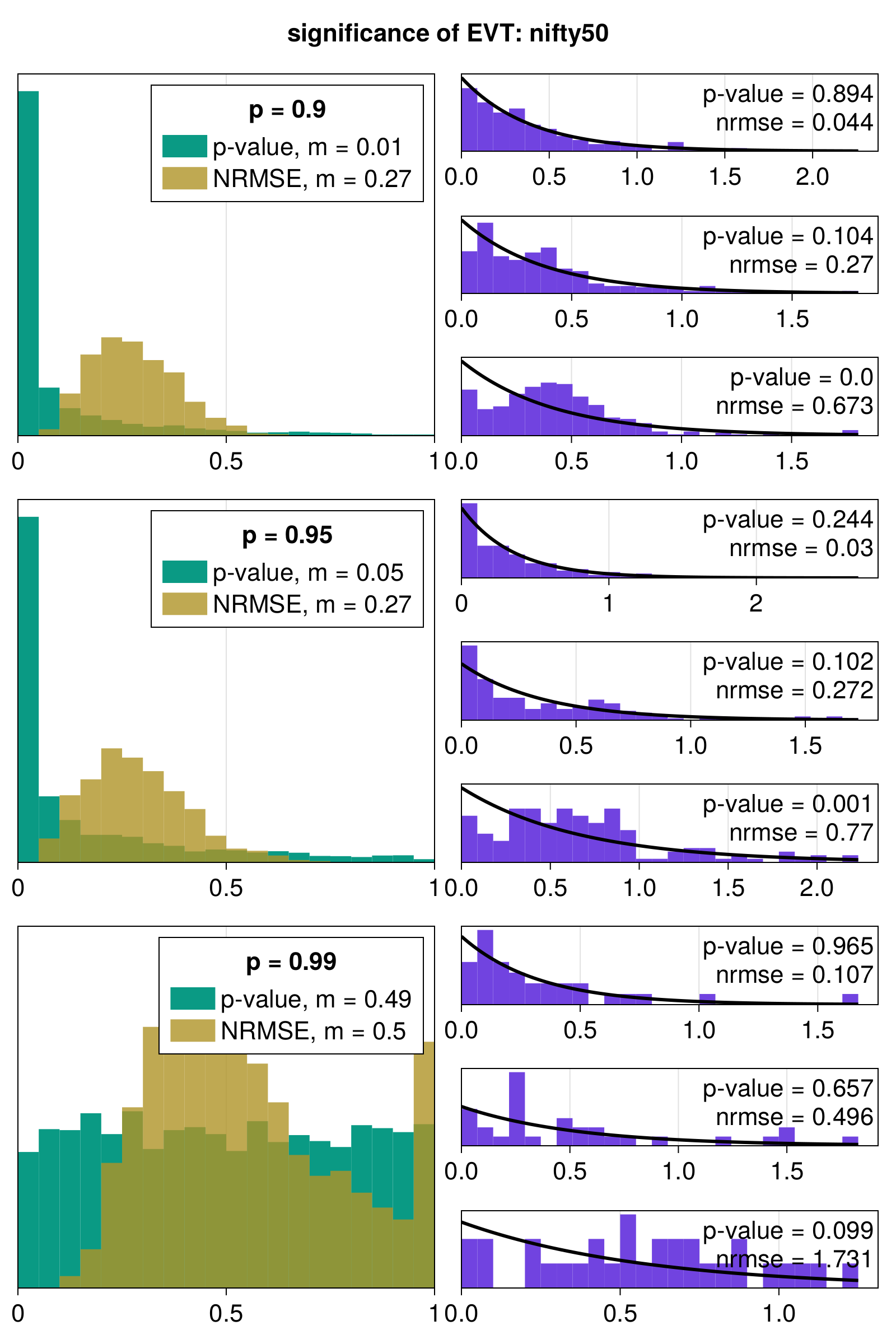}
    \caption{Same as Fig.~\ref{fig:evt_significance} but now for the ``nifty50'' stock market timeseries (embedded in 6 dimensional space). Note that this timeseries is only 3125 samples long, hence we used much smaller $p$ when compared to other plots. The proportion $N(1-p)$ however remains similar.}
    \label{fig:evt_significance3}
\end{figure}

See Figures \ref{fig:evt_significance2} and  \ref{fig:evt_significance3}.

\section{Increasing embedding dimension of real world data}
\label{app:embedding_real}

\begin{figure}[t]
    \centering
    \includegraphics[width=\columnwidth]{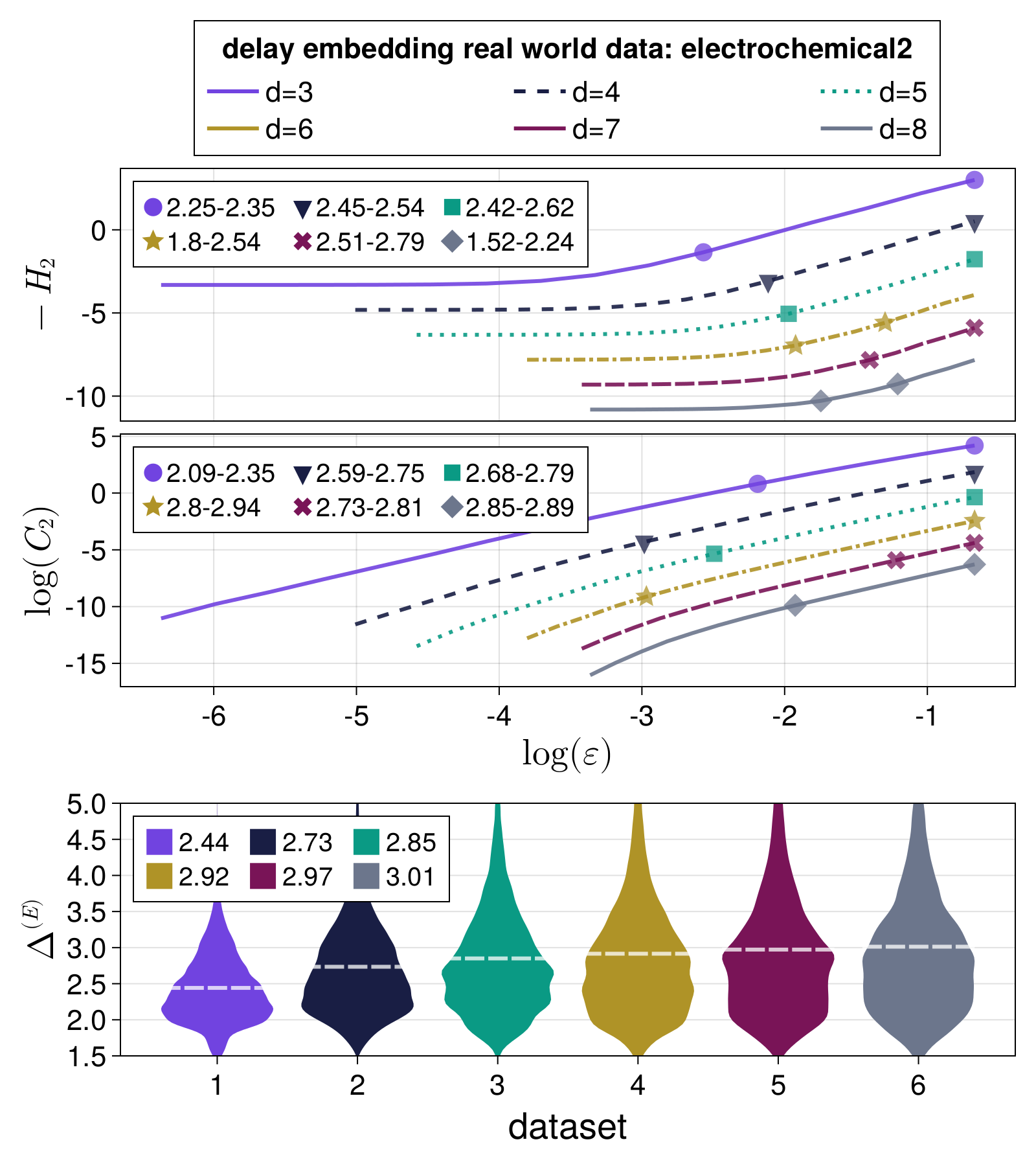}
    \caption{Analysis of behavior of FD estimators as we increase embedding dimension of real world data (delay time was estimated as the minimum of self-mutual information). Here for the ``electrochemical 2'' dataset.}
    \label{fig:embedding_experimental1}
\end{figure}

\begin{figure}[t]
    \centering
    \includegraphics[width=\columnwidth]{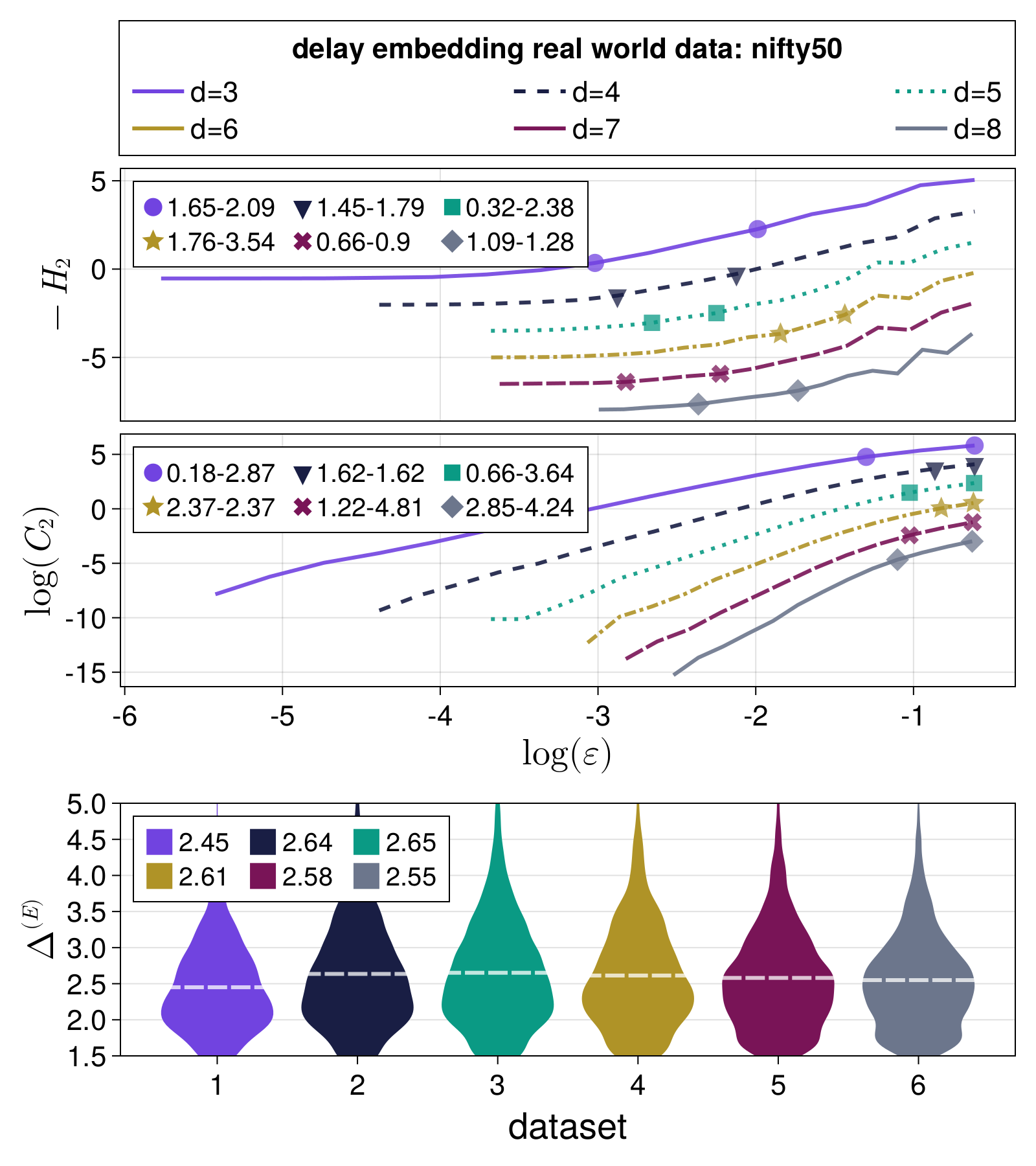}
    \caption{Same as in Fig.~\ref{fig:embedding_experimental1} but for the ``nifty50'' dataset. Note the convergence of the EVT FD into a small value of $\Delta^{(E)} \approx 2.5$ already for $d \ge 5$. Practically identical results are obtained with the ``vostok'' dataset: $C_2$ fails to converge to any FD with increasing $d$, while EVT converges to $\Delta^{(E)} \approx 3.3$ for $d \ge 5$.}
    \label{fig:embedding_experimental2}
\end{figure}

In Figures \ref{fig:embedding_experimental1} and \ref{fig:embedding_experimental2} we perform what is known as standard practice when estimating FDs: increasing the embedding dimension iteratively until a convergence of FD appears at the largest scales of $\varepsilon$~\cite{KantzSchreiber2003}. Especially for this subsection we estimate the slope of the \emph{right-most} linear scaling region (i.e., the one at the largest $\varepsilon$), as opposed to the slope of the \emph{largest} linear region. That is because in real data it is this slope that should indicate the FD of the underlying deterministic dynamics, if any exist (see Sec.~\ref{sec:noise}).

Indeed, for the ``electrochemical 2'' dataset, the convergence of $\Delta^{(C)}_2$ becomes apparent very quickly. On the other hand, for ``nifty50'' there is no convergence (we computed $C_2$ up to $d=13$, not shown). The results of $H_2$ for ``nifty50'' are inaccurate because the timeseries has only 3125 points; they should not be trusted (but anyways they do not show any convergence either).

We also perform the same analysis for the EVT approach, which once again re-enforces that the method fails to understand the the stock market timeseries should not have a convergent dimension. Instead, the dimension estimates converge very rapidly with increasing $d$.

\section*{Acknowledgements}

We thank: Ignacio Del Amo for an initial draft code implementation of the extreme value theory dimension estimator, and for providing the simple proof of non-equivalence between the box counting and Hausdorff dimensions; three independent reviewers for constructive criticism that greatly improved the quality of the manuscript; Nils Bertschinger for helpful discussions regarding distributions of p-values; Gabrielle Messori and Davide Faranda for discussions and clarifications regarding the usage of extreme value theory for estimating a fractal dimension, including its p-value test of Ref.~\cite{Faranda2017}.

\section*{REFERENCES}
\bibliography{REFERENCES}

\end{document}